\documentclass[
  a4paper,        
  twoside,        
  BCOR=25mm,      
  DIV=16,         
  open=right,     
]{scrreprt}

\usepackage[a-2b]{pdfx}

\hypersetup{
  bookmarksnumbered   = true,        
  bookmarksopen       = true,        
  bookmarksopenlevel  = 0,           
  bookmarksdepth      = 3,           
  pdfpagemode         = UseOutlines, 
  pdfstartview        = Fit,         
  plainpages          = false,       
  pdfpagelabels       = true,        
  colorlinks          = true,        
  linkcolor           = black,       
  citecolor           = black,       
  urlcolor            = black,       
  filecolor           = black,
}

\usepackage[main=english,ngerman]{babel}

\usepackage{microtype}  

\usepackage{graphicx}
\usepackage{pdfcomment}

\usepackage{amssymb}
\usepackage{mathtools}
\usepackage{physics}

\usepackage[]{todonotes}

\usepackage[automark,headsepline=false, footsepline=false]{scrlayer-scrpage}
\newcommand{\currentreprint}{}  
\defpairofpagestyles{reprint}{%
  \clearpairofpagestyles
  \ofoot*{\pagemark}
  \cfoot*{\footnotesize\currentreprint}
}

\usepackage{cleveref}

\usepackage{caption}
\usepackage{makecell}

\usepackage{pdfpages}

\usepackage[numbers,sort&compress]{natbib}

\makeatletter
\g@addto@macro\bfseries{\boldmath}
\makeatother

\usepackage[indent=10pt]{parskip}

\newcommand{\m}[1]{\mathrm{#1}}

\let\phi\varphi
\let\epsilon\varepsilon

\begin{document}


\begin{titlepage}
\thispagestyle{empty}
\centering

\vspace*{1cm}

{\LARGE\bfseries Doubly heavy tetraquark spectroscopy from lattice QCD and efficient distillation using position-space sampling\par}

\vfill

{\Large\bfseries DISSERTATION\par}
\vspace{0.5cm}
zur Erlangung des akademischen Grades\\[0.2cm]
{\large doctor rerum naturalium\\(Dr. rer. nat.)\par}
\vspace{0.3cm}
im Fach Physik: Spezialisierung: Theoretische Physik\\[0.2cm]
eingereicht an der\\[0.2cm]
Mathematisch-Naturwissenschaftlichen Fakultät\\
der Humboldt-Universität zu Berlin

\vspace{0.8cm}

von\\[0.3cm]
{\large\bfseries M.\,Sc.\ Andres Stump\par}

\vfill

\begin{flushleft}
Präsidentin der Humboldt-Universität zu Berlin:\\
Prof.\ Dr.\ Julia von Blumenthal\\[0.3cm]

Dekan der Mathematisch-Naturwissenschaftlichen Fakultät:\\
Prof. Dr. Emil List-Kratochvil\\[0.4cm]

Gutachter/innen:
\begin{enumerate}\setlength{\itemsep}{0pt}\setlength{\parskip}{0pt}
  \item Prof.\ Dr.\ Agostino Patella
  \item Prof.\ Dr.\ Rainer Sommer
  \item Prof.\ Dr.\ Saša Prelovšek
\end{enumerate}

\vspace{0.3cm}
Tag der mündlichen Prüfung: \textit{28.\ Oktober 2026}
\end{flushleft}

\end{titlepage}

\thispagestyle{empty}
\cleardoublepage

\pagenumbering{roman}
\setcounter{page}{1}

\cleardoublepage

\null\vfill\itshape

\begin{flushright}
	I dedicate this thesis to Jesus Christ, the source of all wisdom, strength, health, and knowledge, who carried me through every challenge of this journey.
\end{flushright}
\thispagestyle{empty}
\upshape\cleardoublepage


\selectlanguage{english}
\chapter*{Abstract}
\addcontentsline{toc}{chapter}{Abstract}

Various exotic hadrons have been discovered in recent decades. Of particular interest are doubly heavy tetraquarks, such as the experimentally observed $T_{cc}$, with minimal quark content $cc\overline{u}\overline{d}$, and its as-yet-unobserved bottom counterpart $T_{bb}$. They appear in $DD^*$ and $BB^*$ scattering, respectively, with quantum numbers $I(J^P) = 0(1^+)$.

At the low energies relevant for hadrons, the strong interaction cannot be treated perturbatively. Lattice QCD provides the only known first-principles approach in this regime. It formulates the theory on a discrete spacetime lattice in a finite volume, which allows it to be evaluated numerically. The finite volume renders the energy spectrum discrete, and Lüscher's finite-volume quantization condition relates this spectrum to the infinite-volume scattering amplitude, from which the properties of near-threshold states can be extracted. Obtaining an accurate finite-volume spectrum using the variational method requires a broad basis of interpolating operators. For four-quark systems such as $DD^*$ and $BB^*$, this includes both bilocal two-meson operators, resembling weakly bound ``meson molecules'', and local tetraquark operators, better suited to compact, deeply bound states.

Distillation is an efficient framework for computing two-point functions of bilocal operators. The computational cost of local multiquark operators, however, scales with a high power of the physical volume and becomes prohibitive in large volumes. To address this, we developed a position-space sampling method, in which the contraction is carried out in position space and the required smeared propagators are evaluated on randomly shifted sparse grids rather than on the full spatial lattice. The random offset yields an unbiased estimator of the two-point function, while the regular grid keeps the variance low. This makes local multiquark operators with four or more (anti)quarks affordable within distillation.

We applied this method on two gauge ensembles at the $\m{SU}(3)$-flavour-symmetric point, corresponding to a pion mass of $m_\pi \approx 420\;\m{MeV}$, treating the bottom quark with a relativistic heavy-quark action. For the $DD^*$ system, our operator-importance analysis demonstrates that local tetraquark operators are important, alongside bilocal two-meson operators, for obtaining an accurate finite-volume spectrum, and thus an accurate scattering amplitude. A preliminary $s$-wave Lüscher analysis identifies the $T_{cc}$ as a virtual bound state $17(3)\;\m{MeV}$ below threshold. In an exploratory study of the $BB^*$ system, we obtain $T_{bb}$ binding energies of around $55\;\m{MeV}$ and find evidence for an additional shallow virtual bound state about $2\;\m{MeV}$ below threshold.


\selectlanguage{ngerman}
\chapter*{Zusammenfassung}
\addcontentsline{toc}{chapter}{Zusammenfassung}

In den letzten Jahrzehnten wurden verschiedene exotische Hadronen entdeckt. Von besonderem Interesse sind doppelt schwere Tetraquarks wie das experimentell beobachtete $T_{cc}$ mit minimalem Quarkinhalt $cc\overline{u}\overline{d}$ und sein bislang nicht beobachtetes Bottom-Gegenstück $T_{bb}$. Sie treten in der $DD^*$- bzw.\ $BB^*$-Streuung mit den Quantenzahlen $I(J^P) = 0(1^+)$ auf.

Bei den für Hadronen relevanten niedrigen Energien lässt sich die starke Wechselwirkung nicht perturbativ behandeln. Die Gitter-QCD bietet den einzigen bekannten Ab-initio-Zugang in diesem Bereich. Sie formuliert die Theorie auf einem diskreten Raumzeitgitter in einem endlichen Volumen, wodurch sie numerisch ausgewertet werden kann. Das endliche Volumen führt zu einem diskreten Energiespektrum, und Lüschers Quantisierungsbedingung verknüpft dieses Spektrum mit der Streuamplitude im unendlichen Volumen, aus der sich die Eigenschaften schwellennaher Zustände ableiten lassen. Um mit der Variationsmethode ein akkurates Spektrum im endlichen Volumen zu erhalten, ist eine divese Basis von Interpolationsoperatoren erforderlich. Für Vier-Quark-Systeme wie $DD^*$ und $BB^*$ umfasst dies sowohl bilokale Zwei-Meson-Operatoren, die schwach gebundenen „Mesonmolekülen“ ähneln, als auch lokale Tetraquark-Operatoren, die besser für kompakte, tief gebundene Zustände geeignet sind.

Die Distillation ist ein effizientes Verfahren zur Berechnung von Zweipunktfunktionen bilokaler Operatoren. Der Rechenaufwand für lokale Multiquark-Operatoren skaliert jedoch mit einer hohen Potenz des physikalischen Volumens und wird bei großen Volumina prohibitiv hoch. Um dieses Problem zu lösen, entwickelten wir ein Ortsraum-Sampling Verfahren, bei dem die Kontraktion im Ortsraum durchgeführt wird und die erforderlichen geschmierten Propagatoren nicht auf dem vollständigen räumlichen Gitter, sondern auf zufällig verschobenen dünn besetzten Gittern berechnet werden. Die zufällige Verschiebung liefert einen erwartungstreuen Schätzer der Zweipunktfunktion, während das regelmäßige Gitter die Varianz gering hält. Dadurch werden lokale Multiquark-Operatoren mit vier oder mehr (Anti-)Quarks im Rahmen der Distillation rechnerisch handhabbar.

Wir wendeten diese Methode auf zwei Eichfeld-Ensembles am $\m{SU}(3)$-flavoursymmetrischen Punkt an, was einer Pionmasse von $m_\pi \approx 420\;\m{MeV}$ entspricht, wobei wir das Bottom-Quark mit einer relativistische Schwere-Quark-Wirkung behandelten. Für das $DD^*$-System zeigt unsere Wichtigkeitsanalyse der Operatoren, dass lokale Tetraquark-Operatoren neben bilokalen Zwei-Meson-Operatoren wichtig sind, um ein genaues Spektrum im endlichen Volumen und damit eine genaue Streuamplitude zu erhalten. Eine vorläufige $s$-Wellen-Lüscher-Analyse identifiziert das $T_{cc}$ als einen virtuellen gebundenen Zustand $17(3)\;\m{MeV}$ unterhalb der Schwelle. In einer explorativen Untersuchung des $BB^*$-Systems erhalten wir $T_{bb}$-Bindungsenergien von etwa $55\;\m{MeV}$ und finden Hinweise auf einen zusätzlichen, schwach gebundenen virtuellen Zustand etwa $2\;\m{MeV}$ unterhalb der Schwelle.


\cleardoublepage

\selectlanguage{english}
\cleardoublepage

\tableofcontents

\cleardoublepage

\pagenumbering{arabic}
\setcounter{page}{1}

\chapter{Introduction}

Quantum chromodynamics (QCD) is the quantum field theory of the strong interaction, one of the four fundamental forces of nature. It describes the interaction between quarks and gluons, which are the elementary constituents of nuclear matter. In QCD, quarks carry a charge called colour (red, green or blue) and interact by exchanging gluons, the massless gauge bosons of the theory. Since QCD is a non-Abelian gauge theory based on $\m{SU}(3)$, the gluons themselves carry colour charge. This gives rise to a self-coupling of the gluons, which results in asymptotic freedom and confinement. As a consequence, quarks and gluons are confined into colour-neutral composite particles called hadrons. These are the states that can be observed in experiments.

Since QCD is asymptotically free, the strong coupling constant $\alpha_s$ decreases as the energy scale is increased. At large momentum transfers $Q \gg 1\;\mathrm{GeV}$, $\alpha_s(Q)$ is small, and perturbation theory in $\alpha_s$ can be applied to make precise predictions from first principles. The theory has been tested extensively in this regime through collider experiments, successfully describing, for example, the running of the coupling and deep-inelastic-scattering cross-sections~\cite{ParticleDataGroup:2024cfk}. At low energies, however, $\alpha_s$ grows large, and perturbation theory breaks down. This is the regime in which quarks and gluons are confined into hadrons, and in which describing the spectrum and properties of hadrons from the QCD Lagrangian is a highly non-trivial task. Progress in this domain has historically been driven largely by experimental discoveries, with theoretical understanding trailing behind.

A practical framework for organizing the hadron spectrum at low energies is the quark model, proposed independently by Gell-Mann and Zweig~\cite{Gell-Mann:1964ewy,Zweig:1964ruk}. In its simplest form, it classifies hadrons as composite objects built from quarks. Exploiting the approximate $\m{SU}(3)$ flavour symmetry between the three lightest quarks $u$, $d$, and $s$, hadrons can be organized into multiplets according to their flavour quantum numbers. In the minimal quark model, mesons are described as quark-antiquark ($q\overline{q}$) pairs and baryons as three-quark ($qqq$) states. This scheme has been remarkably successful: it famously predicted the existence and properties of the $\Omega^-$ baryon, which was subsequently discovered~\cite{Barnes:1964pd}. A more refined treatment is provided by the constituent quark model~\cite{Entem:2025bqt}, in which the current quarks of the QCD Lagrangian acquire a much larger constituent mass from their surrounding gluon cloud. This approach has enabled quantitative descriptions of the meson and baryon spectra, nucleon-nucleon interactions, and the magnetic properties of hadrons.

Despite the success of the quark model, QCD in principle allows for a far richer spectrum. The only hard constraint from the theory is that observable particles must be colour-neutral (colour-singlet) objects. Any combination of quarks and gluons satisfying this requirement is, in principle, allowed. This includes states with gluonic degrees of freedom --- hybrid mesons and purely gluonic bound states called glueballs --- as well as states with quark content beyond $q\overline{q}$ or $qqq$. The latter include tetraquarks ($qq\overline{q}\overline{q}$), pentaquarks ($qqqq\overline{q}$), and multiquark states with more than five quarks. Such particles are collectively referred to as exotic hadrons~\cite{Husken:2024rdk,Chen:2022asf}, and their existence has been discussed since the inception of the quark model~\cite{Gell-Mann:1964ewy,Zweig:1964ruk}. While experimental evidence was lacking for decades, the landscape has changed dramatically over the past twenty years.

The era of exotic hadron spectroscopy began in 2003, when the Belle collaboration discovered the $\chi_{c1}(3872)$~\cite{Belle:2003nnu}, a charmonium-like state, i.e.\ one containing a charm-anticharm pair. Its unusual proximity to the $D^0\overline{D}^{*0}$ threshold and its small decay width were difficult to reconcile with a simple $c\overline{c}$ interpretation, suggesting a more complex internal structure. In the following years, various similar states containing $c\overline{c}$ or $b\overline{b}$ pairs were discovered at $B$-factories and at the LHC, none of which fitted naturally into the conventional quark model. Their properties suggest, however, that they are tetraquarks consisting of two heavy and two light (anti)quarks. In 2015, the LHCb collaboration reported the first observation of pentaquark states with hidden charm, i.e.\ also containing a $c\overline{c}$ pair~\cite{LHCb:2015yax}. Further exotic structures with open charm, i.e.\ with nonzero charm quantum number, followed: in 2020 LHCb announced the observation of the first tetraquarks with single open charm~\cite{LHCb:2020bls,LHCb:2020pxc} and double open charm~\cite{LHCb:2021vvq}. The latter is called the doubly charmed tetraquark $T_{cc}(3875)^+$. In 2011, the WASA-at-COSY experiment reported evidence for the $d^*(2380)$, a dibaryon resonance that is an excited state of the deuteron~\cite{PhysRevLett.106.242302}. Various other exotic hadrons have been added to the list since then.

A particularly interesting example among these discoveries is the aforementioned $T_{cc}^+$, which was studied extensively by LHCb~\cite{LHCb:2021vvq,LHCb:2021auc}. It appears in the $D^0D^0\pi^+$ mass spectrum just below the $D^{*+}D^0$ threshold, with quantum numbers $I(J^P) = 0(1^+)$ and minimal quark content $cc\overline{u}\overline{d}$. The $T_{cc}^+$ is remarkable for being extraordinarily narrow: its measured width is approximately $48\;\mathrm{keV}$, making it a particularly long-lived exotic hadron. This narrowness arises from the strongly restricted decay phase space into $D^0D^0\pi^+$: since its mass lies slightly below the $D^{*+}D^0$ threshold, the decay cannot proceed via a real, on-shell $D^{*+}$, and must instead go through a virtual $D^{*+}$. As a consequence, the strong decay is highly suppressed. These properties make the $T_{cc}^+$ a particularly well-suited system for both experimental and theoretical investigation, and it serves as a central focus of this thesis.

A central open question concerning the $T_{cc}^+$, and exotic hadrons more broadly, is that of its inner structure. Its proximity to the $D^{*+}D^0$ threshold suggests an interpretation as a loosely bound hadronic ``molecule'', in which two colour-singlet mesons are held together by a weakly attractive force. An alternative picture is that of a compact diquark-antidiquark state, in which a tightly bound $cc$ diquark and a $\overline{u}\overline{d}$ antidiquark are confined within a single hadron. These two scenarios are not mutually exclusive, and the physical state may be a superposition of both.

The doubly bottom analogue of the $T_{cc}$, denoted $T_{bb}$, has the same quantum numbers $I(J^P) = 0(1^+)$ and minimal quark content $\overline{b}\overline{b}ud$, but has not yet been observed experimentally. The large bottom-quark mass is expected to produce significantly stronger binding than in the charm sector: theoretical predictions, using a variety of methods, place the $T_{bb}$ $50$--$190\;\mathrm{MeV}$ below the $BB^*$ threshold~\cite{Bicudo:2012qt,Francis:2016hui,Leskovec:2019ioa,Hudspith:2023loy,Aoki:2023nzp,Alexandrou:2024iwi,Colquhoun:2024jzh,Tripathy:2025vao,Vujmilovic:2025czt,Brown:2012tm,Bicudo:2015vta,Hoffmann:2026mcz,Bicudo:2016ooe}. Discovering this state experimentally would require high collision energies and the reconstruction of complex decay signatures involving semileptonic decays and bottom-quark hadrons. Theoretical calculations can nevertheless make definitive predictions for its properties, and such investigations of the $T_{bb}$ and of $BB^*$ scattering form another major component of this thesis. 

The existence and properties of exotic hadrons provide direct tests of QCD in the strongly coupled regime. Reliable, model-independent theoretical predictions are essential for interpreting experimental discoveries and guiding future particle searches.

The only known method for making first-principles, non-perturbative predictions from QCD is lattice QCD~\cite{Wilson:1974sk}. In lattice QCD, the theory is formulated on a discrete Euclidean spacetime lattice with lattice spacing $a$ and a finite volume. On this lattice, the path integral of QCD becomes finite-dimensional and is evaluated numerically using Markov chain Monte Carlo methods. This approach makes no model assumptions and is, in principle, systematically improvable. However, the continuum limit $a \to 0$ has to be taken, finite-volume effects have to be accounted for, and quantities computed in Euclidean spacetime have to be related to their Minkowski-spacetime counterparts. Lattice QCD has achieved impressive successes, for example in precisely determining the strong coupling constant, calculating quark masses, and computing the properties of single hadrons that are stable under the strong interaction. These properties include the hadron spectrum, decay constants, and form factors. 

One reason why the investigation of such QCD-stable hadrons, such as pions and nucleons, has been particularly successful is the following. Finite-volume effects for these states are exponentially suppressed as $e^{-m_\pi L}$, i.e.\ by the product of the box size $L$ and the smallest mass scale, the pion mass $m_\pi$~\cite{Luscher:1985dn}. Consequently, by choosing a sufficiently large box size, these effects can be neglected. Most hadrons, however, are not stable under the strong interaction; they manifest as resonances or bound states of multi-hadron systems. For such states, the finite-volume effects decay only with inverse powers of the box size and have to be taken into account~\cite{Luscher:1986pf}. Moreover, there is no longer a clear one-to-one correspondence between the finite-volume and infinite-volume states.

Within hadron spectroscopy, the rigorous method for investigating such bound states or resonances is to analyse the poles in the corresponding scattering amplitude. However, extracting this amplitude from lattice QCD data is difficult, as it is inherently a dynamical, infinite-volume quantity. The Euclidean signature prohibits the direct investigation of dynamical properties, and the finite volume does not allow the preparation of asymptotic states. Instead, the states in a finite volume form a discrete spectrum as a consequence of the quantization. An infinite-volume extrapolation, which would in any case be computationally prohibitive, does not resolve the issue of the lack of asymptotic states.

One way of dealing with this limitation is to use Lüscher's finite-volume quantization conditions~\cite{Luscher:1986pf,Luscher:1990ux}. These relate the low-lying finite-volume energies to the scattering amplitude in infinite volume. In their original form, these relations applied to elastic two-particle scattering in the rest frame. Since then, the formalism has been extended to moving frames~\cite{Rummukainen:1995vs,Kim:2005gf,Leskovec:2012gb,Gockeler:2012yj}, to coupled channels with particles of nonzero spin~\cite{He:2005ey,Liu:2005kr,Guo:2012hv,Briceno:2014oea}, and to three-particle systems~\cite{Briceno:2012rv,Hammer:2017uqm,Hammer:2017kms}. These tools allow one to rigorously extract resonance poles and near-threshold bound states from finite-volume lattice QCD data.

Applying the Lüscher method requires accurate determinations of the finite-volume energy levels from lattice QCD correlation functions. For single-particle systems, one can extract the ground-state energy from the exponential decay of a two-point function at large Euclidean times, where the excited-state contributions to its spectral decomposition have died out. Multi-hadron systems --- in this work we deal only with systems of two mesons --- have a much denser spectrum, resulting in stronger excited-state contamination. In addition, one is often interested in extracting excited states as well. The state-of-the-art method for extracting the low-lying energy levels is the variational method~\cite{Luscher:1990ck,Blossier:2009kd}. It uses a basis of interpolating operators, which carry the quantum numbers of the hadronic system of interest, to construct a correlator matrix. By solving a generalized eigenvalue problem (GEVP) for this matrix, optimized operators are constructed as linear combinations of the basis operators. These optimized operators predominantly overlap with a single energy level, removing the contamination from other levels. With this method, one can in principle extract as many energy levels as there are operators in the basis. In practice, however, one uses a larger operator basis than the number of levels one wishes to extract, in order to reliably suppress excited-state contamination.

The choice of operators is a delicate task, as the inner structure of the finite-volume states is not well known, especially for exotic hadrons. A variety of operators with different spin, colour and spatial structures should therefore be included in the basis to ensure a strong overlap with the low-lying states. This includes nonlocal scattering operators, which resemble weakly bound ``molecular'' states, as well as local multiquark operators, which are better suited to more deeply bound states. Failing to include the relevant operators may result in an inaccurate finite-volume spectrum~\cite{Wilson:2015dqa}.

Two-point functions of nonlocal scattering operators have a numerical structure very different from those of local operators, which makes combining the two types into a single correlator matrix challenging. An efficient framework for computing two-point functions of nonlocal scattering operators in lattice QCD is distillation~\cite{HadronSpectrum:2009krc}. By applying Laplacian Heaviside (LapH) smearing, the quark fields are projected onto a low-dimensional subspace spanned by the lowest eigenvectors of the spatial gauge-covariant Laplacian. This makes the computation of the spatial all-to-all propagators, required for two-point functions of nonlocal scattering operators, tractable. Two-point functions of local multiquark operators with four or more (anti)quarks, on the other hand, can be computed more easily using standard lattice QCD approaches such as the point-to-all method. Combining both operator types efficiently within a single framework is therefore a non-trivial problem.

The first original contribution of this thesis is the development of a position-space sampling method within the distillation framework. It allows the efficient computation of two-point functions involving both nonlocal scattering operators and local multiquark operators in distillation. The second contribution is the application of this method to two physically interesting systems. The first is the $DD^*$ scattering system with $I(J^P) = 0(1^+)$, in which the $T_{cc}$ tetraquark emerges. There, we investigate the importance of local tetraquark operators alongside bilocal meson-meson operators for extracting an accurate finite-volume spectrum. This yields improved finite-volume energy levels, on which we perform a Lüscher analysis to extract the $s$-wave scattering phase shift. Our position-space sampling method and the results for the $DD^*$ system have been published in~\cite{Stump:2025owq}. The second system is the $BB^*$ system with the same $I(J^P)$ quantum numbers, the results for which are presented here for the first time. We apply the position-space sampling method both to obtain the binding energy of the deeply bound $T_{bb}$ and to investigate $BB^*$ scattering close to threshold. To our knowledge, only one other, methodologically limited, study of $BB^*$ scattering close to threshold on the lattice exists, in contrast to the deeply bound $T_{bb}$, for which a number of binding-energy determinations already exist.

This thesis is organized as follows. Chapter~\ref{chapter:Lattice_QCD} provides a basic overview of the formulation of QCD on a lattice. In Chapter~\ref{chapter:Hadron_spectroscopy} we review hadron spectroscopy in lattice QCD, including the analysis of scattering amplitudes, the motivation and application of the Lüscher method, the computation of finite-volume energy levels and the construction of interpolating operators. Chapter~\ref{chapter:Distillation_method} describes the distillation framework and how it enables efficient spectroscopy calculations in lattice QCD. In Chapter~\ref{chapter:Position-space_sampling_method} we present the position-space sampling method that we developed within the distillation framework, which overcomes the shortcomings of distillation regarding local multiquark operators. In Chapter~\ref{chapter:Results} we present the results of this thesis: an investigation of the efficiency of our position-space sampling method, a study of the importance of local operators for the $DD^*$ system and the application of the position-space sampling method to the $DD^*$ and $BB^*$ systems to obtain scattering amplitudes. Finally, in Chapter~\ref{chapter:Conclusion} we conclude and outline directions for future work.
\chapter{Lattice QCD}
\label{chapter:Lattice_QCD}

This chapter provides an overview of the formulation of QCD on a discrete spacetime lattice, the theoretical framework underlying all calculations presented in this thesis. We begin with the continuum QCD Lagrangian, which we Wick-rotate to Euclidean spacetime. We then describe the discretization of the gauge action and the quark action, including the improvement terms needed to reduce discretization artefacts. Next we discuss the relativistic heavy quark action used to treat the bottom quark. Finally, we turn to the evaluation of the lattice QCD path integral via Monte Carlo methods, before briefly mentioning the setting of the lattice scale and the tuning of the bare parameters.

\section{QCD Lagrangian}
\label{sec:QCD_Lagrangian}

In the continuum, QCD is defined by the Minkowski-space Lagrangian density
\begin{equation}
  \label{eq:QCD_Lagrangian}
  \mathcal{L}(\psi,\overline{\psi},A) = \sum_{f=1}^{N_f}\,\overline{\psi}^{(f)}
    \left(\m{i}\gamma^\mu D_\mu - m_f\right)\psi^{(f)}
    - \frac{1}{2}\tr\!\left[F_{\mu\nu}F^{\mu\nu}\right],
\end{equation}
which describes the dynamics of the quark field $\psi$ and the gluon field $A_\mu$.

The second term in Eq.~\eqref{eq:QCD_Lagrangian} is the gauge sector. The gluon field $A_\mu$ is a Hermitian, traceless $3\times3$ matrix, conveniently expanded in the basis of Gell-Mann matrices. Its dynamics are governed by the field-strength tensor
\begin{equation}
  \label{eq:field_strength}
  F_{\mu\nu} = \partial_\mu A_\nu - \partial_\nu A_\mu + \m{i}g\left[A_\mu, A_\nu\right],
\end{equation}
where $g$ is the strong coupling constant. The commutator term $\m{i}g[A_\mu, A_\nu]$ is a direct consequence of the non-Abelian nature of the gauge group and gives rise to the self-interaction of the gluons.

The first term in Eq.~\eqref{eq:QCD_Lagrangian} is the fermion sector. The quark fields $\psi^{(f)}_{\substack{\alpha\\a}}$ carry a Dirac index $\alpha=1, 2, 3, 4$ and a colour index $a=1, 2, 3$, and they come in $N_f$ flavours (denoted by $f$), which have bare masses $m_f$. They couple to the gauge field through the gauge-covariant derivative
\begin{equation}
  \label{eq:covariant_derivative}
  D_\mu = \partial_\mu + \m{i}g A_\mu.
\end{equation}
The entire Lagrangian is invariant under local $\m{SU}(3)$ gauge transformations, where the 3 is the number of colours. In the limit of vanishing quark masses, it possesses an additional chiral symmetry, which is spontaneously broken by the vacuum; at nonzero quark masses, this symmetry is also broken explicitly. This plays a central role in the low-energy dynamics of QCD. Throughout this thesis, we work in the chiral representation of the gamma matrices $\gamma_\mu$, for which we specify the Euclidean variant in Section~\ref{sec:wick_rotation}.

The physical content of the theory is encoded in vacuum expectation values of operators $\mathcal{O}[\psi,\overline{\psi},A]$, which are given by the path integral
\begin{equation}
  \label{eq:vev_minkowski}
  \ev{\mathcal{O}} = \frac{1}{Z}\int\mathcal{D}\psi\,\mathcal{D}\overline{\psi}\,\mathcal{D}A\;
    \mathcal{O}[\psi,\overline{\psi},A]\; \m{e}^{\m{i}S[\psi,\overline{\psi},A]}.
\end{equation}
Here, $S[\psi,\overline{\psi},A] = \int\dd[4]{x}\,\mathcal{L}(\psi,\overline{\psi},A)$ is the action and
\begin{equation}
  \label{eq:partition_minkowski}
  Z = \int\mathcal{D}\psi\,\mathcal{D}\overline{\psi}\,\mathcal{D}A\; \m{e}^{\m{i}S[\psi,\overline{\psi},A]}
\end{equation}
is the partition function. Evaluating the Euclidean version of this path integral nonperturbatively is the central goal of lattice QCD.

\section{Discretization of QCD}
\label{sec:discretization}

Regularizing the theory by formulating it on a discrete spacetime lattice allows us to evaluate the path integral numerically. This requires two preparatory steps: a rotation to Euclidean spacetime, which allows for a probabilistic interpretation of the path integral, and the discretization of the gauge and fermion actions in a way that preserves gauge invariance.

\subsection{Wick rotation to Euclidean spacetime}
\label{sec:wick_rotation}

The Minkowski-space weight $\m{e}^{\m{i}S}$ in Eq.~\eqref{eq:vev_minkowski} is a pure phase and is therefore unsuitable for a probabilistic, Monte Carlo treatment. This issue is solved by a Wick rotation, in which the time coordinate is analytically continued to imaginary values, $t \to -\m{i}t$. Under this rotation, the oscillatory factor $\m{e}^{\m{i}S}$ turns into a real weight $\m{e}^{-S_E}$, where $S_E$ is the Euclidean action. This weight has the form of a Boltzmann factor, which establishes a direct correspondence between Euclidean quantum field theory and classical statistical mechanics. This makes the theory amenable to importance sampling with Monte Carlo methods.

In Euclidean spacetime, the metric becomes trivial, so that there is no longer a distinction between upper and lower Lorentz indices. In this spacetime, the gamma matrices $\gamma_\mu$ ($\mu = 1, 2, 3, 4$) satisfy the anticommutation relation
\begin{equation}
  \label{eq:gamma_anticommutation}
  \{\gamma_\mu,\gamma_\nu\} = 2\delta_{\mu\nu}\mathbf{1}_4,
\end{equation}
where we choose $\mu = 4$ for the time component and use lowercase Latin indices for the spatial components. We work in the chiral representation
\begin{equation}
  \label{eq:gamma_chiral}
  \gamma_i = \pmqty{0 & -\m{i}\sigma_i \\ \m{i}\sigma_i & 0}\;\;(i = 1,2,3), \qquad
  \gamma_4 = \pmqty{0 & -\mathbf{1}_2 \\ -\mathbf{1}_2 & 0}, \qquad
  \gamma_5 = \pmqty{\mathbf{1}_2 & 0 \\ 0 & -\mathbf{1}_2},
\end{equation}
where $\sigma_i$ are the Pauli matrices. We also define the antisymmetric tensor
\begin{equation}
  \label{eq:sigma_munu}
  \sigma_{\mu\nu} = \frac{\m{i}}{2}\left[\gamma_\mu, \gamma_\nu\right],
\end{equation}
which appears in the improvement term of the fermion action introduced below. Furthermore, we define the parity projector and the charge-conjugation matrix
\begin{equation}
  \label{eq:parity_proj_charge_conj_matrix}
	P_\pm = \frac{1}{2}(\mathbf{1}_4 \pm \gamma_4), \quad C = \m{i}\gamma_2\gamma_4,
\end{equation}
respectively, which are used for the construction of operators. For the remainder of this thesis, we work exclusively in Euclidean spacetime.

\subsection{Lattice}
\label{sec:lattice}

We discretize Euclidean spacetime on a hypercubic lattice
\begin{equation}
  \label{eq:lattice}
  \Lambda = \left\{\, an \;\middle|\; n \in \mathbb{Z}_{N_s}^3\times\mathbb{Z}_{N_t} \,\right\},
\end{equation}
where $\mathbb{Z}_n = \{0, 1, \dots, n-1\}$ and $a$ is the lattice spacing. The lattice has a finite four-volume $L^3\times T$ with spatial extent $L = aN_s$ and temporal extent $T = aN_t$. We write $x = (\vb*{x}, t)$ to separate the spatial and temporal components for points $x\in \Lambda$. The finite lattice spacing introduces a momentum cutoff of order $\tfrac{\pi}{a}$, so that the lattice acts as a gauge-invariant ultraviolet regulator.

The quark fields are defined on the lattice sites, whereas the gauge field is associated with the links between neighbouring sites, as described below. In the spatial directions, we impose periodic boundary conditions. In the temporal direction, we use two types of boundary conditions. The first uses periodic and antiperiodic boundary conditions for the gauge and fermion fields, respectively. The second uses open boundary conditions~\cite{Luscher:2011kk}, the motivation for which is briefly discussed in Section~\ref{sec:monte_carlo}.

\subsection{Gauge action}
\label{sec:gauge_action}

\begin{figure}
  \centering
  \includegraphics[width=\textwidth]{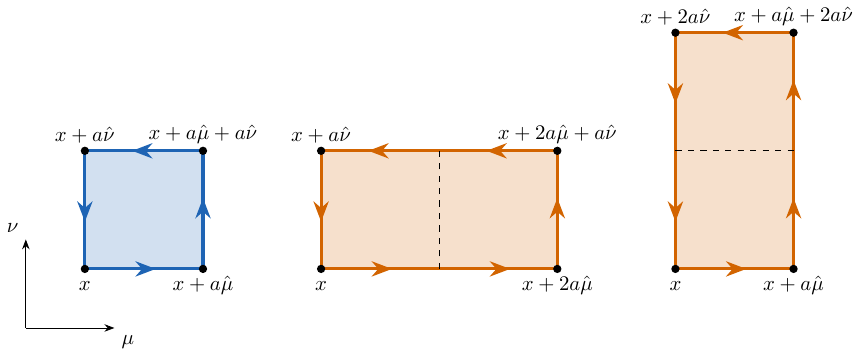}
  \caption{\label{fig:plaquette_rectangle}%
    Loops of link variables entering the improved gauge action: the elementary plaquette $U^{(p)}$ (left) and the planar rectangle $U^{(r)}$ (middle and right).}
\end{figure}

To construct a discrete, gauge-invariant action, it is advantageous to represent the gauge field by the compact link variables
\begin{equation}
  \label{eq:link_variable}
  U_\mu(x) = \exp\!\left(\m{i}ag A_\mu(x)\right),
\end{equation}
which are elements of the gauge group $\m{SU}(3)$ rather than of its algebra. They are associated with the link between $x$ and $x + a\hat{\mu}$, where $\hat{\mu}$ is the unit vector in direction $\mu$. Under a local gauge transformation $\Omega(x)\in \m{SU}(3)$, the link variables transform as
\begin{equation}
  \label{eq:link_transformation}
  U_\mu(x) \rightarrow \Omega(x)\,U_\mu(x)\,\Omega(x+a\hat{\mu})^\dagger.
\end{equation}
The simplest gauge-invariant object that can be built from link variables is the plaquette, the product of links around an elementary square (see Figure~\ref{fig:plaquette_rectangle}). It is given by
\begin{equation}
  \label{eq:plaquette}
  U_{\mu\nu}(x) = U_\mu(x)\,U_\nu(x+a\hat{\mu})\,U_\mu(x+a\hat{\nu})^\dagger\,U_\nu(x)^\dagger.
\end{equation}
From it, one constructs the Wilson gauge action
\begin{equation}
  \label{eq:wilson_gauge_action}
  S_G(U) = \frac{\beta}{6}\sum_{x\in\Lambda}\sum_{\mu,\,\nu=1}^{4}
    \tr\!\left[\mathbf{1} - U_{\mu\nu}(x)\right],
\end{equation}
where $\beta = 6/g^2$ is the bare inverse coupling. An expansion in the lattice spacing shows that this action reduces to the continuum Yang--Mills action up to corrections of order $a^2$.

The $O(a^2)$ artefacts of the Wilson gauge action can be systematically reduced using the Symanzik improvement programme~\cite{Symanzik:1983dc}. In this programme, suitable higher-dimensional operators are added to the action to cancel discretization effects order by order in $a$. Adding the planar rectangles (see Figure~\ref{fig:plaquette_rectangle}) to the simple plaquette action yields the improved gauge action
\begin{equation}
  \label{eq:improved_gauge_action}
  S_G[U] = \frac{\beta}{6}\sum_{x\in\Lambda}\left(
    c_0\sum_{p}\tr\!\left[\mathbf{1} - U^{(p)}(x)\right]
    + c_1\sum_{r}\tr\!\left[\mathbf{1} - U^{(r)}(x)\right]\right).
\end{equation}
The first sum in the bracket runs over all plaquettes $U^{(p)}(x)$ at the site $x$; it corresponds to the term in Eq.~\eqref{eq:wilson_gauge_action}. The second sum runs over all rectangles $U^{(r)}(x)$ at $x$. The coefficients must satisfy the condition $c_0 + 8c_1 = 1$ to recover the correct continuum limit. Since the symmetries of the lattice additionally forbid operators with odd orders in $a$, the leading lattice artefacts of this action are $O(a^4)$. This improvement, however, holds only at tree level; at loop level, $O(g^2a^2)$ artefacts can still appear.

The gauge ensembles used in this work were generated with the tree-level Symanzik-improved Lüscher--Weisz action~\cite{Luscher:1984xn}, which uses
\begin{equation}
  \label{eq:LW_coefficients}
  c_0 = \frac{5}{3}, \qquad c_1 = -\frac{1}{12}.
\end{equation}

\subsection{Quark action}
\label{sec:quark_action}

For a single quark flavour, the fermion action takes the bilinear form
\begin{equation}
  \label{eq:fermion_action}
  S_F(\psi,\overline{\psi},U) = a^4\sum_{x,y\in\Lambda}\overline{\psi}(x)\,D(x,y)\,\psi(y),
\end{equation}
where $D$ is the lattice Dirac operator. A naive discretization replaces the continuum covariant derivative by a symmetric finite difference, giving~\cite{Gattringer:2010zz}
\begin{equation}
  \label{eq:naive_dirac}
  D(x,y) = \gamma_\mu\nabla^\m{sym}_\mu(x,y) + m\,\delta_{x,y},
\end{equation}
where the gauge-covariant symmetric derivative is
\begin{equation}
  \label{eq:sym_derivative}
  \nabla^\m{sym}_\mu(x,y) = \frac{1}{2a}\left(U_\mu(x)\,\delta_{x+a\hat{\mu},y}
    - U_\mu(y)^\dagger\,\delta_{x-a\hat{\mu},y}\right),
\end{equation}
and a summation over the repeated index $\mu$ is understood.

This naive discretization suffers from the fermion-doubling problem: the lattice Dirac operator describes not one but $2^4 = 16$ degenerate fermion species, the so-called doublers. This is not an artefact of this particular discretization but a manifestation of the Nielsen--Ninomiya theorem, which states that no lattice Dirac operator can simultaneously be local, preserve chiral symmetry, and be free of doublers. One must therefore give up at least one of these properties.

Wilson's solution is to give up chiral symmetry by adding the Wilson term $-\tfrac{a}{2}\nabla^*_\mu\nabla_\mu$, built from the gauge-covariant forward and backward derivatives
\begin{align}
  \label{eq:forward_derivative}
  \nabla_\mu(x,y) &= \frac{1}{a}(U_\mu(x)\,\delta_{x+a\hat{\mu},y} - \delta_{x,y}), \\
  \label{eq:backward_derivative}
  \nabla^*_\mu(x,y) &= \frac{1}{a}(\delta_{x,y} - U_\mu(y)^\dagger\,\delta_{x-a\hat{\mu},y}).
\end{align}
The Wilson term gives the doublers a mass of order $1/a$, so that they decouple from the theory in the continuum limit $a\to0$. The price is an explicit breaking of chiral symmetry, which introduces discretization errors of order $a$.

\begin{figure}
  \centering
  \includegraphics[width=0.65\textwidth]{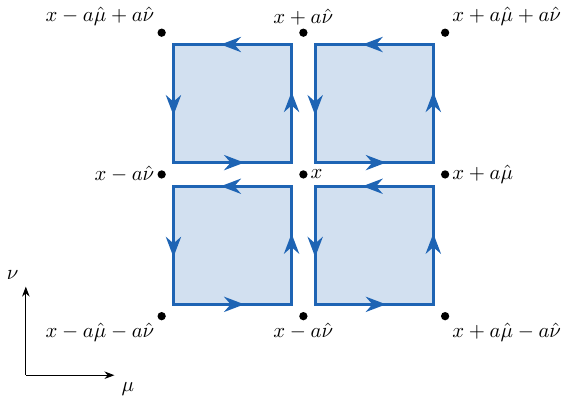}
  \caption{\label{fig:clover}%
    The four loops of link variables used to construct the clover discretization $\hat{F}_{\mu\nu}$ of the field-strength tensor at site $x$.}
\end{figure}

These $O(a)$ effects can be removed by applying the Symanzik improvement programme to the fermion action as well. On shell, i.e.\ when the classical equations of motion are satisfied, the leading improvement is achieved by adding the Sheikholeslami--Wohlert, or clover, term
\begin{equation}
  \label{eq:clover_term}
  c_\m{sw}\,\frac{\m{i}a}{4}\sigma_{\mu\nu}\hat{F}_{\mu\nu}(x)\,\delta_{x,y},
\end{equation}
where $\hat{F}_{\mu\nu}$ is a lattice discretization of the field-strength tensor. It is defined as
\begin{equation}
    \hat{F}_{\mu\nu}(x) = \frac{1}{8a^2} (Q_{\mu\nu}(x) - Q_{\nu\mu}(x)),
\end{equation}
where $Q_{\mu\nu}$ is the sum of link loops
\begin{equation}
\begin{split}
  Q_{\mu\nu}(x)
    &= U_\mu(x)\,U_\nu(x+a\hat{\mu})\,
       U_\mu(x+a\hat{\nu})^\dagger\,U_\nu(x)^\dagger \\
    &\quad + U_\nu(x)\,U_\mu(x-a\hat{\mu}+a\hat{\nu})^\dagger\,
             U_\nu(x-a\hat{\mu})^\dagger\,U_\mu(x-a\hat{\mu}) \\
    &\quad + U_\mu(x-a\hat{\mu})^\dagger\,U_\nu(x-a\hat{\mu}-a\hat{\nu})^\dagger\,
             U_\mu(x-a\hat{\mu}-a\hat{\nu})\,U_\nu(x-a\hat{\nu}) \\
    &\quad + U_\nu(x-a\hat{\nu})^\dagger\,U_\mu(x-a\hat{\nu})\,
             U_\nu(x+a\hat{\mu}-a\hat{\nu})\,U_\mu(x)^\dagger
\end{split}
\end{equation}
displayed in Figure~\ref{fig:clover}. 
At tree level, i.e.\ to lowest order in the gauge coupling $g$, the choice $c_\m{sw} = 1$ removes the $O(a)$ artefacts. Beyond tree level, the coefficient receives corrections, $c_\m{sw} = 1 + O(g^2)$, and has to be tuned, ideally nonperturbatively. The gauge ensembles used in this work employ a nonperturbatively determined value of $c_\m{sw}$~\cite{Bulava:2013cta}.

Combining these ingredients gives the Wilson-clover Dirac operator (for flavour $f$),
\begin{equation}
  \label{eq:wilson_clover_dirac}
  D_f = \gamma_\mu\nabla^\m{sym}_\mu - \frac{a}{2}\nabla^*_\mu\nabla_\mu
    + c_\m{sw}\,\frac{\m{i}a}{4}\sigma_{\mu\nu}\hat{F}_{\mu\nu} + m_f.
\end{equation}
In practice, the bare quark mass is usually expressed in terms of the hopping parameter
\begin{equation}
  \label{eq:hopping_parameter}
  \kappa_f = \frac{1}{2(am_f + 4)}.
\end{equation}

The Wilson-clover formulation is one of several discretizations of the fermion action. Other commonly used formulations include staggered, twisted-mass and domain-wall fermions. These either preserve chiral symmetry while being more expensive (domain-wall fermions), are computationally cheaper but have doublers (staggered fermions), or offer advantages regarding the cutoff effects for operators while breaking other symmetries (twisted-mass fermions). Overall, Wilson-clover fermions offer a good compromise, presenting few technical difficulties while remaining computationally affordable.

\subsection{Relativistic heavy quark action}
\label{sec:RHQ_action}

The Wilson-clover action is well suited to light quarks, but it is difficult to use for heavy quarks such as the bottom quark. To see why, consider the dispersion relation of a free Wilson-clover fermion, which for small momenta takes the form
\begin{equation}
  \label{eq:dispersion_M1_M2}
  E(\vb*{p}) = M_1 + \frac{\vb*{p}^2}{2M_2} + O(\vb*{p}^4),
\end{equation}
or, equivalently,
\begin{equation}
  \label{eq:dispersion_M1_M2_squared}
  E(\vb*{p})^2 = M_1^2 + \frac{M_1}{M_2}\vb*{p}^2 + O(\vb*{p}^4),
\end{equation}
where $M_1$ and $M_2$ are referred to as the ``rest mass'' and the ``kinetic mass'', respectively~\cite{El-Khadra:1996wdx}. For light quarks and small lattice spacings, where $am_f \ll 1$, both masses satisfy
\begin{equation}
  aM_i = am_f + O\qty((am_f)^2) \quad (i = 1, 2)
\end{equation}
so that the continuum dispersion relation $E(\vb*{p})^2 = m_f^2 + \vb*{p}^2$ is recovered up to corrections of order $am_f$ and $\vb*{p}^4$. On typical lattices, however, the bottom quark has $am_b \sim 1$--$2$, so that the powers of $am_b$ in the expansion are no longer small and $M_1 \neq M_2$. As a consequence, the lattice dispersion relation deviates significantly from the continuum one, leading to a large and uncontrolled systematic error. The discrepancy in $M_1$ can always be absorbed into a redefinition of the mass. However, the inequality $M_1 \neq M_2$ results in a modified dispersion relation
\begin{equation}
  \label{eq:dispersion_csq}
  E(\vb*{p})^2 = m_b^2 + c^2\vb*{p}^2
\end{equation}
for small $\vb*{p}$, with a speed-of-light parameter $c^2 = M_1/M_2$ that differs from unity.\footnote{Note that this speed-of-light parameter is defined differently from the physical speed of light.} Reaching the regime $am_b \ll 1$ directly would require an extremely fine lattice spacing $a \lesssim 0.01\;\m{fm}$, which is computationally prohibitive and, moreover, worsens the problem of critical slowing down discussed in Section~\ref{sec:monte_carlo}.

Several approaches have been developed to treat heavy quarks on the lattice. One example is heavy-quark effective theory (HQET)~\cite{Eichten:1987xu,Georgi:1990um}, which expands in powers of $1/m_b$. The extreme case is the static approximation~\cite{Grinstein:1990mj}. Another is non-relativistic QCD (NRQCD)~\cite{Thacker:1990bm}, a non-relativistic effective theory which, being a cutoff theory, does not possess a continuum limit. In this thesis we use a different approach, known as the relativistic heavy quark (RHQ) action~\cite{El-Khadra:1996wdx}.

The central idea behind the RHQ approach is that a heavy quark is, to a good approximation, static: it propagates much further in time than in space. This asymmetry between the temporal and spatial dynamics is reflected in the dispersion relation and motivates an action in which the temporal and spatial terms of the Dirac operator carry different coefficients. Formally, since the action must accommodate this heavy mass scale, its couplings distinguish the temporal from the spatial directions, reducing the symmetry of the action from the hypercubic group $H(4)$ to the octahedral group $O_h$ of spatial rotations. This allows one to construct an anisotropic action with more freedom to absorb the $am_b$ effects. This is done systematically by employing the Symanzik improvement programme.

Restricting to spatial rotational symmetry, each of the three terms of the Wilson-clover Dirac operator --- the symmetric derivative $\gamma_\mu\nabla^\m{sym}_\mu$, the Wilson term $\nabla^*_\mu\nabla_\mu$, and the clover term $\sigma_{\mu\nu}\hat{F}_{\mu\nu}$ --- splits into a temporal and a spatial part. Each of them may carry an independent coefficient. We can absorb the coefficient of the temporal derivative term $\gamma_4\nabla^\m{sym}_4$ by a field redefinition, which amounts to a redefinition of the mass. One then arrives at the RHQ Dirac operator for the bottom quark~\cite{El-Khadra:1996wdx,Aoki:2001ra},
\begin{equation} \label{eq:RHQ_dirac}
\begin{split}
    D_b^\m{RHQ} = \gamma_4\nabla^\m{sym}_4 + \nu\,\gamma_i\nabla^\m{sym}_i
    - r_t\,\frac{a}{2}\nabla^*_4\nabla_4 - r_s\,\frac{a}{2}\nabla^*_i\nabla_i
    + c_E\,\frac{\m{i}a}{2}\sigma_{4i}\hat{F}_{4i}
    + c_B\,\frac{\m{i}a}{4}\sigma_{ij}\hat{F}_{ij} + m_b.
\end{split}
\end{equation}
Here, the index $4$ denotes the temporal direction, the Latin indices $i, j$ run over the three spatial directions, and a summation over repeated indices is understood.

The RHQ action thus contains the parameters $m_b$, $r_t$, $r_s$, $\nu$, $c_E$, and $c_B$. Instead of the bare mass $m_b$, the quark-mass dependence can equivalently be described by the (generalized) hopping parameter
\begin{equation}
  \kappa_b = \frac{1}{2(am_b + r_t + 3r_s)}.
\end{equation}
Following common practice, we set $r_t = 1$~\cite{El-Khadra:1996wdx}, which leaves a five-parameter action whose coefficients must be tuned. This can be done nonperturbatively, for example, as briefly described in Section~\ref{sec:scale_setting}. Alternative formulations of this action impose the restrictions $r_s = \nu$ and, additionally, $c_E = c_B$, resulting in four- and three-parameter actions, respectively~\cite{Lin:2006ur,Christ:2006us}. Since the values of the tuned parameters depend on the gauge and light-quark actions of the underlying ensembles, the tuning has to be performed separately for each discretization. For the ensembles used in this work, a nonperturbative tuning is available for the five-parameter action~\cite{Hudspith:2026vxt}, but not for the restricted versions, which have been tuned for other discretizations. In this work, we therefore use the full five-parameter action. The nonperturbative tuning of Ref.~\cite{Hudspith:2026vxt} moreover yields parameter values that are compatible with neither $c_E = c_B$ nor $r_s = \nu$.

\section{Lattice QCD path integral}
\label{sec:LQCD_path_integral}

Now that we have discretized the QCD action on the lattice, we turn to the evaluation of the lattice QCD path integral. In terms of this path integral, the vacuum expectation value of an operator $\mathcal{O}$ is given by
\begin{equation}
  \label{eq:vev_euclidean}
  \ev{\mathcal{O}} = \frac{1}{Z}\int\mathcal{D}\psi\,\mathcal{D}\overline{\psi}\,\mathcal{D}U\;
    \mathcal{O}[\psi,\overline{\psi},U]\; \m{e}^{-S_F[\psi,\overline{\psi},U] - S_G[U]},
\end{equation}
where $S_F$ and $S_G$ are the discretized fermion and gauge action, respectively. The partition function $Z$ is defined in the usual way:
\begin{equation}
  \label{eq:partition_euclidean}
  Z = \int\mathcal{D}\psi\,\mathcal{D}\overline{\psi}\,\mathcal{D}U\;
    \m{e}^{- S_F[\psi,\overline{\psi},U] - S_G[U]}.
\end{equation}
Because of the lattice regularization, these path integrals are finite-dimensional: they reduce to integrals over the fields $\psi(x)$, $\overline{\psi}(x)$, and the link variables $U_\mu(x)$ at every lattice site. The former are Grassmann integrals and the latter is an integral with respect to the Haar measure of $\m{SU}(3)$.

As the quark fields are anticommuting Grassmann variables, it is difficult to represent them directly on a computer. They are therefore integrated out analytically. Performing the Grassmann (Gaussian) integration contracts pairs of quark fields in the operator $\mathcal{O}$ via Wick's theorem, resulting in quark propagators $S_f=D_f^{-1}$~\cite{Gattringer:2010zz}. In the simplest case, we obtain
\begin{equation}
  \label{eq:quark_propagator}
  \ev*{\psi^{(f)}(x)\,\overline{\psi}^{(f)}(y)}_F = \frac{1}{a^4}S_f(x,y),
\end{equation}
where spin and colour indices have been suppressed, and $\ev{\dots}_F$ denotes the fermionic part of the vacuum expectation value. This procedure replaces the original operator by a Wick-contracted operator $\tilde{\mathcal{O}}[U]$ that depends only on the gauge fields. Additionally, it produces a fermion determinant $\det(D_f[U])$ for each quark flavour. The remaining expectation value is a purely bosonic integral over the gauge field,
\begin{equation}
  \label{eq:vev_gauge}
  \ev*{\tilde{\mathcal{O}}}_G = \frac{1}{Z}\int\mathcal{D}U\;
  \tilde{\mathcal{O}}[U]\, \det(D[U])\, \m{e}^{-S_G[U]},
\end{equation}
where $\det(D[U]) = \prod_{f=1}^{N_f}\det(D_f[U])$.

The fermion determinant encodes the effects of the dynamical, or sea, quarks: sampling gauge configurations according to the weight $\det(D[U])\, \m{e}^{-S_G[U]}$ amounts to including the sea quarks explicitly in the simulation. The Wilson-clover Dirac operator, like most lattice Dirac operators --- including that of the RHQ action --- satisfies $\gamma_5$-hermiticity,
\begin{equation}
  \label{eq:gamma5_hermiticity}
  \gamma_5 D \gamma_5 = D^\dagger.
\end{equation}
This guarantees that $\det(D)$ is real. For a degenerate pair of light quarks, the determinant is then $\det(D_l)^2 \geq 0$ and can be interpreted as part of a probability weight. For this reason, the $u$ and $d$ quarks are treated as mass-degenerate in most simulations. In practice, the determinant is not computed directly but represented by an auxiliary bosonic field~\cite{Fucito:1980fh}. Using
\begin{equation}
  \label{eq:pseudofermions}
  \det(D_l)^2 = {\det}(D_l^\dagger D_l)
    = \int\mathcal{D}\phi\,\mathcal{D}\phi^\dagger\;
      \m{e}^{-\phi^\dagger (D_l^\dagger D_l)^{-1}\phi},
\end{equation}
the determinant of the degenerate light pair is rewritten as a Gaussian integral over pseudofermion fields $\phi$, $\phi^\dagger$. The term $-\phi^\dagger (D_l^\dagger D_l)^{-1}\phi$ is added to the action when sampling the gauge configurations to include the determinant.

For nondegenerate quarks such as the strange quark, the contribution to the weight involves a single fermion determinant $\det(D_f) = {\det}(D_f^\dagger D_f)^{1/2}$. To include it, one uses a rational approximation of the fractional power $(D_f^\dagger D_f)^{1/2}$, so that the determinant, written as a Gaussian integral over pseudofermion fields,
\begin{equation} \label{eq:pseudofermions_rational}
  \det(D_f) = \int\mathcal{D}\phi\,\mathcal{D}\phi^\dagger\;
      \m{e}^{-\phi^\dagger (D_f^\dagger D_f)^{-1/2}\phi},
\end{equation}
can also be included in the sampling process. For heavy quarks such as the charm and the bottom, the corresponding fermion determinant is close to unity and can be neglected. This treatment is called the quenched approximation.

The gauge ensembles used in this work were generated with dynamical $u$, $d$ and $s$ quarks, which were included as described above. Such ensembles are generally referred to as $N_f = 2 + 1$ ensembles, which reflects the degeneracy of the $u$ and the $d$. More specifically, we work at the $\m{SU}(3)$-flavour-symmetric point, where the $u$, $d$ and $s$ quarks are all degenerate. The heavier $c$ and $b$ quarks were included nondynamically, i.e.\ just as valence quarks without their fermion determinants.

\section{Monte Carlo methods}
\label{sec:monte_carlo}

The path integral in Eq.~\eqref{eq:vev_gauge} is an integral over an enormous number of degrees of freedom and cannot be evaluated directly. Instead, it is estimated using Markov chain Monte Carlo (MCMC) methods. The key idea is importance sampling: rather than performing the Monte Carlo integral using random gauge links $U$ distributed according to the Haar measure, one generates an ensemble of gauge configurations distributed according to the probability weight $\propto\det(D[U])\, \m{e}^{-S_G[U]}$. Expectation values are then estimated as sample means over the ensemble. The statistical error of the estimate decreases as $1/\sqrt{N}$ in the number of configurations $N$. This error scaling is independent of the number of dimensions, although the prefactor strongly depends on the operator $\mathcal{O}$.

The ensemble is generated as a Markov chain using the Metropolis--Hastings algorithm: starting from a given configuration, the gauge links are updated according to some update scheme, and the proposed configuration is then accepted or rejected. For the chain to converge to the desired distribution, the update scheme must satisfy detailed balance and ergodicity. The former means that the probability of being at configuration $U$ and moving to $U'$ equals that of the reverse process. The latter means that every configuration is reachable by the update scheme, and that the chain does not become trapped in a cycle.

The standard algorithm for dynamical fermions is the hybrid Monte Carlo (HMC) algorithm~\cite{Duane:1987de}. In the HMC, one introduces auxiliary, Gaussian-distributed momenta conjugate to the gauge fields, together with a fictitious simulation time. These define a fictitious classical Hamiltonian for the gauge and pseudofermion fields, according to which the gauge fields are evolved in configuration space. This is done by numerically integrating the corresponding equations of motion. This preserves the Hamiltonian, which ensures that the resulting gauge configurations follow the desired distribution. However, the integration scheme introduces numerical errors due to the finite integration step size. A Metropolis accept/reject step at the end of each trajectory corrects for these errors.

As mentioned in the previous section, the fermion determinants are included using pseudofermion fields. These are sampled according to the weight $\m{e}^{-\phi^\dagger (D_f^\dagger D_f)^{-1/n}\phi}$ with $n = 1, 2$ for degenerate and nondegenerate quarks, respectively. When evolving the gauge fields $U$, these random fields are used in the pseudofermion part $-\phi^\dagger (D_f^\dagger D_f)^{-1/n}\phi$ of the action. For nondegenerate quarks ($n=2$), a rational approximation of $(D_f^\dagger D_f)^{1/2}$ is used. Therefore, the corresponding HMC is called the rational hybrid Monte Carlo (RHMC) algorithm~\cite{Clark:2003na,Clark:2006fx}.

The evolution of the gauge fields requires computing the force coming from the pseudofermions. This amounts to inverting $(D_f^\dagger D_f)$ on the pseudofermion field at each integration step. Inverses of the Dirac operators also appear when performing the Wick contraction to integrate out the quarks in the path integral. These inversions are performed by solving the linear equation $D_f\psi = \eta$ or $(D_f^\dagger D_f)\psi = \eta$ on source fields $\eta$. This is the dominant computational cost, both in generating the gauge configurations and in computing the gauge-field expectation value of an operator (called the measurement). It becomes increasingly expensive as the quark masses decrease, where the Dirac operator becomes ill-conditioned.

To tackle this difficult task, Krylov solvers such as the conjugate gradient, generalized conjugate residual, and multigrid solvers~\cite{Saad:2003}, in combination with preconditioners, are used. The numerical data used in this work were produced with the latter two solvers. For the bottom quark, we instead used a hopping-parameter-expansion solver, which exploits the fact that the Dirac operator can be written as $D \propto \mathbf{1} - \kappa H$ in terms of the hopping matrix $H$~\cite{Gattringer:2010zz}. The inverse $D^{-1}$ is then given by a geometric series in $\kappa H$, which can be truncated after a moderate number of terms because the hopping parameter $\kappa$ is small for heavy quarks.

A further important subtlety is autocorrelation: successive configurations in a Markov chain are not statistically independent. To reduce the correlation between data points, one saves only an equally spaced subset of the configurations in the chain. The remaining autocorrelation must be accounted for in the error analysis to avoid underestimating the statistical uncertainty. In this work we use the $\Gamma$-method, which estimates the autocorrelation explicitly~\cite{WOLFF2004143}.

Autocorrelations grow rapidly as the continuum limit is approached, a phenomenon known as critical slowing down. As $a\to0$, the barriers between different topological sectors, labelled by the topological charge, become so large that the HMC is increasingly unlikely to cross them. Consequently, the algorithm tends to become trapped and, in the worst case, loses ergodicity. This can be mitigated by using open boundary conditions in time instead of (anti)periodic ones~\cite{Luscher:2011kk}, which allow topological charge to flow into and out of the lattice through its temporal boundaries. The drawback is that translational invariance in time is lost near the boundary, as the gauge fields create spurious boundary states there. As a result, measurements must be performed sufficiently far from the boundaries to avoid boundary effects. In this work, we used gauge ensembles generated by the Coordinated Lattice Simulations (CLS) collaboration, which uses the methods described above~\cite{Bruno:2014jqa}.

\section{Scale setting and tuning}
\label{sec:scale_setting}

All quantities computed on the lattice are dimensionless, expressed in units of the lattice spacing $a$. To convert them to physical units, the value of $a$ must be determined, a procedure known as scale setting. In addition, the bare parameters of the action --- in particular the quark masses --- must be tuned so that the simulation describes the desired physical (or, where intended, heavier-than-physical) point. Both steps require experimental input.

Scale setting proceeds by computing a chosen reference quantity in lattice units and comparing it to its known physical value. Common choices include hadronic quantities such as the decay constants $f_\pi$ and $f_K$ or the mass of the $\Omega$ baryon, and theoretical scales derived from the gradient flow, such as $t_0$~\cite{Luscher:2010iy}. For the CLS ensembles used in this work, the scale setting was performed using the gradient-flow scale $t_0$~\cite{Bruno:2016plf}; we use the updated lattice spacings of Ref.~\cite{Strassberger:2021tsu}.

The bare parameters are tuned as follows. The bare inverse coupling $\beta = 6/g^2$ controls the lattice spacing $a$, while the light and strange hopping parameters $\kappa_l$ and $\kappa_s$ control the corresponding quark masses. At the $\m{SU}(3)$-flavour-symmetric point, $\kappa_l$ and $\kappa_s$ are set to the same value. These hopping parameters must be tuned before generating the final gauge ensemble, since the light and strange quarks are included dynamically in the simulation, i.e.\ they are explicitly included in the fermion determinants. The same is true for the improvement coefficient $c_\m{sw}$ in the Wilson-clover action. The charm and bottom quarks, on the other hand, are included nondynamically; consequently, the corresponding parameters can be tuned after generating the ensemble. For the charm quark, simulated using the Wilson-clover action, this amounts to tuning the hopping parameter $\kappa_c$ to reproduce, e.g., the $D$-meson mass. For the bottom quark, the five parameters of the RHQ action must all be tuned simultaneously. This was done by collaborators following the procedure detailed in Ref.~\cite{Hudspith:2021iqu}, applied to the bottom quark~\cite{Hudspith:2026vxt}. As physical input, they used the masses of several bottomonium states, and the action parameters were tuned such that the lattice dispersion relation $E(\vb*{p})^2 = m^2 + c^2\vb*{p}^2$ matches the continuum one, i.e.\ $c^2 = 1$.
\chapter{Hadron spectroscopy}
\label{chapter:Hadron_spectroscopy}

The central goal of this thesis is to compute properties of hadrons from lattice QCD. On the lattice, the theory is necessarily formulated in a finite volume, so that the energy spectrum is discrete. For hadrons that are well below any strong-decay threshold, such as the pion or the nucleon, the finite-volume corrections to the mass are exponentially suppressed in the box size, and the mass is simply given by the finite-volume ground-state energy. This simple picture breaks down, however, for states that lie near a multi-hadron threshold, e.g.\ resonances or shallow bound states. In either case, the finite-volume energies can no longer be identified directly with the mass. The rigorous, model-independent way to define the state's properties is through the poles of the corresponding scattering amplitude. In the case of the $T_{cc}^+$, for example, it is the $DD^*$ amplitude. This chapter reviews the tools required for this analysis. We first summarize the relevant properties of scattering amplitudes and their analytic structure. We then introduce Lüscher's finite-volume formalism, which relates the discrete finite-volume spectrum to the infinite-volume scattering amplitude. Finally, we discuss the extraction of the finite-volume energies from Euclidean correlation functions and the construction of interpolating operators.

\section{Scattering amplitudes, bound states and resonances}
\label{sec:amplitude_analysis}

In this section we introduce the basic concepts of amplitude analysis, following the corresponding review of the Particle Data Group~\cite{ParticleDataGroup:2024cfk}. The goal is to establish how bound states and resonances are encoded in the scattering amplitude, and how the amplitude can be parametrized in a way that is consistent with the fundamental properties of the $S$-matrix.

Resonances and bound states appear as enhancements in scattering cross-sections. These enhancements originate from poles of the scattering amplitude that appear when we analytically continue the amplitude into the complex energy plane. The state-of-the-art way of defining the mass $m$ of a hadron --- and its width $\Gamma$, if it is a resonance --- is through the position of such a pole at $m \pm \m{i}\tfrac{\Gamma}{2}$. For a stable bound state, the width vanishes, $\Gamma = 0$, and the pole lies on the real energy axis, whereas for a resonance the poles appear in complex-conjugate pairs. For an isolated, narrow resonance appearing as an enhancement above threshold, the amplitude can be described by a Breit--Wigner parametrization~\cite{ParticleDataGroup:2024cfk}. This simple description, however, breaks down for more complicated systems featuring several nearby poles or coupled channels. In general, one instead constrains the amplitude using the unitarity and analyticity properties of the $S$-matrix, which allows for a rigorous parametrization.

\subsection{$S$-matrix and partial-wave decomposition}
\label{sec:smatrix_partial_wave}

We restrict the discussion to the single-channel (elastic) scattering of two spin-zero particles with masses $m_1$ and $m_2$, occasionally commenting on the multi-channel (inelastic) case. The central object describing a scattering process is the $S$-matrix. It maps incoming asymptotic states in the far past to outgoing asymptotic states in the far future. Its matrix elements give the probability for a given initial state to evolve into a given final state. As is customary, we define the scattering matrix $\hat{T}$ as the nontrivial part of $\hat{S}$ via $\hat{S} = 1 + \m{i}\hat{T}$. Finally, the scattering amplitude $\mathcal{M}$ is defined in terms of $\hat{T}$ as
\begin{equation}
  \label{eq:transfer_matrix}
  \langle p_1', p_2' \,|\, \m{i}\hat{T} \,|\, p_1, p_2 \rangle
    = (2\pi)^4\,\delta^{(4)}(p_1' + p_2' - p_1 - p_2)\;
      \m{i}\mathcal{M}(p_1', p_2'; p_1, p_2),
\end{equation}
where $|\,p_1, p_2\rangle$ are asymptotic two-particle states. The amplitude $\mathcal{M}$ is a Lorentz-invariant function of the momenta and is the quantity most directly related to physical observables such as cross-sections and decay rates. It is conveniently parametrized in terms of the Mandelstam variables
\begin{equation}
  \label{eq:mandelstam}
  s = (p_1 + p_2)^2, \qquad t = (p_1 - p_1')^2,
\end{equation}
where $s$ is the square of the energy in the centre-of-momentum frame, $E_\m{cm}^2$, and $t$ is related to the angle between the incoming and outgoing particles. Thus, the scattering amplitude can equivalently be parametrized by $s$ and the scattering angle $\theta$ between $\vb*{p}_1$ and $\vb*{p}_1'$ in the c.m.\ frame. It is furthermore convenient to change from the basis $|p_1, p_2\rangle$ to a basis $|P, \hat{\vb*{q}}_\m{cm}\rangle$ labelled by the total momentum $P = p_1 + p_2$ and the direction $\hat{\vb*{q}}_\m{cm}$ of the c.m.\ scattering momentum $\vb*{q}_\m{cm} = \vb*{p}_1 = -\vb*{p}_2$, defined in the c.m.\ frame. Its magnitude $q_\m{cm} = \abs{\vb*{q}_\m{cm}}$ is a function of $s = E_\m{cm}^2$, determined by
\begin{equation}
  \label{eq:qcm_implicit}
  \sqrt{s} = \sqrt{m_1^2 + q_\m{cm}^2} + \sqrt{m_2^2 + q_\m{cm}^2},
\end{equation}
with the solution
\begin{equation}
  \label{eq:qcm_explicit}
  q_\m{cm} = \sqrt{\frac{1}{4}s - \frac{1}{2}(m_1^2 + m_2^2) + \frac{(m_1^2 - m_2^2)^2}{4s}}.
\end{equation}

Since angular momentum is conserved, the $S$-matrix is diagonal in the angular-momentum basis. This basis is obtained by projecting onto states of definite orbital angular momentum,
\begin{equation}
  \label{eq:angular_momentum_basis}
  |P, \ell, m\rangle = \frac{\rho}{\sqrt{2\pi}}
    \int \m{d}^2\hat{\vb*{q}}_\m{cm}\; Y_{\ell m}(\hat{\vb*{q}}_\m{cm})\; |P, \hat{\vb*{q}}_\m{cm}\rangle,
\end{equation}
where $\rho(s) = \tfrac{q_\m{cm}}{8\pi\sqrt{s}}$ is the two-body phase-space factor. The prefactor ensures the normalization
\begin{equation}
  \label{eq:am_normalization}
  \langle P', \ell', m' \,|\, P, \ell, m \rangle
    = (2\pi)^4\,\delta^{(4)}(P' - P)\;\delta_{\ell'\ell}\,\delta_{m'm}.
\end{equation}
In this basis, the scattering amplitude reads
\begin{equation}
  \label{eq:amplitude_am_basis}
  \langle P', \ell', m' \,|\, \m{i}\hat{T} \,|\, P, \ell, m \rangle
    = (2\pi)^4\,\delta^{(4)}(P' - P)\;\delta_{\ell'\ell}\,\delta_{m'm}\;
      2\m{i}\rho\,\mathcal{M}_\ell(s),
\end{equation}
where $\mathcal{M}_\ell(s)$ is the partial-wave decomposition of the amplitude. The partial-wave-projected amplitude is related to the full amplitude by
\begin{equation}
  \label{eq:partial_wave_expansion}
  \mathcal{M}(s, \cos\theta) = \sum_{\ell=0}^{\infty} (2\ell + 1)\,\mathcal{M}_\ell(s)\,P_\ell(\cos\theta),
\end{equation}
in terms of the Legendre polynomials $P_\ell$. Because we have removed the angular dependence, $\mathcal{M}_\ell$ depends only on $s$. Compared to Eq.~\eqref{eq:transfer_matrix} we obtain an additional prefactor of $2\rho$, which consequently also appears in the $S$-matrix element
\begin{equation}
  \label{eq:S_matrix_element}
  S_\ell = 1 + 2\m{i}\rho(s)\,\mathcal{M}_\ell(s)
\end{equation}
in this basis. This prefactor is important, as it reveals analytic properties of $\mathcal{M}_\ell$. Unitarity of the $S$-matrix implies $|S_\ell| = 1$, so that $S_\ell$ can be written in terms of a real phase shift $\delta_\ell(s)$ as $S_\ell(s) = \m{e}^{2\m{i}\delta_\ell(s)}$. The partial-wave amplitude then takes the form
\begin{equation}
  \label{eq:amplitude_phase_shift}
  \mathcal{M}_\ell(s) = \frac{1}{2\m{i}\rho(s)}\left(\m{e}^{2\m{i}\delta_\ell(s)} - 1\right)
    = \frac{1}{\rho(s)\left(\cot\delta_\ell(s) - \m{i}\right)}.
\end{equation}
As a consequence, it is convenient to describe the scattering amplitude in terms of $\cot{\delta_\ell(s)}$.

In the presence of several channels --- for example the coupled $DD^*$ and $D^*D^*$ channels --- the partial-wave $S$-matrix becomes a matrix. Labelling the channels by an index $a$, each channel has its own masses $m_{1,a}$ and $m_{2,a}$, its own c.m.\ momentum $q_{\m{cm},a}$, and its own phase-space factor $\rho_a(s) = \tfrac{q_{\m{cm},a}}{8\pi\sqrt{s}}$. The $S$-matrix then becomes
\begin{equation}
  \label{eq:S_matrix_coupled}
  S_{\ell,ab} = \delta_{ab} + 2\m{i}\,\sqrt{\rho_a(s)}\,\mathcal{M}_{\ell,ab}(s)\,\sqrt{\rho_b(s)},
\end{equation}
and its parametrization is correspondingly more involved. A convenient parametrization that automatically respects unitarity is provided by the $K$-matrix, in terms of which
\begin{equation}
  \label{eq:K_matrix}
  \mathcal{M}_\ell(s) = \frac{1}{\sqrt{\rho}\left(K_\ell(s)^{-1} - \m{i}\right)\sqrt{\rho}},
\end{equation}
with the generalized phase-space matrix $\rho(s) = \m{diag}(\rho_1(s), \rho_2(s), \dots)$. Unitarity and time-reversal invariance of the $S$-matrix imply that $K_\ell$ is real and symmetric. In the single-channel case, this reduces to
\begin{equation}
  \label{eq:K_matrix_single}
  K_\ell(s) = \tan\delta_\ell(s).
\end{equation}

\subsection{Analytic structure and Riemann sheets}
\label{sec:analytic_structure}

The location of the amplitude's poles in the complex energy plane determines the nature of the corresponding state, so we now examine the analytic structure of the partial-wave amplitude $\mathcal{M}_\ell(s)$. From Eq.~(\ref{eq:amplitude_phase_shift}) we see that the inverse of this amplitude is
\begin{equation}
  \label{eq:amplitude_phase_shift2}
  \mathcal{M}_\ell(s)^{-1} = \rho(s)\cot\delta_\ell(s) - \m{i}\rho(s),
\end{equation}
which exposes the origin of its structure. The term $\rho\cot\delta_\ell$ depends only on even powers of $q_\m{cm}$ and therefore introduces no nonanalyticities for physical scattering, as will become apparent from the effective range expansion in the next section. The term $-\m{i}\rho(s) = -\m{i}\,\tfrac{q_\m{cm}}{8\pi\sqrt{s}}$, by contrast, depends on $s$ through square roots, which give rise to branch points and their associated branch cuts. The explicit factor of $\sqrt{s}$ has no branch point in the physical scattering region. The square root contained in $q_\m{cm}$, cf.\ Eq.~(\ref{eq:qcm_explicit}), however, produces a branch point at the two-particle threshold $s_\m{thr} = (m_1 + m_2)^2$. By convention, the associated branch cut is chosen to run along the real axis towards positive infinity, and is referred to as the right-hand cut. In the multi-channel case, there is such a branch point at each channel threshold $s_{\m{thr},a} = (m_{1,a} + m_{2,a})^2$.

\begin{figure}[h]
  \centering
  \includegraphics[width=\textwidth]{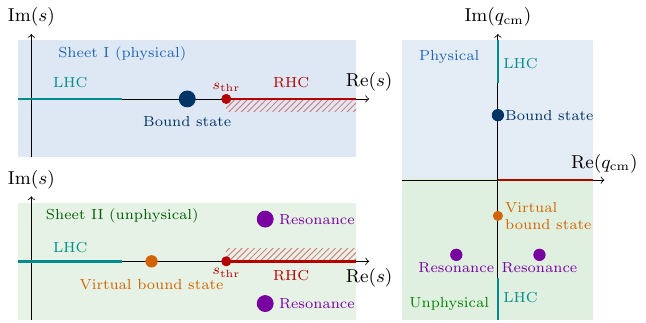}
  \caption{\label{fig:amplitude_sheets}%
    Analytic structure of the single-channel partial-wave amplitude $\mathcal{M}_\ell$ in the complex $s$ plane (left) and the complex $q_\m{cm}$ plane (right). In $s$, the amplitude consists of two sheets due to the right-hand cut: the physical sheet (top) and the unphysical sheet (bottom). On these two sheets, the right-hand cut starting at the threshold $s_\m{thr}$ is shown together with the possible locations of bound-state, virtual-bound-state, and resonance poles. The shaded red areas indicate on which side of the physical scattering axis the discontinuity from the cut lies. In the $q_\m{cm}$ plane, the right-hand cut opens up, and both sheets are mapped to one plane. The locations of the same amplitude poles are also displayed in this plane. In addition, a left-hand cut (LHC) is shown in both planes.}
\end{figure}

The branch cut connects two Riemann sheets. The physical sheet is the one on which $\mathcal{M}_\ell(s)$ is connected to the physical scattering region, approached from $\m{Im}(s) > 0$. The second sheet is referred to as the unphysical sheet. Owing to causality, locality, and vacuum stability, the physical sheet is free of singularities except for poles on the real axis below threshold. These poles correspond to bound states. The unphysical sheet, however, can contain poles both on the real axis below threshold and off the real axis. The former correspond to virtual bound states, the latter to resonances. Owing to the reflection principle, resonance poles always appear in complex-conjugate pairs, and if they lie close to the physical region they strongly influence the amplitude. In addition to the right-hand cut, the amplitude can possess a left-hand cut below threshold, arising, for example, from one-pion exchange. Such exchanges manifest themselves as poles in the full amplitude that become branch cuts in the partial-wave-projected amplitude. The left panel of Figure~\ref{fig:amplitude_sheets} illustrates the analytic structure of $\mathcal{M}_\ell(s)$ in the single-channel case, showing the physical and unphysical sheets and potential nonanalyticities.

It is often more transparent to describe the analytic structure directly in terms of $q_\m{cm}$. As a function of $q_\m{cm}$, the scattering amplitude contains no square root. Consequently, the two-body threshold at $q_\m{cm} = 0$ is no longer a branch point, and the two Riemann sheets in $s$ map onto the upper and lower half of the complex $q_\m{cm}$ plane. In this variable, a bound state and a virtual bound state correspond to a pole on the positive and negative imaginary axis, respectively. A resonance appears as a pair of poles in the lower half of the complex plane. This is illustrated in the right panel of Figure~\ref{fig:amplitude_sheets}.

\subsection{Parametrization of the amplitude: the effective range expansion}
\label{sec:ERE}

To extract the pole positions, one needs a functional form for the amplitude, e.g.\ for the phase shift $\delta_\ell$ in terms of $q_\m{cm}$ or $s$. Restricting again to the single-channel case, a widely used near-threshold approximation is the effective range expansion (ERE). It expands $q_\m{cm}^{2\ell+1}\cot\delta_\ell$ in powers of $q_\m{cm}^2$ around threshold,
\begin{equation}
  \label{eq:ERE}
  q_\m{cm}^{2\ell+1}\cot\delta_\ell = -\frac{1}{a_\ell} + \frac{r_\ell}{2}\,q_\m{cm}^2 + O(q_\m{cm}^4),
\end{equation}
in terms of the scattering length $a_\ell$ and the effective range $r_\ell$. The higher-order coefficients are called shape parameters. From this expression we see that the $q_\m{cm}\cot{\delta_\ell}$ term appearing in the scattering amplitude (cf.\ Eq.~\eqref{eq:amplitude_phase_shift2}) depends only on even powers of $q_\m{cm}$, making it real for physical scattering and free of branch cuts.

In the simplest case, where only the $s$-wave ($\ell = 0$) is retained, the amplitude becomes
\begin{equation}
  \label{eq:swave_amplitude_ERE}
  \mathcal{M}_0(s) = \frac{8\pi\sqrt{s}}{-\dfrac{1}{a_0} + \dfrac{r_0}{2}q_\m{cm}^2 - \m{i}q_\m{cm}}.
\end{equation}
The poles of this amplitude in $q_\m{cm}$ are located at
\begin{equation}
  \label{eq:ERE_pole_positions}
  q_\m{cm}^\pm = \frac{1}{r_0}\left(\m{i} \pm \sqrt{-1 + 2\frac{r_0}{a_0}}\,\right).
\end{equation}
Consequently, the ratio $\tfrac{r_0}{a_0}$ determines what type of states the poles correspond to. For $\tfrac{r_0}{a_0} > \tfrac{1}{2}$ the pole positions have a nonzero real part, and we thus have a resonance. For $\tfrac{r_0}{a_0} < \tfrac{1}{2}$ the term in the square root is negative, and we obtain a purely imaginary pole position; this corresponds to a pair of (virtual) bound states. Thus, when using this parametrization, we have either two or zero (virtual) bound states.\footnote{In the case $\tfrac{r_0}{a_0} = \tfrac{1}{2}$ we obtain a single pole of order two.} However, in the limit $r_0 \rightarrow 0$ the pole $q_\m{cm}^+$ moves to $\m{i}\infty$ or $-\m{i}\infty$ and is consequently no longer relevant. In that case we have a single bound or virtual bound state with pole position $q_\m{cm} = \frac{\m{i}}{a_0}$. The former case appears when $a_0 > 0$ and the latter when $a_0 < 0$.

When studying a two-particle system, it is instructive to investigate how the pole structure evolves when, e.g., the mass of the particles is changed. One possible evolution is the following. The two particles bind at large masses with $a_0 > 0$ and $r_0 \approx 0$. When the mass is decreased, $a_0$ moves to a negative value, resulting in a virtual bound state. When the mass is decreased further, $r_0$ acquires a negative value, which causes the poles to move away from the imaginary axis, resulting in a resonance. This behaviour has been found for the $T_{cc}$ tetraquark when varying the charm-quark mass. For smaller-than-physical charm masses the $T_{cc}$ is unbound, while at the physical charm-quark mass it is bound~\cite{Collins:2024sfi}. The opposite holds for the up- and down-quark masses: the $T_{cc}$ is a virtual bound state if they are heavier than physical.

\subsection{Systems with spin}
\label{sec:spin_amplitude}

The discussion so far has assumed spinless particles. When the scattering particles carry spin, the description becomes more involved. This is relevant, for example, when investigating $DD^*$ or $BB^*$ scattering. One then works in the basis $|P, J, m_J, L, S\rangle$, where $J$ and $m_J$ are the total angular momentum and its projection onto the $z$-axis, $L$ is the orbital angular momentum of the two particles in the c.m.\ frame, and $S$ is their total spin. There may in addition be several channels. The $S$-matrix is now diagonal only in $J$ and $m_J$ (and the total momentum $P$), but not in $L$ and $S$. As a consequence, the interaction in a given $J$ can generally no longer be described by a single scattering phase shift, since $S_J$ is a matrix mixing different $(L, S)$ combinations. One then resorts to a $K$-matrix parametrization. We return to the consequences of this for the finite-volume analysis in Section~\ref{sec:spin_luscher}.

\section{Lüscher's finite-volume formalism}
\label{sec:luscher}

We now turn to the central methodological question of how a scattering amplitude can be obtained from finite-volume lattice data. Throughout this section, we follow the review of Ref.~\cite{Briceno:2017max}.

\subsection{Finite-volume problem and basic idea}
\label{sec:luscher_idea}

When a theory is placed in a finite spatial volume with periodic boundary conditions, the momenta are quantized as $\vb*{p} = \tfrac{2\pi}{L}\vb*{d}$ with $\vb*{d} \in \mathbb{Z}^3$, and the energy spectrum becomes discrete. This finite-volume setting poses several fundamental difficulties for the study of scattering. There is no one-to-one correspondence between the states in finite and infinite volume. In addition, asymptotic states, needed to define scattering, cannot be constructed in a finite box. Finally, the Euclidean signature does not grant direct access to dynamical, real-time quantities such as the scattering amplitude.

The key insight of the Lüscher formalism is to turn the finite volume from a limitation into a tool. Given some information about the infinite-volume theory, a quantization condition (QC) fixes the finite-volume states and their energies. Lüscher's QC does precisely this: it determines the finite-volume spectrum given the infinite-volume scattering amplitude, and thereby provides a relation between the two. Since the QC depends on the box size, the finite-volume spectrum is volume-dependent, and measuring it on the lattice provides, through the QC, access to the scattering amplitude.

\subsection{Finite-volume two-point function}
\label{sec:FV_two_point}

To derive the QC, we make a number of simplifying assumptions. Throughout this section, we assume that the continuum limit has been taken, i.e.\ that we work in a continuous but finite volume. In addition, we assume an infinite time extent, i.e.\ $T \rightarrow \infty$. This is usually a good assumption, as in most lattice QCD simulations we have $T = 2L$, which is large enough to suppress thermal effects sufficiently. We further assume that low-energy QCD can be described using hadrons as the effective degrees of freedom, as in chiral perturbation theory, although the derivation is largely agnostic to the specific effective theory used. We also restrict ourselves to a single channel and to kinematics for which at most two particles can go on shell. If three particles could go on shell, a three-particle Lüscher formalism would be required.

\begin{figure}
  \centering
  \includegraphics[width=0.9\textwidth]{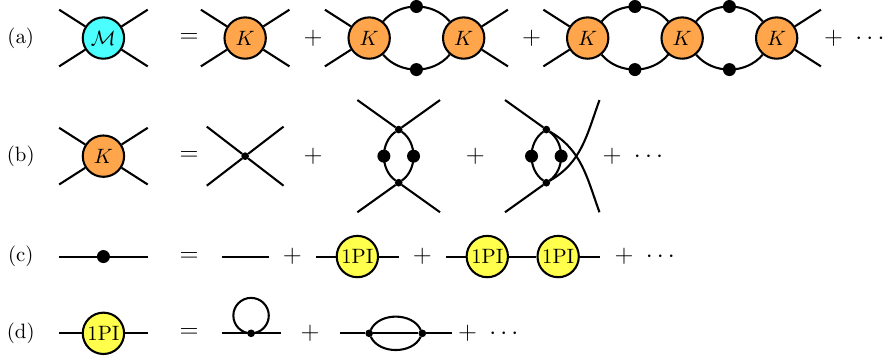}
  \caption{\label{fig:scattering_diagram}%
    Diagrammatic representation of the two-particle scattering amplitude (a) in terms of the Bethe--Salpeter kernel (b) and the fully dressed single-particle propagator (c). The last is given as the sum of all one-particle-irreducible diagrams (d).}
\end{figure}

In this framework, the two-particle scattering amplitude has a diagrammatic representation as the sum of all amputated, on-shell diagrams with two incoming and two outgoing legs. These diagrams can be resummed in a way that will turn out to be convenient for the derivation of Lüscher's QC. First, we write the fully dressed propagator as a geometric series in the self-energy --- the sum of all one-particle-irreducible diagrams --- with successive self-energy insertions joined by free propagators. From these propagators we then construct all two-particle-irreducible diagrams, which together form the Bethe--Salpeter kernel $K$. As a final step, the scattering amplitude is written as a geometric series in the Bethe--Salpeter kernels, in which successive kernels are connected by two fully dressed propagators. This representation is shown in Figure~\ref{fig:scattering_diagram} and holds in a general field theory with four-point vertices.

In a finite volume, two complications arise: asymptotic states cannot be defined, so there is no scattering process in the usual sense, and, because of the periodic boundary conditions, a particle can propagate around the box and interact again. To arrive at the QC, we instead consider a two-point correlation function that can be defined in both finite and infinite volume. In finite volume it reads
\begin{equation}
  \label{eq:FV_correlator}
  C_L(t' - t, \vb*{P}) = \int_L \m{d}^3\vb*{x}' \int_L \m{d}^3\vb*{x}\;
    \m{e}^{-\m{i}\vb*{P}\cdot(\vb*{x}'-\vb*{x})}\,
    \langle \mathcal{A}(x')\,\mathcal{B}^\dagger(x)\rangle_L,
\end{equation}
where $\mathcal{B}^\dagger$ and $\mathcal{A}$ are creation and annihilation operators, respectively, carrying the quantum numbers of the two-hadron channel under study. The spatial integral projects the correlator onto definite total momentum $\vb*{P} = \tfrac{2\pi}{L}\vb*{D}$ ($\vb*{D}\in \mathbb{Z}^3$), and the expectation value is taken in the finite-volume vacuum $|\Omega_L\rangle$. We denote the corresponding two-point function in infinite volume by $C_\infty$. Inserting a complete set of finite-volume energy eigenstates yields the spectral decomposition
\begin{equation}
  \label{eq:FV_spectral}
  C_L(t' - t, \vb*{P}) = L^6 \sum_n \m{e}^{-E_n(t'-t)}\,
    \langle \Omega_L|\mathcal{A}|E_n, \vb*{P}, L\rangle\,
    \langle E_n, \vb*{P}, L|\mathcal{B}^\dagger|\Omega_L\rangle.
\end{equation}
It contains the finite-volume energies $E_n$ of the system, the quantities we ultimately want to relate to the amplitude.

\begin{figure}
  \centering
  \includegraphics[width=0.68\textwidth]{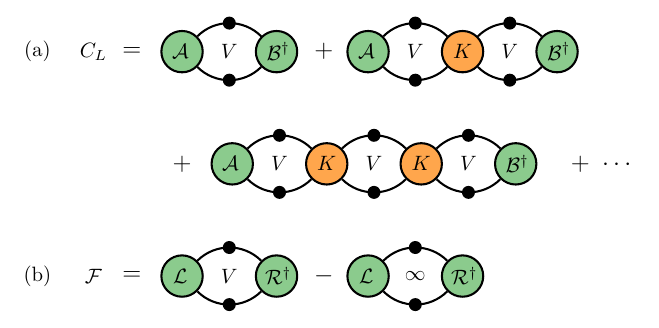}
  \caption{\label{fig:FV_two_point}%
    (a) Diagrammatic representation of the finite-volume two-point function for energies where two-particle states can go on shell. (b) Difference between the finite-volume and infinite-volume two-point loops.}
\end{figure}

Diagrammatically, under the assumptions above --- energies below any three-particle threshold, so that only two-particle states can go on shell, and only a single channel --- the finite-volume correlator has the following shape: it is a geometric series in the Bethe--Salpeter kernel inserted between the operators $\mathcal{A}$ and $\mathcal{B}^\dagger$, with two fully dressed single-particle propagators connecting them~\cite{Kim:2005gf}. This structure is shown in Figure~\ref{fig:FV_two_point}.
In the figure, the $V$ in the particle loop indicates a sum over the allowed discrete momenta rather than a loop integral. We denote the corresponding infinite-volume loop by $\infty$. Qualitatively, only on-shell particles can propagate over large distances and thereby sample the boundaries of the box, while the Bethe--Salpeter kernels describe the local interaction of the particles. As a result, in energy-momentum space the correlator develops pole singularities at the energies of the allowed finite-volume two-particle states. Diagrams in which the intermediate state cannot go on shell are non-singular and therefore do not contribute to these finite-volume poles.

The infinite-volume two-point function $C_\infty$ has the same diagrammatic representation, and we are ultimately interested in the difference $C_L - C_\infty$, which relates infinite- and finite-volume properties. The non-singular, off-shell contributions are common to both and vary smoothly with the loop momentum. As we will see below, such terms cancel in the finite-/infinite-volume difference up to terms that are exponentially suppressed in $L$. We therefore do not track them in what follows.

The reason for considering both $C_L$ and $C_\infty$ is the following. As mentioned above, in energy-momentum space the finite-volume correlator $C_L(P)$ has poles at the energies of the interacting finite-volume two-particle states that carry the quantum numbers of the system. In contrast, the infinite-volume correlator $C_\infty(P)$ has cuts in $E$ from the continuum of states and depends on the scattering amplitude. The difference between the two therefore encodes the relation between the infinite-volume scattering information and the finite-volume spectrum, and our goal is to compute this difference diagrammatically.

\subsection{Quantization condition and its interpretation}
\label{sec:QC}

The central object in the computation of the difference $C_L - C_\infty$ is the difference between the two-particle loop evaluated in finite and in infinite volume. We denote it by $\mathcal{F}$, and it is displayed at the bottom of Figure~\ref{fig:FV_two_point}. In finite volume, the loop momenta are summed rather than integrated, so this difference involves the sum-integral difference
\begin{equation}
  \label{eq:sum_integral_difference}
  \frac{1}{L^3}\sum_{\vb*{k}} - \int \frac{\m{d}^3\vb*{k}}{(2\pi)^3},
\end{equation}
where the sum runs over all $\vb*{k} = \frac{2\pi}{L}\vb*{n}$ with $\vb*{n} \in \mathbb{Z}^3$. A key property, following from the Poisson summation formula, is that this sum-integral difference, applied to a smooth function, decreases exponentially as $\m{e}^{-mL}$, where $m$ is the lightest mass scale (usually the pion mass $m_\pi$) and $L$ is the box size~\cite{Kim:2005gf}. We neglect such terms for $L \gg m_\pi^{-1}$. Only in the region where the two intermediate particles go on shell does the function become non-smooth, and it is this on-shell singularity that produces a power-law finite-volume dependence. Everything else varies smoothly with the loop momentum and hence cancels in $C_L - C_\infty$ up to the exponentially suppressed terms. These residual effects can be understood as virtual pions propagating around the box, which contribute at the level $\m{e}^{-m_\pi L}$.

Carrying out this analysis and neglecting exponentially suppressed terms shows that the on-shell contribution can be written as matrix-vector products in angular-momentum space. For generic functions $\mathcal{L}$ and $\mathcal{R}^\dagger$ in the loop it is given by
\begin{equation}
  \label{eq:FL_factorized1}
  \mathcal{F} = -\mathcal{L}_{\ell' m'}(P)\, F_{\ell' m', \ell m}(P, L)\, \mathcal{R}^\dagger_{\ell m}(P)
    = \mathcal{L}\, F\, \mathcal{R}^\dagger,
\end{equation}
where $\mathcal{L}$ and $\mathcal{R}^\dagger$ are partial-wave expanded in the c.m.\ momentum direction $\hat{\vb*{k}}_\m{cm}$ of the loop momentum. The matrix $F_{\ell'm',\ell m}(P, L)$ is purely kinematic and geometric and depends on the box size and the total momentum $P$.\footnote{In principle, $F$ is not a matrix, as the angular-momentum space is infinite-dimensional. We still use the term matrix, as it is used in the literature.} We defer details of the derivation to Appendix~\ref{app:loop_difference}, where an explicit form of $F$ is given in Eq.~(\ref{eq:F_matrix_appendix}). Because it captures all the non-exponentially-suppressed finite-volume dependence, $F$ is the only volume-dependent object in $\mathcal{F}$.

We use the same decomposition in angular-momentum space for the finite- and infinite-volume two-particle loop, which gives the matrices $F_L$ and $F_\infty$, respectively, satisfying $F_L - F_\infty = F$. In terms of $F_\infty$, we can write the scattering amplitude as a geometric series
\begin{equation}
  \label{eq:amplitude_geometric}
  \mathcal{M} = \sum_{n=0}^{\infty}(-K F_\infty)^n K,
\end{equation}
using the Bethe--Salpeter kernel (cf.\ Figure~\ref{fig:scattering_diagram}). In this expression, both the Bethe--Salpeter kernel and the amplitude are partial-wave expanded in the incoming and outgoing c.m.\ momentum directions $\hat{q}'_\m{cm}$ and $\hat{q}_\m{cm}$. For $\mathcal{M}$ this expansion is given by
\begin{equation}
  \label{eq:amplitude_kernel_expansion}
  \mathcal{M}(P, \hat{q}'_\m{cm}, \hat{q}_\m{cm})
    = Y_{\ell'm'}(\hat{q}'_\m{cm})^*\,\mathcal{M}_{\ell'm',\ell m}(E_\m{cm})\,Y_{\ell m}(\hat{q}_\m{cm}),
\end{equation}
where $\mathcal{M}_{\ell'm',\ell m}$ now depends only on the c.m.\ energy $E_\m{cm}$. We note that this partial-wave decomposition differs from the convention used in Section~\ref{sec:smatrix_partial_wave}.

The finite- and infinite-volume two-point function can also be written as such a series, multiplied, however, from the left and right by the partial-wave decompositions $A$ and $B^\dagger$ of the operators $\mathcal{A}$ and $\mathcal{B}^\dagger$, respectively. For $C_L$, it is given by
\begin{equation}
  C_L = -A\, F_L \sum_{n=0}^{\infty} (-K F_L)^n\, B^\dagger,
\end{equation}
and analogously for $C_\infty$ using $F_\infty$.

Using these building blocks, we can perform a resummation of the series appearing in $C_L - C_\infty$ to write it in terms of the finite-/infinite-volume difference $F$ and the scattering amplitude $\mathcal{M}$. We carry out this derivation explicitly in Appendix~\ref{app:resummation}. It leads to the compact result
\begin{equation}
  \label{eq:CL_minus_Cinf}
  C_L(P) = C_\infty(P) - \tilde{A}(P)\big[F^{-1}(P, L) + \mathcal{M}(E_\m{cm})\big]^{-1}\tilde{B}^\dagger(P),
\end{equation}
where the end-cap factors $\tilde{A}$ and $\tilde{B}^\dagger$, like $C_\infty$ and $\mathcal{M}$, are independent of the volume. The entire finite-volume dependence resides in $F$.

To extract the finite-volume energies, we transform Eq.~(\ref{eq:CL_minus_Cinf}) to time-momentum space and equate it with the spectral representation, Eq.~(\ref{eq:FV_spectral}),
\begin{align}
  \label{eq:CL_time_momentum}
  C_L(t'-t,\vb*{P}) &= L^3\int \frac{\m{d}P_4}{2\pi}\, \m{e}^{\m{i}P_4(t'-t)}
    \Big(C_\infty(P) - \tilde{A}\big[F^{-1}+\mathcal{M}\big]^{-1}\tilde{B}^\dagger\Big) \nonumber \\
    &= L^6 \sum_n \m{e}^{-E_n(t'-t)}\,
      \langle \Omega_L|\mathcal{A}|E_n,\vb*{P},L\rangle\,
      \langle E_n,\vb*{P},L|\mathcal{B}^\dagger|\Omega_L\rangle.
\end{align}
The integration over $P_4 = \m{i}E$ picks up the energy poles $E_n(L)$. Since the infinite-volume correlator $C_\infty$ does not depend on the finite-volume energies, these poles must arise from the second term in Eq.~(\ref{eq:CL_minus_Cinf}), i.e.\ from the energies at which the matrix $F^{-1} + \mathcal{M}$ is singular. This yields Lüscher's finite-volume QC,
\begin{equation}
  \label{eq:quantization_condition}
  \det\big[F^{-1}(P, L) + \mathcal{M}(E_\m{cm})\big] = 0,
\end{equation}
a determinant equation in angular-momentum space. Its interpretation is as follows: for a given total momentum $\vb*{P}$ and a given scattering amplitude $\mathcal{M}$, expressed in the angular-momentum basis, the discrete finite-volume spectrum $E_n(P, L)$ with the specified quantum numbers is given by all energies at which the determinant vanishes.

\subsection{\texorpdfstring{$s$}{s}-wave case}
\label{sec:swave_case}

The angular-momentum space in which Eq.~(\ref{eq:quantization_condition}) is formulated is infinite-dimensional and must be truncated at some maximal orbital angular momentum $\ell$. This truncation is justified near threshold, where higher partial waves are suppressed as $q_\m{cm}^\ell$. In the simplest case, only the $\ell = 0$ ($s$-wave) contribution is retained, and the determinant condition reduces to a single scalar equation,
\begin{equation}
  \label{eq:swave_QC}
  q_\m{cm}\cot\delta_0(q_\m{cm}) = \frac{2}{\sqrt{\pi}L\gamma}\,
    \mathcal{Z}_{00}\!\left(\frac{L}{2\pi}\vb*{P}, \gamma, \left(\frac{Lq_\m{cm}}{2\pi}\right)^2\right),
\end{equation}
where $\gamma = \tfrac{E}{E_\m{cm}}$ is the Lorentz factor relating the moving and c.m.\ frames, and $\mathcal{Z}_{00}$ is a generalized zeta function. We follow the notation of~\cite{Morningstar:2017spu} for $\mathcal{Z}$. This function encapsulates the finite-volume sum over the momenta and can be evaluated numerically.

\begin{figure}
  \centering
  \includegraphics[width=0.5\textwidth]{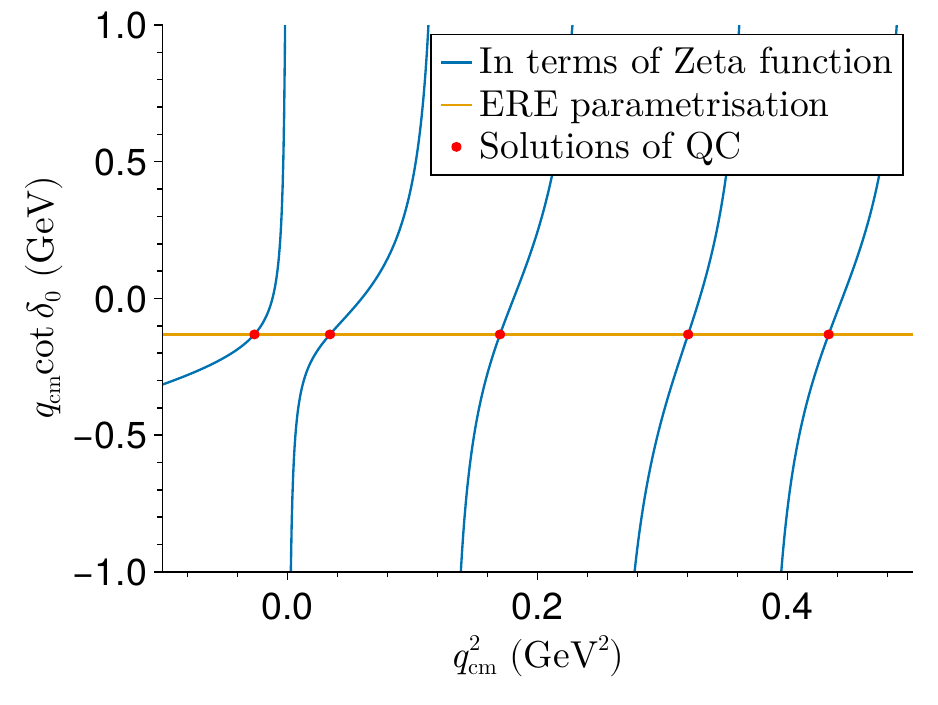}%
  \includegraphics[width=0.5\textwidth]{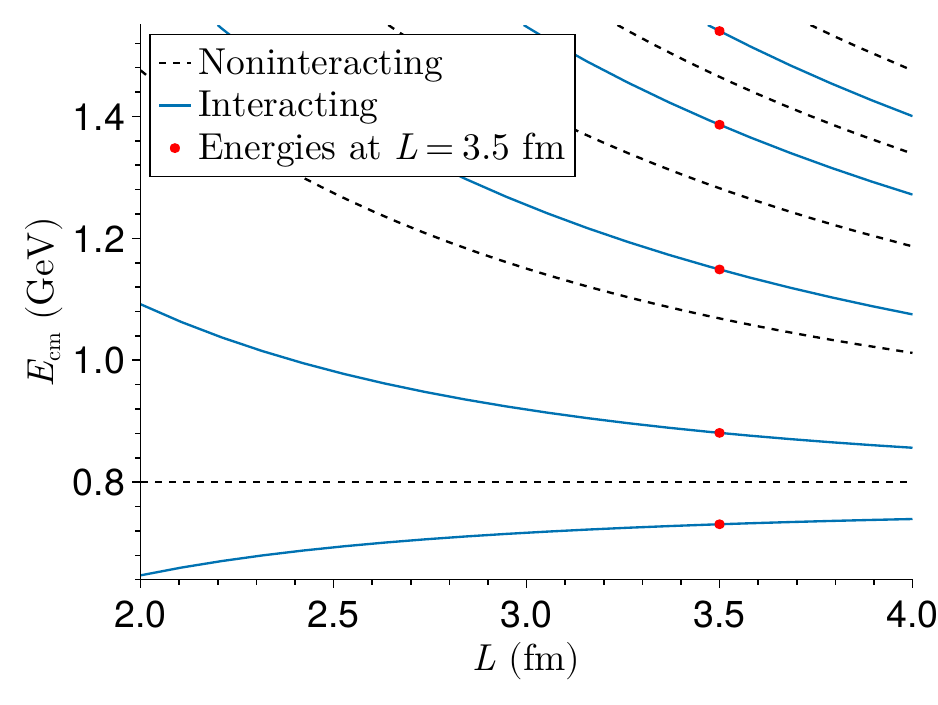}
  \caption{\label{fig:mock_luscher_analysis}%
    Left: graphical solution of the $s$-wave QC, Eq.~(\ref{eq:swave_QC}), in the rest frame. The generalized zeta function (right-hand side) is shown together with the constant-ERE phase shift $q_\m{cm}\cot\delta_0 = -1/a_0$; their intersections give the finite-volume energy levels. The curves use mock data with $m = 400\;\m{MeV}$, $L = 3.5\;\m{fm}$, and $a_0 = 1.5\;\m{fm}$. Right: finite-volume c.m.\ energy levels $E_\m{cm}$ as a function of the box size $L$, obtained from the $s$-wave QC as displayed in the left figure. The dashed black lines show the noninteracting two-particle energies, and the solid blue curves show the interacting levels, lowered by the attractive interaction. The red points are at $L = 3.5\;\m{fm}$, corresponding to the intersections in the left figure.}
\end{figure}

The way in which the QC determines the spectrum is most easily illustrated in this $s$-wave case using the leading-order ERE, i.e.\ a constant $q_\m{cm}\cot\delta_0 = -1/a_0$. For a given value of the scattering length $a_0$, solving Eq.~(\ref{eq:swave_QC}) yields the discrete finite-volume energy levels $E_\m{cm}$. This is shown in the left panel of Figure~\ref{fig:mock_luscher_analysis}, where the right-hand side of Eq.~(\ref{eq:swave_QC}) and the constant $-1/a_0$ are plotted as functions of $q_\m{cm}^2$; their intersections are the solutions of the QC and thus determine the $E_\m{cm}$.
The resulting volume dependence of the spectrum is shown in the right panel of Figure~\ref{fig:mock_luscher_analysis}. The noninteracting two-particle energies are $E_\m{cm} = 2\sqrt{m^2 + \vb*{p}^2}$ for back-to-back relative momenta with $\vb*{p}^2 = \left(\tfrac{2\pi}{L}\right)^2 n$, $n \in \mathbb{Z}_{\geq 0}$. The interacting levels are shifted relative to these, here lowered by the attractive interaction.

The two-particle Lüscher method has a limited range of validity. It applies only below the lowest three-particle threshold and above any left-hand cut, such as the one from one-pion exchange. In practice, these two bounds can substantially restrict the number of finite-volume energy levels that may be used in the analysis.

\subsection{Systems with spin}
\label{sec:spin_luscher}

As anticipated in Section~\ref{sec:spin_amplitude}, the formalism becomes more complicated when the scattering particles carry spin, but it has been worked out in the literature, and we use it in this work since the $DD^*$ and $BB^*$ systems have total spin $S = 1$. In this case, the matrices $F$ and $\mathcal{M}$ are formulated in the $|J, m_J, L, S\rangle$ basis, and the QC can be rewritten in terms of a $K$-matrix (cf.\ Section~\ref{sec:smatrix_partial_wave}) and a box matrix $B$ as
\begin{equation}
  \label{eq:QC_box_matrix}
  \det\big[K^{-1} - B\big] = 0,
\end{equation}
as done in~\cite{Morningstar:2017spu}. In this form, $K$ takes over the role of the scattering amplitude and $B$ encodes the finite-volume kinematics. For spin $S = 0$, the box matrix is related to $F$ by $B = \m{i}\rho + F$.

In a cubic finite volume, the full rotational symmetry of the continuum is reduced to that of the lattice. For a system with total momentum $\vb*{P}$, the relevant symmetry is then the little group of the lattice rotation group with respect to $\vb*{P}$, i.e.\ the subgroup of lattice rotations that leave $\vb*{P}$ invariant. Consequently, the finite-volume states and interpolating operators are classified by their irreducible representations (irreps), as discussed in more detail in Section~\ref{sec:lattice_irreps}.

This classification organizes the QC. While the $K$-matrix is diagonal in $J$ and $m_J$, the box matrix $B$ respects only the reduced lattice symmetry and mixes different values of $J$, though not different $S$. Grouping the QC according to the little-group irreps block-diagonalizes it, so that the different irreps decouple. One can then restrict the analysis to a single irrep, associated with the $J$ of interest~\cite{Morningstar:2017spu}. A truncation in $L$ is still required --- the $s$-wave analysis uses only $L = 0$ --- and is justified by the strong suppression of higher partial waves as $q_\m{cm}^L$.

The $s$-wave analysis nonetheless remains simple: at $L=0$ the only allowed total angular momentum $J$ is $J=S$. As a result, the $K$-matrix and the box matrix reduce to single numbers. Consequently, Eq.~(\ref{eq:QC_box_matrix}) again provides the one-to-one relation, Eq.~\eqref{eq:swave_QC}, between the finite-volume energy and the scattering phase shift, as in the spinless $s$-wave case.

\subsection{Extracting the amplitude from the finite-volume spectrum}
\label{sec:extracting_amplitude}

The QC determines the finite-volume spectrum from a given scattering amplitude, whereas in practice we wish to invert this relation: we measure the spectrum and want to determine the amplitude. The strategy for doing so is the following. One first chooses a parametrization of the scattering amplitude, such as the ERE of Eq.~(\ref{eq:ERE}). By solving the QC for given parameters, one obtains the low-lying finite-volume energies as a function of these parameters. The model parameters are then optimized by a least-squares fit of the predicted spectrum to the measured one.

\section{Extraction of finite-volume energies}
\label{sec:fv_energy_extraction}

Having established how the finite-volume spectrum is related to the scattering amplitude, we now describe how the low-lying finite-volume energy levels are extracted from lattice QCD data, as required for the Lüscher analysis. These energies are obtained from Euclidean two-point functions of suitable interpolating operators, and the variational method is used to resolve the excited states in the multi-hadron spectrum.

\subsection{Euclidean two-point functions}
\label{sec:euclidean_two_point}

The finite-volume energies are encoded in the Euclidean two-point function $C(t) = \langle\mathcal{O}(t)\mathcal{O}(0)^\dagger\rangle$ of an operator $\mathcal{O}$ carrying the quantum numbers of interest, for example those of the pion. Inserting a complete set of energy eigenstates gives the spectral decomposition
\begin{equation}
  \label{eq:spectral_decomposition}
  C(t)
    = \sum_{n=0}^{\infty}
      \langle\Omega|\mathcal{O}|n\rangle\langle n|\mathcal{O}^\dagger|\Omega\rangle \,\m{e}^{-E_n t}
    = \sum_{n=0}^{\infty} A_n\, \m{e}^{-E_n t},
\end{equation}
where $|\Omega\rangle$ is the finite-volume vacuum and $A_n = \abs{\langle\Omega|\mathcal{O}|n\rangle}^2$ are the overlap factors, up to normalization. In general, the operator $\mathcal{O}$ does not create just a single state. Instead, it couples to all states $|n\rangle$ with the same quantum numbers, i.e.\ to all energy levels arising from the quantization of the system in the finite volume. For the pion, these include, in addition to the single-pion state, a three-pion state, a pion-proton-antiproton state, and so on. We order their energies such that $E_n < E_{n+1}$.

This form of the spectral decomposition assumes an infinite time extent and a Hermitian, positive transfer matrix. In practice, both assumptions are only approximately satisfied: for periodic boundary conditions, finite-time effects are suppressed for $t \ll T/2$ or can be accounted for explicitly, and for open boundary conditions they are suppressed for operators placed far from the temporal boundaries. The violations of positivity caused by improvement terms in the action are negligible for $t \gg a$.

At large Euclidean times, the sum in Eq.~(\ref{eq:spectral_decomposition}) is dominated by the ground state,
\begin{equation}
  \label{eq:large_t_behaviour}
  \langle\mathcal{O}(t)\mathcal{O}(0)^\dagger\rangle
    = A_0\, \m{e}^{-E_0 t}\left[1 + O\!\left(\m{e}^{-(E_1 - E_0)t}\right)\right],
\end{equation}
so that the ground-state energy $E_0$ can be extracted from the exponential decay. A convenient way to do so is via the effective energy
\begin{equation}
  \label{eq:effective_energy}
  E_\m{eff}(t) = -\frac{1}{a}\log\!\left(\frac{C(t+a)}{C(t)}\right)
    \xrightarrow[t \to \infty]{} E_0.
\end{equation}
It is the discrete logarithmic derivative of the two-point function $C(t)$, from which we obtain $E_0$, e.g.\ by a plateau fit at large times. In the rest frame, $E_\m{eff}$ is also called the effective mass, denoted by $m_\m{eff}$, because $E_0$ is then the mass of the particle corresponding to the ground state.

As a concrete example, we consider the charged pion, with quantum numbers $J^P = 0^-$ (a pseudoscalar). An interpolating operator carrying these quantum numbers is
\begin{equation}
  \label{eq:pion_operator}
  \mathcal{O}^\pi(\vb*{P}, t) = \sum_{\vb*{x} \in \Lambda_3}
    \m{e}^{-\m{i}\vb*{P}\cdot\vb*{x}}\,(\overline{d}\,\gamma_5\, u)(\vb*{x}, t),
\end{equation}
projected onto (total) momentum $\vb*{P}$, where $\Lambda_3$ is the spatial lattice. This is the operator for the $\pi^+$, but we suppress the charge label, as it is not relevant in pure QCD. Applying the Wick contractions to the two-point function yields
\begin{equation}
  \label{eq:pion_correlator}
  C^\pi(t) = \sum_{\vb*{x}', \vb*{x} \in \Lambda_3} \m{e}^{-\m{i}\vb*{P}\cdot(\vb*{x}'-\vb*{x})}\,
    \ev{\tr\!\left[\gamma_5 S_u(\vb*{x}', t; \vb*{x}, 0)\;
      \gamma_5 S_d(\vb*{x}, 0; \vb*{x}', t)\right]}_G,
\end{equation}
which involves the full inverse of the Dirac operator, $S_f = D_f^{-1}$. Computing this inverse in its entirety is prohibitively expensive, so one instead resorts to one of several methods to evaluate it. In the point-to-all method, one exploits spatial translation invariance and $\gamma_5$-hermiticity so that only twelve columns of the inverse Dirac operator --- one for each colour and spin index --- are required. These columns are obtained by inverting the Dirac operator on point sources. In the stochastic-estimator approach, one introduces noise vectors $\eta_i$ satisfying $\langle \eta_i \eta_j^\dagger \rangle = \delta_{ij}$ and inserts the stochastic estimator $\tfrac{1}{N}\sum_{i=1}^N\eta_i \eta_i^\dagger$ for the identity between the two Dirac operators. Solving the Dirac equation on the noise vectors then allows one to estimate the trace in the two-point function stochastically for sufficiently large $N$.

A generic difficulty is the signal-to-noise problem: for most systems, the signal decreases exponentially faster than the noise as the Euclidean time $t$ increases. The pion at rest, $\vb*{P} = \vb*{0}$, is a notable exception. Consequently, one cannot go to arbitrarily large $t$, and the signal is often lost before a clear plateau emerges in the effective energy. This calls for alternative extraction methods that mitigate this problem. Furthermore, one is typically interested in the excited-state energies as well. For a two-particle system, for instance, the few lowest levels in the finite-volume spectrum are needed to constrain the scattering amplitude reliably through the L\"{u}scher quantization condition. Both of these considerations motivate the use of the variational method.

\subsection{Variational method}
\label{sec:variational_method}

If the effective energy of a single operator does not plateau early enough, the ground-state overlap factor $A_0$ is too small relative to the others, i.e.\ the operator does not predominantly create the ground state. This can be improved by using a basis of operators $\mathcal{O}_i$, $i = 1, 2, \dots, N$, and constructing an optimized operator for the $n$-th level as a linear combination
\begin{equation}
  \label{eq:optimized_operator}
  \mathcal{O}_{\m{opt},n}(t) = \sum_{i=1}^{N} v_{n,i}^*\,\mathcal{O}_i(t).
\end{equation}
The factors $v_{n,i}^*$ must be tuned such that the overlap factors $A_m$ are small for $m \neq n$. In that way, the two-point function is dominated by a single state, $\langle\mathcal{O}_{\m{opt},n}(t)\,\mathcal{O}_{\m{opt},n}(0)^\dagger\rangle \approx \tilde{A}_n\, \m{e}^{-E_n t}$.

Determining the coefficient vectors $v_n = (v_{n,1}, \dots, v_{n,N})$ amounts to solving the generalized eigenvalue problem (GEVP)
\begin{equation}
  \label{eq:GEVP}
  C(t)\, v_n(t, t_0) = \lambda_n(t, t_0)\; C(t_0)\, v_n(t, t_0)
\end{equation}
for a reference time $t_0 < t$, where
\begin{equation}
  \label{eq:correlator_matrix}
  C_{ij}(t) = \ev*{\mathcal{O}_i(t)\,\mathcal{O}_j(0)^\dagger}
\end{equation}
is the correlator matrix of the operator basis. We refer to $\mathcal{O}_j(0)$ as the source operator at source time ${t=0}$ and to $\mathcal{O}_i(t)$ as the sink operator. The generalized eigenvalues satisfy $\lambda_n(t, t_0) \approx \m{e}^{-E_n(t - t_0)}$, so that the energy levels can be extracted either from an exponential fit to $\lambda_n$ or, as in the single-operator case, from a plateau fit to the effective energy
\begin{equation}
  \label{eq:GEVP_effective_energy}
  E_{\m{eff},n}(t) = -\frac{1}{a}\log\!\left(\frac{\lambda_n(t+a, t_0)}{\lambda_n(t, t_0)}\right).
\end{equation}
The correlator matrix is Hermitian by construction, and in practice we choose the operators such that it is real and symmetric. Assuming a positive transfer matrix\footnote{Since positivity is only approximately satisfied for improved actions (see Section~\ref{sec:euclidean_two_point}), this can only be expected at sufficiently large Euclidean times.} and linearly independent operators, the matrices $C(t)$ and $C(t_0)$ are positive definite, and the eigenvalues $\lambda_n$ are positive. The ground-state effective energy $E_{\m{eff},0}(t)$ then approaches its plateau from above. For excited states, such an approach from above is not guaranteed.

If we additionally choose $t_0 \geq t/2$, it is ensured that~\cite{Blossier:2009kd}
\begin{equation}
  \label{eq:GEVP_suppression}
  E_{\m{eff},n}(t) = E_n + O\!\left(\m{e}^{-(E_N - E_n)t}\right),
\end{equation}
i.e.\ the excited-state contamination is suppressed such that, for a basis of $N$ operators, the leading contamination comes from the $N$-th excited state. The effective energies are therefore expected to reach their plateaus quickly, which mitigates the signal-to-noise problem. This method requires the full correlator matrix of all operators in the basis, which can be computationally expensive, as discussed in later sections.

In practice, the method fails if all operators in the basis have a particularly small overlap with a given energy level. That level then cannot be resolved well, and the resulting finite-volume spectrum may be qualitatively wrong~\cite{Wilson:2015dqa}. The choice of the operator basis is therefore crucial. An example relevant to this work is the $BB^*$ system, where diquark-antidiquark operators, in addition to bilocal two-meson operators, are needed to resolve the spectrum reasonably well~\cite{Prelovsek:2025vbr}. As a general guideline, the basis should contain operators with a variety of spin, colour, and spatial structures. For the $DD^*$ and $BB^*$ systems, in particular, one wants both bilocal two-meson operators and local tetraquark operators. Their construction is the subject of the next section.

Before describing procedures for constructing such operators, we mention another method for improving the signal in two-point functions: source-time averaging. Owing to the translational invariance of the lattice, the two-point function $\langle\mathcal{O}_i(t+t_\m{src})\mathcal{O}_j(t_\m{src})^\dagger\rangle$ can be computed at multiple source times $t_\m{src}$ and then averaged. Time-reversal symmetry additionally allows one to average over positive and negative $t$, since, in the limit of infinite statistics, the correlators at $t$ and $-t$ are equal up to a possible sign flip. We perform this average as $\frac{1}{2}\!\left(C_{ij}(t) + \eta_i\eta_jC_{ji}(-t)\right)$, where $\eta_i=\pm 1$ is the sign acquired by the operator $\mathcal{O}_i$ under time reversal. Finally, since we construct the operators such that the correlator matrix is real in expectation, its imaginary part contains only statistical noise, and we therefore discard it.

One has to be cautious, however, when performing the source-time and $\pm t$ averages with open temporal boundary conditions. Close to the boundaries, translational invariance in time is lost, and the correlator receives additional contributions from boundary states that are not captured by Eq.~(\ref{eq:spectral_decomposition}). The source times must therefore be placed such that all source times $t_\m{src}$ and sink times $\pm t+t_\m{src}$ used in the analysis are sufficiently far from the boundaries. This can substantially restrict the range of usable time slices, and for source times near a boundary, only the time direction pointing away from that boundary can be used. We determined how close to the boundaries the source times can be placed by varying the number of source times included in the average, choosing this number such that the energy estimates are stable under the variation. Boundary effects are thus negligible within the statistical uncertainties, and Eq.~(\ref{eq:spectral_decomposition}) remains applicable.

\section{Construction of interpolating operators}
\label{sec:operator_construction}

To use the variational method, we have to construct a basis of interpolating operators that create the states of interest. On the lattice, these operators must be organized according to the symmetries of the discretized theory, which differ from those in the continuum. In this section we first discuss how the continuum angular-momentum content is subduced into lattice irreducible representations (irreps), and then construct the single-meson, tetraquark, and bilocal two-meson operators used in this work.

\subsection{Lattice irreps and the subduction of angular momentum}
\label{sec:lattice_irreps}

In the continuum, hadronic states are classified by quantum numbers, such as the total angular momentum $J$, parity $P$, isospin $I$, and other flavour quantum numbers. On the lattice, the translational and rotational symmetries of continuous spacetime are broken down to discrete translations and rotations. In a cubic box, momentum remains a good quantum number, but angular momentum does not, so that states of different partial waves mix. The residual rotational symmetry, moreover, depends on the total momentum, since one cannot boost to the rest frame on the lattice. As a consequence, the continuum $J^P$ representations are in general no longer irreps of the relevant rotation group on the lattice. Instead, they subduce to a sum of lattice irreps. In the following, we restrict ourselves to mesons and therefore consider only integer spin.

To label the lattice irreps, we use the Mulliken symbols~\cite{cotton1991chemical}, the standard notation for the irreps of point groups in lattice QCD spectroscopy. They consist of a capital letter, potentially decorated with a subscript 1 or 2 and a superscript $+$ or $-$. The letter denotes the dimension of the irrep: $A$ and $B$ for one-dimensional, $E$ for two-dimensional, and $T$ for three-dimensional representations, with further symbols that are not relevant here. The subscript --- and the letter $A$ or $B$ in the one-dimensional case --- distinguishes irreps according to their transformation properties under certain rotations and reflections. In the rest frame, the superscript $+$ or $-$ denotes parity.\footnote{An alternative notation uses $g$ and $u$ for $+$ and $-$, respectively.} In moving frames, parity is not part of the little group, so that states of opposite parity mix and the superscript is dropped.

The relevant residual rotational symmetries are the little groups of the total momenta used in the analysis, listed in Table~\ref{tab:point_groups}.
\begin{table}
  \centering
  \renewcommand{\arraystretch}{1.2}
  \begin{tabular}{lcccc}
    \hline \hline
    $\vb*{D}$ & $(0, 0, 0)$ & $(0, 0, n)$ & $(0, n, n)$ & $(n, n, n)$ \\
    \hline
    Point group & $O_h$ & $C_{4v}$ & $C_{2v}$ & $C_{3v}$ \\
    \hline \hline
  \end{tabular}
  \caption{\label{tab:point_groups}%
    Little groups (point groups) of the residual rotational symmetry for the total momenta $\vb*{P} = \tfrac{2\pi}{L}\vb*{D}$, with non-negative integer $n$ (usually $n = 0, 1, 2$). Character tables of these groups can be found, e.g., in~\cite{dresselhaus2007group, altmann_point-group_1994}.}
\end{table}
The subduction of the lowest continuum $J^P$ representations into these lattice irreps is, for the rest frame and for the $\vb*{D} = (0,0,1)$ frame,
\begin{equation}
  \label{eq:subduction_examples}
  \begin{aligned}
    \vb*{D} = \vb*{0}: \quad &0^\pm \cong A_1^\pm
      &\vb*{D} = (0,0,1): \quad &0^+ \cong A_1 \\
    &1^\pm \cong T_1^\pm
      & &1^+ \cong A_2 \oplus E \\
    &2^\pm \cong E^\pm \oplus T_2^\pm
      & &2^+ \cong A_1 \oplus B_1 \oplus B_2 \oplus E
  \end{aligned}
\end{equation}
where for the moving frame the results for negative parity, $J^+ \rightarrow J^-$, are obtained by the exchanges $A_1 \leftrightarrow A_2$ and $B_1 \leftrightarrow B_2$. As anticipated, the two parities mix for $\vb*{D} \neq \vb*{0}$.

To reduce the mixing of continuum total angular momenta as far as possible, and to be able to relate the lattice results to the continuum theory, one constructs operators that transform in a definite lattice irrep. Suppose an operator $\mathcal{O}$ transforms under a rotation or reflection $R \in \m{O}(3)$ according to a representation $D(R)$,
\begin{equation}
  \label{eq:operator_transformation}
  \mathcal{O} \xrightarrow{R} D(R)\,\mathcal{O},
\end{equation}
where, for now, we suppress the transformation of the momenta carried by $\mathcal{O}$. Given a total momentum $\vb*{P} = \tfrac{2\pi}{L}\vb*{D}$, we wish to subduce this operator into an irrep $\Lambda$ of the little group $G_{\vb*{P}}$ of $\vb*{P}$, with representation matrices $D^{\Lambda}_{\mu\nu}(R)$ of dimension $d_\Lambda$. After first restricting the operator to the lattice --- i.e.\ using the allowed lattice momenta and discretizing any covariant derivatives --- one can project onto the row $\mu$ of the irrep $\Lambda$ using the projector~\cite{dresselhaus2007group}
\begin{equation}
  \label{eq:projector}
  P^\Lambda_{\mu\nu} = \frac{d_\Lambda}{|G_{\vb*{P}}|}\sum_{R \in G_{\vb*{P}}} D^\Lambda_{\mu\nu}(R)^*\,D(R),
\end{equation}
where $|G_{\vb*{P}}|$ is the order of the little group. The projected operator is
\begin{equation}
  \label{eq:projected_operator}
  \mathcal{O}^\Lambda_\mu = P^\Lambda_{\mu\nu}\,\mathcal{O}.
\end{equation}
For a fixed target row $\mu$, one tests the different values of $\nu$ until a non-vanishing operator is obtained; if the result vanishes for all $\nu$, the operator does not subduce into the irrep $\Lambda$. We use this construction throughout the following.

For the orbital and spin degrees of freedom, we work with spherical harmonics in the Cartesian basis, in which the basis for angular momentum $\ell$ is formed by symmetric, traceless rank-$\ell$ tensors. Under an $\m{O}(3)$ transformation $R$, each tensor index is rotated by one factor of $R$, so that the tensor transforms in the spin-$\ell$ representation. The spherical harmonics for the lowest values of $\ell$ are listed in Table~\ref{tab:spherical_harmonics}.
\begin{table}
  \centering
  \renewcommand{\arraystretch}{1.2}
  \begin{tabular}{lccc}
    \hline \hline
    Angular momentum & $\ell = 0$ & $\ell = 1$ & $\ell = 2$ \\
    \hline
    Spherical harmonic & $1$ & $p_i$ & $p_ip_j - \tfrac{1}{3}\delta_{ij}\,p_kp_k$ \\
    \hline \hline
  \end{tabular}
  \caption{\label{tab:spherical_harmonics}%
    Spherical harmonics in the Cartesian basis as functions of the momentum $\vb*{p}$, for the lowest values of $\ell$ and up to normalization.}
\end{table}

\subsection{Single-meson operators}
\label{sec:single_particle_operators}

For a single meson, the degrees of freedom relevant for the rotational symmetry are spin and orbital angular momentum coming from covariant derivatives inserted into the quark bilinear. As an example, we consider the $D$ meson, a pseudoscalar. We build bilinear operators of the form
\begin{equation}
  \label{eq:single_meson_operator}
  \mathcal{O}_\mu(\vb*{P}, t) = \sum_{\vb*{x} \in \Lambda_3} \m{e}^{-\m{i}\vb*{P}\cdot\vb*{x}}\,
    (\overline{d}\,\Gamma_\mu\, c)(\vb*{x}, t) = (\overline{d}\,\Gamma_\mu\, c)(\vb*{P}, t),
\end{equation}
where $\Gamma_\mu$ encapsulates the spin structure and any covariant derivatives, and $\mu$ is the row index of the corresponding irrep. In the last step, we introduced a shorthand notation for the momentum projection. Under a little-group rotation $R \in G_{\vb*{P}}$, which by definition satisfies $R\vb*{P} = \vb*{P}$, the operator is supposed to transforms as
\begin{equation}
  \label{eq:single_meson_transformation}
  \mathcal{O}_\mu(\vb*{P}, t) \xrightarrow{R} D_{\mu\nu}^\Lambda(R)\,\mathcal{O}_\nu(\vb*{P}, t)
\end{equation}
for a lattice irrep $\Lambda$, where a summation over repeated indices is understood. The relevant lattice irreps for the $D$ meson for the lowest two momenta are $A_1^-$ (rest frame) and $A_2$ ($\vb*{D} = (0,0,1)$). Restricting to at most one covariant derivative in $\Gamma$, one can show that the possible continuum operators that subduce to these irreps are those with the Dirac structures listed in Table~\ref{tab:single_particle_ops}, organized by the spin-parity $S^P$ they carry.
\begin{table}
  \centering
  \renewcommand{\arraystretch}{1.2}
  \begin{tabular}{lccc}
    \hline \hline
    $S^P$ & $0^-$ & $1^+$ & $2^-$ \\
    \hline
    $\Gamma$
      & $\begin{array}{l}\gamma_5 \\ 
        D_i\gamma_i\gamma_5 \\ 
        \end{array}$
      & $\begin{array}{l}\gamma_i\gamma_5 \\ 
          D_i\gamma_5 \\ 
        \epsilon_{ijk}D_j\gamma_k \\ 
        \end{array}$
      & $\begin{array}{l}(D_i\gamma_j - \tfrac{1}{3}\delta_{ij}D_k\gamma_k)\gamma_5 
        \end{array}$ \\
    \hline \hline
  \end{tabular}
  \caption{\label{tab:single_particle_ops}%
    Possible Dirac structures $\Gamma$ for the continuum quark bilinear $\overline{d}\,\Gamma\, c$ with up to one covariant derivative $D_i$, organized by the spin-parity $S^P$. The complete set is given by these structures and their products with $\gamma_4$, which does not change $S^P$. A summation over repeated indices ($i,j,k=1,2,3$) is understood.}
\end{table}
To subduce these into the lattice irreps, we first replace the continuum covariant derivative $D_i$ by its lattice discretization $\nabla^\m{sym}_i$.

In the rest frame, $0^-$ subduces into $A_1^-$, and of the structures in Table~\ref{tab:single_particle_ops} only the $0^-$ operators subduce into this irrep (see Eq.~\eqref{eq:subduction_examples}). Since $A_1^-$ is one-dimensional with representation $D^{A_1^-}(R) = \det(R)$, the projector is trivial --- the two factors of $\det(R)$ cancel --- and one simply uses the $0^-$ operators of the table, giving four operators. For the $\vb*{D} = (0,0,1)$ frame, the construction is analogous: applying the projector into the $A_2$ irrep, the $S = 0$ structure subduces trivially, while for $S = 1$ only the $i = 3$ component and for $S = 2$ only the $i = j = 3$ component survive. This yields twelve $A_2$ operators. Similarly, we can construct vector operators for the $D^*$ meson.

\subsection{Tetraquark operators}
\label{sec:tetraquark_operators}

Before turning to two-meson operators, we consider local tetraquark operators. These behave similarly to single-meson operators but offer more freedom in their spin and colour structure. However, we do not use covariant derivatives in their construction here. As an example, we take the $I(J^P) = 0(1^+)$ channel relevant for the $DD^*$ system, and use input from the quark model to guide the choice of operators.

In the continuum, one can place two mesons at the same point and couple their spins to reach the $1^+$ representation. This gives, for example, the local $DD^*$ operator
\begin{equation}
  \label{eq:local_DDstar}
  T_i^{DD^*}(t) = (\overline{u}\,\gamma_5\, c\;\overline{d}\,\gamma_i\, c)(\vb*{P}, t)
    - \{u \leftrightarrow d\},
\end{equation}
and the local $D^*D^*$ operator
\begin{equation}
  \label{eq:local_DstarDstar}
  T_i^{D^*D^*}(t) = \epsilon_{ijk}\,(\overline{u}\,\gamma_j\, c\;\overline{d}\,\gamma_k\, c)(\vb*{P}, t),
\end{equation}
with $i = 1, 2, 3$, where all quarks are jointly projected onto the total momentum $\vb*{P}$, making the operators local. We suppress the dependence on the total momentum $\vb*{P}$ in the operator names. Further operators can be obtained by changing the colour structure. Combining the $cc$ pair and the $\overline{u}\overline{d}$ pair into a diquark and an antidiquark, respectively, a colour singlet can be formed either from an antitriplet and a triplet, $(\vb*{\overline{3}}_c \otimes \vb*{3}_c)_{\vb*{1}_c}$, or from a sextet and an antisextet, $(\vb*{6}_c \otimes \vb*{\overline{6}}_c)_{\vb*{1}_c}$. We use the former, motivated by the attractive interaction in the $\vb*{\overline{3}}_c$ diquark and the $\vb*{3}_c$ antidiquark found within the quark model~\cite{Huang:2023jec}. For the $(\vb*{\overline{3}}_c \otimes \vb*{3}_c)_{\vb*{1}_c}$ combination one obtains the diquark-antidiquark operator
\begin{equation}
  \label{eq:diquark_operator}
  T_i^\m{diq}(t) = \left(\epsilon_{abc}\, c_b^T\, C\gamma_i\, c_c\;
    \epsilon_{ade}\,\overline{u}_d\, C\gamma_5\, \overline{d}_e^T\right)(\vb*{P}, t),
\end{equation}
where $C$ is the charge-conjugation matrix and $a$, $b$, $c$, $d$ and $e$ are colour indices.

Using Fierz transformations, the diquark-antidiquark operator $T_i^\m{diq}$ can be rewritten as a linear combination of products of two colour-singlet currents of the form
\begin{equation}
  \label{eq:fierz_form}
  (\overline{u}\,\Gamma_A\, c\;\overline{d}\,\Gamma_B\, c)(\vb*{P}, t),
\end{equation}
for various combinations of Clifford-algebra elements $\Gamma_A$ and $\Gamma_B$~\cite{Nieves:2003in, Padmanath:2015era}. This relates $T_i^\m{diq}$ to the local $DD^*$ and $D^*D^*$ operators. Nevertheless, unless one uses a complete basis of operators of the type given in Eq.~\eqref{eq:fierz_form}, the diquark-antidiquark operator remains linearly independent of the local $DD^*$ and $D^*D^*$ operators. Subducing these operators to the lattice is trivial, since there is a one-to-one correspondence between the $1^+$ and $T_1^+$ irreps. In the simulation of the $DD^*/T_{cc}$ system, we use all three of these tetraquark operators.

\subsection{Bilocal two-meson operators}
\label{sec:bilocal_operators}

We now construct bilocal two-meson operators, built from two single-meson operators that transform irreducibly under a lattice irrep $\Lambda_J$ of the little group $G_{\vb*{P}}$ for a given total momentum $\vb*{P}$. In contrast to the single-meson case, these operators carry an additional degree of freedom that transforms under rotations: the orbital angular momentum coming from the relative momentum $\vb*{p}$ between the two mesons. We can combine this degree of freedom with spin, which allows us to define a variety of operators.

If several operators are to be used for the same single meson, as for the $D$, it is best to first construct an optimized single-meson operator using the GEVP and to build the two-meson operators from that single operator. Otherwise, the size of the operator basis for the two-particle system grows rapidly. This has been done, e.g., in Ref.~\cite{Whyte:2024ihh}.

There are several ways to construct these operators, of which we mention two. In the first, one constructs two-meson operators in the infinite-volume continuum that transform according to the desired continuum irrep, combining orbital angular momentum and spin with Clebsch--Gordan (CG) coefficients. One then restricts them to the lattice --- i.e.\ restricting to lattice momenta and replacing derivatives with finite differences --- and subduces them into the lattice irrep with the projector. In the second strategy, the single-meson operators are first coupled to definite spin in the continuum and subduced into a lattice spin irrep $\Lambda_S$. They are then combined with the lattice orbital-angular-momentum irreps using the ``lattice'' CG coefficients to obtain operators in the desired irrep. We use the second strategy, as it makes the origin of each operator more explicit.

Following this approach, we first couple two single-meson operators to definite spin, with total momentum $\vb*{P}$ and relative momenta $\vb*{p}$ and $\vb*{P} - \vb*{p}$ for the two particles. We then subduce the spin into the lattice spin irrep $\Lambda_S$. This gives an operator $\mathcal{O}^{\Lambda_S}_\mu(\vb*{p}, t)$, where $\mu$ is the row index of the irrep. The dependence on the total momentum $\vb*{P}$ is suppressed in the notation. This operator transforms as
\begin{equation}
  \label{eq:spin_operator_transformation}
  \mathcal{O}^{\Lambda_S}_\mu(\vb*{p}, t) \xrightarrow{R}
    D^{\Lambda_S}_{\mu\nu}(R)\,\mathcal{O}^{\Lambda_S}_\nu(R^{-1}\vb*{p}, t)
\end{equation}
under $R \in G_{\vb*{P}}$.

For the orbital angular momentum, we take the spherical harmonics in the Cartesian basis and subduce them into the lattice irreps; the resulting polynomials can be found, e.g., in~\cite{dresselhaus2007group, altmann_point-group_1994}. We denote them by $T^{\Lambda_L}_\mu(\hat{\vb*{p}})$, where $\hat{\vb*{p}}$ is the unit vector of $\vb*{p}$ and $\Lambda_L$ is the corresponding lattice irrep, and they satisfy
\begin{equation}
  \label{eq:orbital_harmonic_transformation}
  T^{\Lambda_L}_\mu(R\hat{\vb*{p}}) = D^{\Lambda_L}_{\mu\nu}(R)\,T^{\Lambda_L}_\nu(\hat{\vb*{p}}).
\end{equation}
An operator $\mathcal{O}(\vb*{p}, t)$ can then be projected onto definite orbital angular momentum as
\begin{equation}
  \label{eq:orbital_projection}
  \sum_{\vb*{p} \in \mathcal{S}_n} T^{\Lambda_L}_\mu(\hat{\vb*{p}})\,\mathcal{O}(\vb*{p}, t),
\end{equation}
where $\mathcal{S}_n = \{\vb*{p} = \tfrac{2\pi}{L}\vb*{d} \mid \vb*{d} \in \mathbb{Z}^3,\, \vb*{d}^2 = n\}$ is the $n$-th momentum shell.

Finally, we combine the spin and orbital parts using the CG coefficients of the little group $G_{\vb*{P}}$. To that end, we search for tensor products $\Lambda_L \otimes \Lambda_S$ that subduce into the desired irrep $\Lambda_J$. We then combine $T^{\Lambda_L}_{\mu_L}(\hat{\vb*{p}})$ and $\mathcal{O}^{\Lambda_S}_{\mu_S}(\vb*{p}, t)$ with the CG coefficients $\langle \Lambda_L \mu_L\, \Lambda_S \mu_S | \Lambda_J \mu_J \rangle$, which describe the change of basis from $(\Lambda_L \otimes \Lambda_S)_{\Lambda_J}$ to $\Lambda_J$. The resulting operator on the $n$-th momentum shell and for the row index $\mu_J$ is
\begin{equation}
  \label{eq:bilocal_operator}
  \mathcal{O}^{(\Lambda_L \otimes \Lambda_S)_{\Lambda_J},\, n}_{\mu_J}(t)
    = \sum_{\vb*{p} \in \mathcal{S}_n} \langle \Lambda_L \mu_L\, \Lambda_S \mu_S | \Lambda_J \mu_J \rangle\,
      T^{\Lambda_L}_{\mu_L}(\hat{\vb*{p}})\,\mathcal{O}^{\Lambda_S}_{\mu_S}(\vb*{p}, t),
\end{equation}
with a summation over $\mu_L$ and $\mu_S$ understood. The CG coefficients can be found in~\cite{altmann_point-group_1994}. This operator has the correct transformation properties under a little-group rotation $R$,
\begin{equation}
  \label{eq:bilocal_transformation}
  \mathcal{O}^{(\Lambda_L \otimes \Lambda_S)_{\Lambda_J},\, n}_{\mu}(t) \xrightarrow{R}
    D^{\Lambda_J}_{\mu\nu}(R)\,\mathcal{O}^{(\Lambda_L \otimes \Lambda_S)_{\Lambda_J},\, n}_{\nu}(t).
\end{equation}
Depending on the momentum shell and the irreps involved, the resulting operator may vanish identically, in which case it does not yield a valid interpolating operator.

In a lattice simulation, one wants to use as many two-particle operators as there are (degenerate) noninteracting energy levels up to a given momentum shell in the system. To do this systematically, one can follow the strategy below. Given a total momentum $\vb*{P}$ and a target irrep $\Lambda_J$, one chooses the particle combinations --- for example $DD^*$ and $D^*D^*$ --- whose spins subduce into irreps $\Lambda_S$ from which $\Lambda_J$ can be constructed. One then chooses how high up to go in the relative-momentum shells; that is, for each particle combination one chooses a maximum $n$ for the relative momentum $\vb*{p}$ with $\vb*{p}^2 = (2\pi/L)^2 n$. One can, for example, go up to the lowest three-particle threshold in the system. Next, for each $\Lambda_S$ one searches for all orbital-angular-momentum irreps $\Lambda_L$ that allow a $\Lambda_J$ operator to be built on each momentum shell. Finally, given all combinations of $\Lambda_S$, $\Lambda_L$ and $n$, one constructs the operators $\mathcal{O}^{(\Lambda_L \otimes \Lambda_S)_{\Lambda_J},\, n}_{\mu}(t)$ as described above. Since the index $\mu$ denotes equivalent operators under little-group rotations, we can average over $\mu$ at the level of the correlator. The same can be done for equivalent total momenta.

We illustrate the construction with the examples relevant for the $DD^*/T_{cc}$ system, i.e.\ $I(J^P) = 0(1^+)$ with flavour content $cc\overline{u}\overline{d}$. We do this in the rest frame, where the relevant irrep is $T_1^+$. For the $DD^*$ operators we start from
\begin{equation}
  \label{eq:bilocal_DDstar_start}
  \mathcal{O}^{DD^*,\, T_1^+}_i(\vb*{p}, t)
    = (\overline{u}\,\gamma_5\, c)(\vb*{p}, t)\;(\overline{d}\,\gamma_i\, c)(-\vb*{p}, t)
    - \{u \leftrightarrow d\},
\end{equation}
with $i = 1, 2, 3$, whose spin is in the $T_1^+$ irrep. For the orbital angular momentum, we may choose the trivial $A_1^+$ irrep, subduced from $0^+$. Since $A_1^+ \otimes T_1^+ \cong T_1^+$, the corresponding CG coefficients form the identity matrix, and we obtain
\begin{equation}
  \label{eq:bilocal_DDstar_trivial}
  \mathcal{O}^{DD^*,\, (A_1^+ \otimes T_1^+)_{T_1^+},\, n}_i(t)
    = \sum_{\vb*{p} \in \mathcal{S}_n} \mathcal{O}^{DD^*,\, T_1^+}_i(\vb*{p}, t),
\end{equation}
for the momentum shells $n = 0, 1, \dots$. The same construction applies to the $D^*D^*$ operator
\begin{equation}
  \label{eq:bilocal_DstarDstar_start}
  \mathcal{O}^{D^*D^*,\, T_1^+}_i(\vb*{p}, t)
    = \epsilon_{ijk}\,(\overline{u}\,\gamma_j\, c)(\vb*{p}, t)\;(\overline{d}\,\gamma_k\, c)(-\vb*{p}, t).
\end{equation}
A nontrivial example arises from $E^+ \otimes T_1^+ \cong T_1^+ \oplus T_2^+$, where we choose $E^+$ for the orbital angular momentum and $T_1^+$ for the spin to build a $T_1^+$ operator. Starting from the $DD^*$ operator in Eq.~\eqref{eq:bilocal_DDstar_start}, the result is
\begin{equation}
  \label{eq:bilocal_DDstar_nontrivial}
  \mathcal{O}^{DD^*,\, (E^+ \otimes T_1^+)_{T_1^+},\, n}_i(t)
    = \sum_{\vb*{p} \in \mathcal{S}_n} \left(p_i^2 - \frac{1}{3}\vb*{p}^2\right)
      \mathcal{O}^{DD^*,\, T_1^+}_i(\vb*{p}, t),
\end{equation}
up to normalization. Note that this operator vanishes on the lowest momentum shell $n = 0$, where all momenta are zero.

Finally, we comment on the difficulties specific to exotic hadrons such as tetraquarks. Since their inner structure is not known a priori, one wishes to include both bilocal two-meson scattering operators, which resemble ``molecular'' configurations, and local tetraquark operators, which are better suited to more compact, deeply bound structures. Moreover, suppressing the excited-state contamination well requires a large operator basis, i.e.\ going high up in the momentum shells for the two-meson operators. Computing the full correlator matrix for such a basis is expensive with traditional methods, which motivates the use of the distillation method introduced in the next chapter.

\chapter{Quark smearing and distillation method}
\label{chapter:Distillation_method}

An essential ingredient for obtaining a clean signal in hadron spectroscopy is quark-field smearing. A widely used smearing method is Laplacian-Heaviside smearing, which underlies the distillation method. In this chapter, we introduce smearing for quarks and describe the distillation framework. The latter allows computing correlation functions of nonlocal multi-hadron operators efficiently. Local multiquark operators with four or more (anti)quarks are, however, expensive in this framework. We investigate how the computational cost scales for these two types of operators and discuss why the latter are expensive.

\section{Quark smearing}
\label{sec:quark_smearing}

Smearing reduces the coupling of a hadron operator to high-energy states and thereby decreases the excited-state contamination. It can be applied to both the gauge and the quark fields. Here we focus on the latter. A smeared quark field is defined in terms of a smearing kernel $K(t)$ that acts on a spinor field $\psi$ as
\begin{equation}
  \label{eq:smeared_field}
  \psi_\m{sm}(\vb*{x}, t) = \sum_{\vb*{y} \in \Lambda_3} K(\vb*{x}, t; \vb*{y}, t)\,\psi(\vb*{y}, t).
\end{equation}
The kernel $K$ acts on the spin, colour, and position indices but does not mix different time slices, since this would alter the form of the spectral decomposition. Moreover, $K$ should be chosen such that it preserves the symmetries of the system.

A convenient operator from which we can construct smearing kernels is the gauge-covariant Laplacian
\begin{equation}
  \label{eq:laplacian}
  \Delta(x,y) = 
  \frac{1}{a^2}\sum_{i=1}^3\left(U_i(x)\,\delta_{x+a\hat{i},y} - 2\delta_{x,y} + U_i(y)^\dagger\,\delta_{x-a\hat{i},y}\right).
\end{equation}
It is proportional to the spatial part of the Wilson term in the Wilson-clover Dirac operator in Eq.~(\ref{eq:wilson_clover_dirac}). The Laplacian is covariant under gauge transformations, lattice translations and rotations, parity and charge conjugation, so that no lattice symmetry is broken. Additionally, it is diagonal in the spin index, which allows operators to be defined in terms of smeared quark fields with well-defined spin.

Since the Laplacian depends on the gauge field, it is time-dependent and inherits the ultraviolet fluctuations of the links. To reduce these fluctuations, one first smears the gauge fields that enter $\Delta$. For this, we use stout smearing~\cite{Morningstar:2003gk}, an analytic smearing procedure that replaces each link variable by a weighted average of its surrounding staples, with a projection ensuring that the smeared link remains in the $\m{SU}(3)$ gauge group.

A widely used quark-smearing method is Gaussian smearing, also known as Wuppertal smearing, which employs the kernel $\m{e}^{\sigma^2\Delta/2}$. In numerical simulations, it is implemented iteratively as
\begin{equation}
  \label{eq:gaussian_smearing}
  \psi_\m{sm} = \left(1 + \alpha\Delta\right)^n\psi,
\end{equation}
which approximates the exponential for large $n$, with smearing parameter $\alpha = \frac{\sigma^2}{2n}$ and the number of iterations $n$~\cite{Gusken:1989ad, Gusken:1989qx}.

Another quark-smearing method, and the one underlying distillation, is Laplacian-Heaviside (LapH) smearing~\cite{HadronSpectrum:2009krc, Morningstar:2011ka}. It uses the kernel
\begin{equation}
  \label{eq:laph_kernel}
  K = \Theta\!\left(\sigma_\m{LapH}^2 + \Delta\right),
\end{equation}
where $\Theta$ is the Heaviside function and $\sigma_\m{LapH}^2$ a cutoff parameter. This kernel is to be understood in terms of the eigenvectors $v^{(k)}(t)$ of $\Delta(t)$, which we call the Laplace modes, and their corresponding eigenvalues $\lambda^{(k)}(t)$, which are all negative. We call the index $k$ the Laplace-mode index. The kernel $K$ projects onto the space spanned by the Laplace modes with $-\lambda^{(k)}(t) < \sigma_\m{LapH}^2$. This imposes a hard cutoff, in contrast to the smooth suppression of the high modes in Gaussian smearing. For a given threshold $\sigma_\m{LapH}^2$, numerical simulations have shown that the number of eigenvectors $N_v$ below that threshold varies only little across time slices~\cite{Morningstar:2011ka}. As a consequence, in practical applications one works with a constant $N_v$. The number of eigenvectors is typically a few tens to a few hundred.

To construct $K$ explicitly, one orders the eigenvalues as
\begin{equation}
  \label{eq:eigenvalue_ordering}
  -\lambda^{(1)}(t) \leq -\lambda^{(2)}(t) \leq -\lambda^{(3)}(t) \leq \dots,
\end{equation}
and collects the lowest $N_v$ eigenvectors into the $3|\Lambda_3| \times N_v$ matrix
\begin{equation}
  \label{eq:V_matrix}
  V(t) = \left(v^{(1)}(t), v^{(2)}(t), \dots, v^{(N_v)}(t)\right).
\end{equation}
The factor of three counts the colour index, and $|\Lambda_3|$ is the number of spatial lattice sites. On the time slice $t$, the kernel is then $K(t) = V(t)V(t)^\dagger$, which acts trivially in spin space. A modified variant of LapH smearing weights the Laplace modes by their eigenvalues when constructing $K(t)$, for example using a Gaussian profile $\m{e}^{-\sigma^2{\lambda^{(k)}}^2/2}$~\cite{Dudek:2010wm, Knechtli:2022bji}. This allows for more flexibility when tuning the smearing width of an operator for a specific hadron.

\section{Distillation method}
\label{sec:distillation}

Smearing the quark fields with $K$ turns the quark propagator into the smeared propagator
\begin{equation}
  \label{eq:smeared_propagator}
  S_{f,\,\m{sm}}(t', t) = K(t')\, S_f(t', t)\, K(t)^\dagger
    = V(t')V(t')^\dagger \, S_f(t', t)\, V(t)V(t)^\dagger,
\end{equation}
which is given in terms of the unsmeared propagator $S_f = D_f^{-1}$ for flavour $f$. As noted previously, it is prohibitively expensive to invert $D_f$ fully, as it is a $12|\Lambda| \times 12|\Lambda|$ matrix. The key observation of distillation is that, on a given time slice $t$, $K(t)$ is a projector onto the space spanned by the $N_v$ lowest Laplace modes. We call this vector space the LapH subspace; it is $4N_v$-dimensional once the spin index is included. Within this subspace, the Dirac operator can be fully inverted. This inversion yields the so-called perambulator
\begin{equation}
  \label{eq:perambulator}
  \tau_{f,\alpha\beta}(t', t) = V(t')^\dagger \, S_{f,\alpha\beta}(t', t)\, V(t),
\end{equation}
a $4N_v \times 4N_v$ matrix, where $\alpha$ and $\beta$ are the spin indices and the Laplace-mode indices are left implicit. It is computed by solving the Dirac equation for each Laplace mode and spin index $\beta$ at the source time $t$ and projecting the result onto the Laplace modes at the sink time $t'$. In terms of the perambulator, the smeared propagator reads
\begin{equation}
  \label{eq:smeared_propagator_peram}
  S_{f,\,\m{sm},\alpha\beta}(t', t) = V(t')\, \tau_{f,\alpha\beta}(t', t)\, V(t)^\dagger,
\end{equation}
with the colour and position indices left implicit.

The distillation method~\cite{HadronSpectrum:2009krc, Morningstar:2011ka} uses LapH smearing to express correlation functions in the LapH subspace rather than in position space. As an example, consider again the charged-pion two-point function in Eq.~\eqref{eq:pion_correlator}. For general source and sink times $t$ and $t'$, it is
\begin{equation}
  \label{eq:pion_corr_distillation}
  C(t', t) = \sum_{\vb*{x}', \vb*{x} \in \Lambda_3}
    \m{e}^{-\m{i}\vb*{P}\cdot(\vb*{x}'-\vb*{x})}\,
    \ev{\tr\!\left[S_u(\vb*{x}', t'; \vb*{x}, t)\,
      S_d(\vb*{x}', t'; \vb*{x}, t)^\dagger\right]}_G,
\end{equation}
where we have used $\gamma_5$-hermiticity $\gamma_5 S_f \gamma_5 = S_f^\dagger$. Replacing the propagators by their smeared counterparts, we obtain the smeared two-point function
\begin{equation}
  \label{eq:pion_corr_smeared}
  C_\m{sm}(t', t) = \ev{\tr\!\left[
    \Phi(\vb*{P}, t')\, \tau_u(t', t)\,
    \Phi(\vb*{P}, t)^\dagger\, \tau_d(t', t)^\dagger\right]}_G.
\end{equation}
Here, we have collected the momentum projection into the mode doublets
\begin{equation}
  \label{eq:mode_doublets}
  \Phi^{(k, l)}(\vb*{P}, t) = \sum_{\vb*{x} \in \Lambda_3}
    \m{e}^{-\m{i}\vb*{P}\cdot\vb*{x}}\,
    v_a^{(k)}(\vb*{x}, t)^*\, v_a^{(l)}(\vb*{x}, t),
\end{equation}
which form $N_v \times N_v$ matrices carrying the Laplace-mode indices $(k, l)$, with a summation over the colour index $a$ and the position $\vb*{x}$. Using the mode doublets in simulations is more convenient than repeatedly recomputing or storing the Laplace modes. Consequently, within distillation, the pion two-point function reduces to the trace of a matrix product in the LapH subspace. An important feature of this construction is that computing the perambulators, while an expensive startup cost, needs to be done only once. The same perambulators can subsequently be reused to compute any correlation function of the given flavours.

The number of Laplace modes $N_v$ is directly related to the smearing radius, which is chosen to reduce the overlap of the operator with high-energy states. A smaller $N_v$ corresponds to a larger smearing radius and therefore generally to a better suppression of the excited states. On the other hand, if $N_v$ is chosen too small, the signal is lost. In practice, $N_v$ is tuned to obtain the fastest convergence of the effective energy to its plateau for a given two-point function, although the computational cost can also be the limiting factor. Once $N_v$ has been fixed, one wishes to keep the physical smearing radius constant as the lattice spacing and the volume are varied. The smearing radius was found to be proportional to the physical spatial volume $V = a^3|\Lambda_3|$~\cite{Morningstar:2011ka}, which means we have to scale $N_v \propto V$. Consequently, a larger volume entails a larger computational cost, both for computing the perambulators and for the contractions needed to compute the two-point function. As we discuss below, the contractions become especially expensive for local multiquark operators containing four or more (anti)quarks.

\section{Multi-hadron and local multiquark operators}
\label{sec:multiquark_distillation}

\begin{figure}
  \centering
  \includegraphics[width=0.62\textwidth]{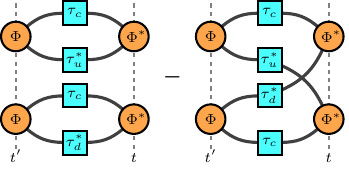}
  \vspace*{1pt}
  \caption{\label{fig:tensor_network_DDstar}%
    Tensor-network diagrams of the two-point function of a bilocal $DD^*$ operator in distillation. It is given in terms of perambulators $\tau_f$ (for flavour $f$) and mode doublets $\Phi$. The two topologically distinct terms arise from the Wick contraction. The spin structure is suppressed.}
\end{figure}

Distillation can equally be applied to baryons. One writes down the unsmeared baryon two-point function, replaces the propagators by their smeared counterparts, and absorbs the Laplace modes and the momentum projection into tensors. Whereas for mesons this gives the rank-2 mode doublets in Eq.~(\ref{eq:mode_doublets}), for baryons one obtains rank-3 tensors called mode triplets, with one Laplace-mode index per quark. They are defined as
\begin{equation}
  \label{eq:mode_triplets}
  \Psi^{(k, l, m)}(\vb*{P}, t) = \sum_{\vb*{x} \in \Lambda_3}
    \m{e}^{-\m{i}\vb*{P}\cdot\vb*{x}}\,
    \epsilon_{abc}\, v_a^{(k)}(\vb*{x}, t)\, v_b^{(l)}(\vb*{x}, t)\, v_c^{(m)}(\vb*{x}, t).
\end{equation}
The method extends further to nonlocal multi-hadron operators and has been used successfully for two- and three-meson scattering~\cite{Wilson:2015dqa, Moir:2016srx, Briceno:2017qmb, Woss:2019hse, Lang:2022elg, Boyle:2024hvv, Yan:2024gwp, Mohler:2013rwa, Fischer:2020yvw, Hansen:2020otl, Fischer:2020jzp, Padmanath:2022cvl, Whyte:2024ihh, Dawid:2025doq, Prelovsek:2025vbr, Shrimal:2025wbu}, for meson-baryon scattering~\cite{Lang:2012db, Hackl:2024whw, Bulava:2022vpq, BaryonScatteringBaSc:2023ori, BaryonScatteringBaSc:2023zvt}, and for baryon-baryon scattering~\cite{Horz:2020zvv, Green:2021qol, Xing:2025uai, BaryonScattering:2025ziz}. In all these cases, the mode doublets and triplets can be reused, since the individual mesons and baryons are each projected onto a separate momentum. The resulting two-point functions can be visualized using tensor-network diagrams, as in Figure~\ref{fig:tensor_network_DDstar} for a bilocal $DD^*$ operator.

The computational cost of the contractions depends on the type of operator. For mesons and bilocal two-meson operators, the contractions are matrix-matrix multiplications in the LapH subspace and scale as $N_v^3$. For baryons and bilocal two-baryon operators, they involve contractions of the perambulator matrices with the rank-3 mode triplets and scale as $N_v^4$. This cost scaling is manageable for the number of Laplace modes used in current simulations.

\begin{figure}
  \centering
  \includegraphics[trim=0 -29 -18 0, width=0.33\textwidth]{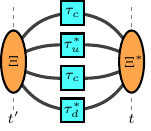}%
  \includegraphics[width=0.67\textwidth]{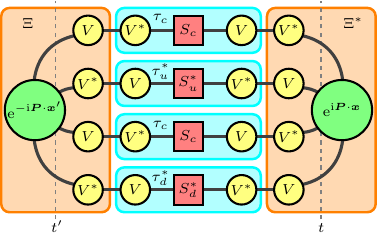}
  \caption{\label{fig:tensor_network_tetraquark_no_pos-space_samp}%
    Tensor-network diagrams of the two-point function of a local tetraquark operator in distillation. In the left diagram, it is displayed in terms of perambulators $\tau_f$ (for flavour $f$) and the rank-4 tensor $\Xi$, whereas in the right diagram all tensors involved in distillation are displayed to visualize the order in which they are contracted. The perambulators $\tau_f$ are constructed by contracting the propagators $S_f$ with the neighbouring $V$ matrices. The remaining $V$'s are contracted with $\m{e}^{-\m{i}\vb*{P}\cdot\vb*{x}'}$, which gives the rank-4 tensor $\Xi$. The spin and colour structure is suppressed in these diagrams.}
\end{figure}

However, for systems such as the $DD^*$, relevant for the $T_{cc}$ tetraquark, we are also interested in the local tetraquark operators introduced in Section~\ref{sec:tetraquark_operators}. Collecting the momentum projection and the Laplace modes into a tensor for these operators yields a rank-4 tensor. For local two-meson operators it reads
\begin{equation}
  \label{eq:Xi_tensor}
  \Xi^{(k,l,m,o)}(\vb*{P}, t) = \sum_{\vb*{x} \in \Lambda_3}
    \m{e}^{-\m{i}\vb*{P}\cdot\vb*{x}}\,
    v_a^{(k)}(\vb*{x}, t)^*\, v_a^{(l)}(\vb*{x}, t)\;
    v_b^{(m)}(\vb*{x}, t)^*\, v_b^{(o)}(\vb*{x}, t),
\end{equation}
while for a diquark-antidiquark operator it has a different colour structure. The corresponding tensor-network diagram of the two-point function is shown in Figure~\ref{fig:tensor_network_tetraquark_no_pos-space_samp}, both in terms of the perambulators and the tensor $\Xi$, and with all tensors involved in distillation displayed explicitly.
The consequence is that the computational cost for two-point functions of local tetraquark operators scales as $N_v^5$. This is very expensive, particularly in a large physical volume. In addition, the tensor $\Xi$ requires a large amount of memory to store. Despite the cost, distillation has been used for local tetraquark operators in this way~\cite{Cheung:2017tnt, Ortiz-Pacheco:2023ble, Prelovsek:2025vbr}.

The problem is even more severe for local hexaquark operators relevant for dibaryons such as the $d^*(2380)$. Such an operator has the generic form
\begin{equation}
  \label{eq:hexaquark_operator}
  \mathcal{O}^H(t) \sim \sum_{\vb*{x} \in \Lambda_3}
    \m{e}^{-\m{i}\vb*{P}\cdot\vb*{x}}\,(uuuddd)(\vb*{x}, t),
\end{equation}
with the spin and colour structure suppressed. In distillation, encapsulating the momentum projection would require a rank-6 tensor, leading to a computational cost that scales as $N_v^7$. This is not feasible even for a small number of Laplace modes, such as $N_v = 32$. It is, however, not clear whether such operators are needed to extract the spectrum of dibaryon systems: they were found to be irrelevant for nucleon-nucleon scattering~\cite{Amarasinghe:2021lqa, Detmold:2024iwz, BaryonScattering:2025ziz}, but they may be relevant for the $d^*$ and the $H$ dibaryon~\cite{Francis:2018qch}.

\section{Cost of local multiquark operators}
\label{sec:cost_multiquark}

The unfavourable cost scaling for local multiquark operators in distillation can be traced back to the order in which the tensor contractions are performed. Summing over the position indices first results in the construction of high-rank tensors such as $\Xi$. This makes the subsequent contractions over the Laplace-mode indices expensive.

A possible remedy is to reverse this order: one first constructs the smeared propagator explicitly, summing over all Laplace-mode indices, and is then left with a double sum over the position indices $\vb*{x}'$ and $\vb*{x}$. This contraction order is illustrated in the next chapter in Figure~\ref{fig:tensor_network_tetraquark_pos-space_samp}. The contraction then consists of the following two steps. First, one computes the smeared propagators from the perambulators via Eq.~(\ref{eq:smeared_propagator_peram}), which requires two matrix-matrix multiplications, giving the cost scalings $N_v^2|\Lambda_3|$ and $N_v|\Lambda_3|^2$. Second, one performs a double sum over the position indices, with cost scaling $|\Lambda_3|^2$. Keeping $N_v \propto V$, the cost at fixed lattice spacing asymptotically scales as $N_v^3$, independent of the number of quarks. This is a substantial improvement over the $N_v^5$ and $N_v^7$ scalings of the traditional distillation approach for local tetraquark and hexaquark operators, respectively.

This approach nonetheless has its own drawbacks. Since $|\Lambda_3| \gg N_v$, the computation remains very expensive, and it would require either writing the full set of Laplace modes to disk or recomputing them for each contraction batch job, both of which are impractical. In this thesis, we instead use a position-space sampling method in addition to distillation, which reduces the computational cost both for computing the smeared propagators and for the double sum, as described in the next chapter.

\chapter{Position-space sampling method}
\label{chapter:Position-space_sampling_method}

In this chapter, we present a position-space sampling method, which substantially improves the volume scaling of the computational cost for local multiquark operators within the distillation framework. A different approach to the same problem was proposed in Ref.~\cite{Lang:2024syy}. The method described here was introduced in our work~\cite{Stump:2025owq}, which we follow closely throughout this chapter. This method yields an estimator for two-point functions that is unbiased and has a low variance compared to other sampling methods.

\section{Sampling method}
\label{sec:pss_method}

The idea of our position-space sampling method is to compute the smeared propagators only on subspaces $\tilde{\Lambda}_3 \subset \Lambda_3$ of the spatial lattice, rather than on the full lattice $\Lambda_3$. This substantially reduces the cost of the double position-space sum in the two-point function of local operators. Two subspaces are needed: one at the sink, $\tilde{\Lambda}_3'$, and one at the source, $\tilde{\Lambda}_3$. One then computes the smeared propagators $S_{f,\,\m{sm}}(\vb*{x}', t'; \vb*{x}, t)$ for all $\vb*{x}' \in \tilde{\Lambda}_3'$ and $\vb*{x} \in \tilde{\Lambda}_3$, which requires, besides the perambulators, the Laplace modes only on these subspaces. That part of the Laplace modes is small enough to be stored on disk. The smeared propagators are then contracted to obtain the two-point function. The two subspaces must be chosen according to a specified scheme for each gauge configuration, and for each source time $t$ if several are used. We describe possible schemes in the following section.

As an example, consider again the charged-pion two-point function in Eq.~(\ref{eq:pion_corr_distillation}). Within position-space sampling, it becomes
\begin{equation}
  \label{eq:pion_corr_pss}
  \tilde{C}_\m{sm}(t', t) = \Big<
    \frac{|\Lambda_3|^2}{|\tilde{\Lambda}'_3|\,|\tilde{\Lambda}_3|}\;
    \sum_{\mathclap{\substack{\vb*{x}' \in \tilde{\Lambda}'_3 \\ \vb*{x} \in \tilde{\Lambda}_3}}}
    \m{e}^{-\m{i}\vb*{P}\cdot(\vb*{x}' - \vb*{x})}\,
    \tr\!\left[S_{u,\,\m{sm}}(\vb*{x}', t'; \vb*{x}, t)\,
      S_{d,\,\m{sm}}(\vb*{x}', t'; \vb*{x}, t)^\dagger\right]
  \Big>_{\!G,\, \tilde{\Lambda}_3^{(\prime)}}.
\end{equation}
The prefactor $|\Lambda_3|^2/\big(|\tilde{\Lambda}'_3||\tilde{\Lambda}_3|\big)$ ensures the correct normalization: it compensates for summing over fewer points, so that the result obtained without position-space sampling is recovered, provided the estimator is unbiased. The two subspaces need not be the same for all gauge configurations and can, for instance, be chosen randomly. Accordingly, the expectation value is taken both over the gauge fields and over the two subspaces. Here we use $\tilde{\Lambda}_3^{(\prime)}$ to collectively denote the two subspaces $\tilde{\Lambda}_3'$ and $\tilde{\Lambda}_3$.

\begin{figure}
  \centering
  \includegraphics[width=0.67\textwidth]{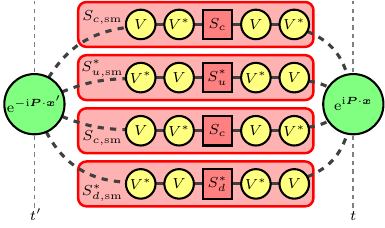}
  \caption{\label{fig:tensor_network_tetraquark_pos-space_samp}%
    Tensor-network diagram of the two-point function of a local tetraquark operator in distillation using the position-space sampling method. Here, the contraction order of the tensors differs from that in Figure~\ref{fig:tensor_network_tetraquark_no_pos-space_samp}: the propagators are contracted with all four $V$'s connected to them, giving the smeared propagators $S_{f,\m{sm}}$ (for flavour $f$). The contraction with the exponentials is then performed only over subspaces of the spatial lattice (indicated by the dashed lines). The spin and colour structure is suppressed in this diagram.}
\end{figure}

For a general local multiquark operator, the procedure is the same, although the way in which the smeared propagators are contracted is more involved. Figure~\ref{fig:tensor_network_tetraquark_pos-space_samp} shows the tensor-network diagram of the two-point function of a local tetraquark operator in the position-space sampling method.
The contraction order differs from that of the traditional distillation method, shown in Figure~\ref{fig:tensor_network_tetraquark_no_pos-space_samp}.

A practical comment on the implementation is in order. Although the subspaces reduce the size of the smeared propagator substantially, storing $S_{f,\,\m{sm}}(\vb*{x}', t'; \vb*{x}, t)$ for all points of the subspaces may still require too much memory in a simulation. In our simulations, we therefore use an explicit loop over the points in the two subspaces, so that the smeared propagator is never constructed in full.

\section{Choice of subspaces}
\label{sec:pss_subspaces}

Various approaches for defining these subspaces have been used in the literature~\cite{Detmold:2019fbk, Li:2020hbj}, though not in combination with distillation. They can be grouped into two categories. The first category comprises sparse grids, in which one uses a regular grid retaining, for example, only every fourth or eighth point in each spatial direction. It has been found empirically that a relatively large separation between the points can be used while keeping the resulting noise from field sparsening insignificant~\cite{Li:2020hbj}. Their disadvantage is that one does not sample the full lattice --- most points are never sampled --- so that the momentum projection is incomplete. This results in a biased estimator, where the correlator receives contamination from other momenta~\cite{Li:2020hbj}. The second category comprises random subspaces, in which the points are chosen fully randomly from the spatial lattice. Since these sample the full lattice, the momentum projection is complete and the estimator is unbiased. The disadvantage of the second type is that it empirically leads to a larger variance than the sparse-grid method~\cite{Li:2020hbj}.

In this thesis, we use the scheme for the subspaces that we introduced in Ref.~\cite{Stump:2025owq}: regular sparse grids that are randomly shifted. One chooses a point separation $N_\m{sep}$, which must divide the spatial extent $N_s$ of the lattice, and defines the subspace as
\begin{equation}
  \label{eq:subspace_definition}
  \tilde{\Lambda}_{3, \tilde{\vb*{x}}} = \left\{a\vb*{n} + \tilde{\vb*{x}}
    \;\middle|\; n_i = 0,\, N_\m{sep},\, 2N_\m{sep},\, \dots,\, N_s - N_\m{sep}
    \;\; (i = 1, 2, 3)\right\},
\end{equation}
with a random offset $\tilde{\vb*{x}} \in \Lambda_3$. It can equivalently be written as
\begin{equation}
  \label{eq:subspace_shift}
  \tilde{\Lambda}_{3, \tilde{\vb*{x}}} = \tilde{\Lambda}_{3, \vb*{0}} + \tilde{\vb*{x}},
\end{equation}
where $\tilde{\Lambda}_{3, \vb*{0}}$ is the subspace that would be used in the sparse-grid method without a random offset, and the addition is understood elementwise. The number of points in the subspace is $|\tilde{\Lambda}_{3, \tilde{\vb*{x}}}| = |\Lambda_3|/N_\m{sep}^3$, which decreases with the third power of $N_\m{sep}$.

To sample from the full lattice, we draw random offsets for each gauge configuration. Additionally, we draw a new offset for each source time $t$ for the sparse grid at the source. At the sink, we reuse the same offset on a given gauge configuration across all sink times $t'$. This preserves correlations between different $t'$, which can lead to more precise effective energies. In contrast to the sparse-grid method without a random offset, the randomly shifted sparse grids yield an unbiased estimator. We show this next.

\section{Unbiasedness of the estimator}
\label{sec:pss_unbiased}

To show that the estimator is unbiased, we note that any two-point function of local operators with definite momentum can be written as
\begin{equation}
  \label{eq:generic_two_point}
  C(t', t) = \sum_{\mathclap{\vb*{x}', \vb*{x} \in \Lambda_3}} \ev{f(\vb*{x}', t'; \vb*{x}, t)}_G,
\end{equation}
where $f$ contains the (smeared) quark propagators and the momentum-projection exponential. Within position-space sampling this becomes
\begin{equation}
  \label{eq:generic_two_point_pss}
  \tilde{C}(t', t) = \frac{|\Lambda_3|^2}{|\tilde{\Lambda}_{3, \vb*{0}}|^2} \big<
    \,\sum_{\mathclap{\substack{\vb*{x}' \in \tilde{\Lambda}_{3, \tilde{\vb*{x}}'} \\ \vb*{x} \in \tilde{\Lambda}_{3, \tilde{\vb*{x}}}}}}
    f(\vb*{x}', t'; \vb*{x}, t)\big>_{G,\, \tilde{\vb*{x}}^{(\prime)}},
\end{equation}
where $\tilde{\vb*{x}}^{(\prime)}$ collectively denotes the random offsets at the sink and the source, and we have used that the number of points $|\tilde{\Lambda}_{3, \tilde{\vb*{x}}}|$ is independent of the offset.

The expectation value over the sparse grids is simply the average over all possible offsets, i.e.\ the average over $\tilde{\vb*{x}}'$ and $\tilde{\vb*{x}}$, so that
\begin{equation}
  \label{eq:offset_average}
  \ev{\dots}_{G,\, \tilde{\vb*{x}}^{(\prime)}}
    = \frac{1}{|\Lambda_3|^2} \sum_{\tilde{\vb*{x}}', \tilde{\vb*{x}} \in \Lambda_3} \ev{\dots}_G.
\end{equation}
Using Eq.~(\ref{eq:subspace_shift}), we can make the dependence on the random offsets explicit and move the offsets out of the index set in the sum and into the argument of $f$. Equation~\eqref{eq:generic_two_point_pss} then becomes
\begin{equation}
  \tilde{C}(t', t) = 
    \frac{1}{|\tilde{\Lambda}_{3, \vb*{0}}|^2}
    \hspace{10pt}\sum_{\mathclap{\substack{\vb*{x}', \vb*{x} \in \tilde{\Lambda}_{3,\vb*{0}} \\
                                 \tilde{\vb*{x}}', \tilde{\vb*{x}} \in \Lambda_3}}}
    \ev{f(\vb*{x}' + \tilde{\vb*{x}}', t'; \vb*{x} + \tilde{\vb*{x}}, t)}_G.
\end{equation}

Using the periodicity of the spatial lattice, we then perform the substitutions
\begin{equation}
  \label{eq:periodicity_substitution}
  \tilde{\vb*{x}}' \rightarrow \tilde{\vb*{x}}' - \vb*{x}', \qquad
  \tilde{\vb*{x}} \rightarrow \tilde{\vb*{x}} - \vb*{x},
\end{equation}
after which the sums over $\vb*{x}'$ and $\vb*{x}$ become trivial and yield a factor $|\tilde{\Lambda}_{3, \vb*{0}}|^2$. This cancels the prefactor, and we finally obtain
\begin{equation}
  \label{eq:unbiased_result}
  \tilde{C}(t', t) = \sum_{\mathclap{\tilde{\vb*{x}}', \tilde{\vb*{x}} \in \Lambda_3}}
    \ev{f(\tilde{\vb*{x}}', t'; \tilde{\vb*{x}}, t)}_G = C(t', t),
\end{equation}
recovering the sum over the full spatial lattice. The position-space sampling estimator is therefore unbiased.

Crucially, it is only because of the random offset that the full lattice is sampled, and the estimator is unbiased. This is not the case for the sparse-grid method without a random offset. At the same time, we use regular grids, which keep the variance small compared to fully random subspaces. Consequently, this position-space sampling method combines the advantages of both approaches.

The derivation can be adapted to mixed two-point functions of a local and a bilocal operator. In that case, one applies position-space sampling only to the local operator, i.e.\ only at the sink or only at the source, and uses full distillation for the bilocal operator. Thus, the estimator is unbiased for mixed two-point functions as well. In fact, in this case a random offset is not required to remove the bias, owing to the periodicity of the lattice and to the use of the full spatial lattice for the bilocal operator.

\section{Choice of point separation and cost scaling}
\label{sec:pss_cost}

The point separation $N_\m{sep}$ controls the variance of the estimator. For $N_\m{sep} = 1$, one recovers the result of distillation without position-space sampling, while increasing $N_\m{sep}$ introduces additional noise from the sampling. The optimal $N_\m{sep}$ would be the one that minimizes the numerical effort required to reach a given total uncertainty. Since determining this optimum exactly is difficult, we instead choose in practice the largest $N_\m{sep}$ for which the uncertainty from the sampling is negligible compared to the Monte Carlo uncertainty. In the following, we refer to this value as the optimal $N_\m{sep}$. To determine this value, one must run simulations for several values of $N_\m{sep}$. The available values are constrained by the requirement that $N_\m{sep}$ divides $N_s$, so that consecutive values are typically related by a factor of two.\footnote{Step sizes other than two are possible in $N_\m{sep}$ if the prime factorization of $N_s$ contains factors other than two.} The cost of the contractions scales as the square of the number of grid points, $|\tilde{\Lambda}_3|^2 \propto N_\m{sep}^{-6}$ (see Section~\ref{sec:cost_multiquark}). Consequently, halving $N_\m{sep}$ increases the cost roughly by a factor of $64$. This steep cost scaling means that only a small range of $N_\m{sep}$ can be scanned in practice: an $N_\m{sep}$ that is too small is prohibitively expensive, while for an $N_\m{sep}$ that is too large the correlator data are likely unusable. Fortunately, as we find in the next chapter, a relatively large separation --- for example $N_\m{sep} = 8$ --- is already sufficient to render the error from the sampling method subdominant. If even this value is too expensive, one can choose a larger $N_\m{sep}$ at the cost of a larger total error.

Once we have determined the optimal point separation for a given operator, we are interested in how it changes when the volume and the lattice spacing are varied. We naively expect the noise from position-space sampling to depend primarily on the physical distance $aN_\m{sep}$ between the points, and thus to be independent of the volume. We therefore keep $aN_\m{sep}$ fixed when varying the lattice spacing and volume. Following the reasoning in Section~\ref{sec:cost_multiquark}, we find that the cost of contractions in position-space sampling scales as $N_v^2|\tilde{\Lambda}_3|$ and $N_v|\tilde{\Lambda}_3|^2$. The number of points in the sparse grid is $|\Lambda_3|/N_\m{sep}^3 = V/(aN_\m{sep})^3$, which is proportional to the physical volume and hence to the number of Laplace modes $N_v$. Consequently, the overall computational cost scales as $N_v^3 \propto V^3$ at fixed lattice spacing, which is much better than the $V^5$ and $V^7$ scalings obtained for local tetraquark and hexaquark operators in the traditional distillation method. Moreover, the cost scaling improves when the continuum limit $a \rightarrow 0$ is taken, since $aN_\m{sep}$ is held fixed. In summary, the position-space sampling method makes local multiquark operators with four or more (anti)quarks feasible within distillation, especially in large physical volumes.
\chapter{Results}
\label{chapter:Results}

In this chapter, we present the results obtained with the position-space sampling method. We first study the efficiency of the method and determine the optimal point separation $N_\m{sep}$ for the two-point functions of single-meson, single-baryon, and local tetraquark operators. We then apply the method to the $DD^*$ and $BB^*$ systems. The $N_\m{sep}$-dependence analysis and the results for the $DD^*$ system have, for the most part, already been presented in our publication~\cite{Stump:2025owq}.

\section{Lattice setup}
\label{sec:Lattice_setup}

We performed our simulations on two CLS gauge ensembles with Wilson-clover fermions (cf.\ Eq.~(\ref{eq:wilson_clover_dirac})) at the $\m{SU}(3)$-flavour-symmetric point, where the quark mass is set approximately to the average of the physical $u$, $d$, and $s$ quark masses. The relevant parameters of the two ensembles are listed in Table~\ref{tab:ensembles}.
\begin{table}
  \centering
  \renewcommand{\arraystretch}{1.2}
  \begin{tabular}{lcccccccc}
    \hline\hline
    Ensemble & $N_s^3 \times N_t$ & $\beta$ & $a$ (fm) & $\kappa_u=\kappa_d=\kappa_s$ & $\kappa_c$ & $N_{\text{cfg}}$ & $N_{\text{src}}$ & $N_v$ \\ 
    \hline
    B450 & $32^3\times64$  & 3.46 & 0.0749 & 0.136890 & 0.126243 & 1612 & 8 & 32 \\ 
    N202 & $48^3\times128$ & 3.55 & 0.0633 & 0.137000 & 0.128423 &  899 & 8 & 68 \\ 
    \hline\hline
  \end{tabular}
  \caption{\label{tab:ensembles}Parameters of the gauge ensembles used for the simulations. The lattice spacing was determined in~\cite{Strassberger:2021tsu}. Here, $N_{\text{cfg}}$ is the number of gauge-field configurations and $N_{\text{src}}$ the number of source times per configuration.}
\end{table}
On both ensembles, the resulting pion mass is $m_\pi \approx 420\;\m{MeV}$. For the valence charm quark, we use the same fermion action as for the light quarks, with the quark mass tuned to reproduce the average of the physical $D^0$, $D^+$, and $D_s^+$ masses, $1901\;\m{MeV}$. On B450 we additionally use a heavier-than-physical charm quark, with hopping parameter $\kappa_c^\m{heavy} = 0.115336$, resulting in a $D$-meson mass of $m_D^\m{heavy} = 2753\;\m{MeV}$. For the $b$ quark, we use the RHQ action introduced in Section~\ref{sec:RHQ_action}, whose parameters are given in Table~\ref{tab:RHQ_params}. They were tuned to reproduce the physical $\eta_b$, $\Upsilon$, $\chi_{b0}$, $\chi_{b1}$, and $h_b$ bottomonium masses, and to reproduce the speed-of-light parameter $c^2 = 1$ (cf.\ Eq.~(\ref{eq:dispersion_csq})).
\begin{table}
  \centering
  \begin{tabular}{lccccc}
    \hline \hline
    Ensemble & $\kappa_b$ & $\nu$ & $r_s$ & $c_E$ & $c_B$ \\
    \hline
    B450 & $0.054727$ & $2.167$ & $1.981$ & $1.947$ & $2.135$ \\
    N202 & $0.068125$ & $1.784$ & $1.631$ & $1.845$ & $1.932$ \\
    \hline \hline
  \end{tabular}
  \caption{\label{tab:RHQ_params}%
    Parameters of the RHQ action for the $b$ quark on the two ensembles as determined in Ref.~\cite{Hudspith:2026vxt}. For the N202 ensemble we used the parameters determined for H200 as it has the same lattice spacing.}
\end{table}

The heavier-than-physical pion mass has the advantage that the $D^*$ meson is QCD-stable, so that the $DD^*$ system is a genuine two-particle system and three-particle decays of the $T_{cc}$ are avoided. The $B^*$ meson is QCD-stable even at the physical pion mass, so that the $BB^*$ system is likewise a two-particle system. To apply the position-space sampling method to the $DD^*$ and $BB^*$ systems, we used the point separation $N_\m{sep}=8$ in the sparse grids on both ensembles.

The lattice spacings are determined with an uncertainty of about $1\%$~\cite{Strassberger:2021tsu}, and the tuned charm hopping parameter reproduces the target $D$-meson mass to better than $0.5\%$. The RHQ parameters were tuned in Ref.~\cite{Hudspith:2026vxt} using the scale setting of Ref.~\cite{Bruno:2016plf}, which differs from the lattice spacings used here by about $2\%$. These uncertainties are not propagated into our results. We estimate their effect on the scattering observables in Sections~\ref{sec:DDstar_luscher} and~\ref{sec:BBstar_swave}.

\section{Data analysis methods}
\label{sec:data_analysis}

Throughout this work, statistical errors are computed using the $\Gamma$-method~\cite{WOLFF2004143}, and errors are always propagated linearly. We perform correlated least-squares fits, i.e.\ we minimize
\begin{equation}
  \label{eq:chi_squared}
  \chi^2(y, \theta) = \sum_{i,j=1}^{n}
    \left(y_i - f(x_i; \theta)\right)\left(C^{-1}\right)_{ij}\left(y_j - f(x_j; \theta)\right),
\end{equation}
for data $x = (x_1, \dots, x_n)$ and $y = (y_1, \dots, y_n)$ with errors only in $y$. Here, $C_{ij} = \m{cov}(y_i, y_j)$ is the covariance matrix of the $y$-data and $f(x; \theta)$ is the model function with fit parameters $\theta$ that is expected to describe the data.

The covariance matrix $C$ is also computed using the $\Gamma$-method. With this method, $C$ is not guaranteed to be positive definite --- which is required by certain minimizers --- and can be very ill-conditioned~\cite{Bruno:2022mfy}. This is particularly problematic for fits to effective masses or correlators, where there is a strong correlation between different times, and therefore has to be addressed. To avoid non-positive-definite and ill-conditioned covariance matrices, we used the following procedure. We first computed the covariance matrix without including the autocorrelation, denoted by $C_0$. We then replace the autocorrelation time of the off-diagonal covariance-matrix elements by the geometric mean of the corresponding diagonal integrated autocorrelation times $\tau_{\m{int},i}$,
\begin{equation}
  \label{eq:autocorr_geometric}
  \tau_{\m{int},ij} = \sqrt{\tau_{\m{int},i}\;\tau_{\m{int},j}}.
\end{equation}
The resulting estimate of the covariance matrix is
\begin{equation}
  \label{eq:covariance_estimate}
  \tilde{C}_{ij} = 2\tau_{\m{int},ij}\;C_{0,ij}.
\end{equation}
This construction ensures that $\tilde{C}$ is positive semi-definite and, empirically, also reduces its condition number. It is implemented, for example, in the \texttt{pyerrors} package~\cite{Joswig:2022qfe}.

Using $\tilde{C}$ instead of $C$ in the fit amounts to performing a fit with a general weight matrix. For such fits, the quality-of-fit measures differ from the standard case: the expectation value of $\chi^2$ is generally not equal to the number of degrees of freedom, and the expression for the $p$-value is likewise modified. We therefore approximate the reduced $\chi^2$, $\chi^2_\m{red} = \chi^2/\ev*{\chi^2}$, and the $p$-value as outlined in Ref.~\cite{Bruno:2022mfy}. To assess the quality of the fits, we required $\chi^2_\m{red} \approx 1$ and a $p$-value $\geq 0.05$. 

\section{$N_\m{sep}$ dependence of the position-space sampling method}
\label{sec:Nsep_dependence}

To study the efficiency of the position-space sampling method, we computed the two-point functions of single-meson, single-baryon, and local tetraquark operators while varying the point separation $N_\m{sep}$ in the sparse grids. We investigated how the variance of the effective energy $E_\m{eff}$ and of the energy obtained from a plateau fit to $E_\m{eff}$ changes with $N_\m{sep}$. The expectation is that, as $N_\m{sep}$ is decreased, the variance decreases until the error is dominated by the Monte Carlo error. Reducing $N_\m{sep}$ beyond that point yields no further reduction in the statistical error. This value is the optimal point separation for the simulations, and finding it for the different operators is the goal of this investigation.

We carried out this investigation mostly on B450 (see Table~\ref{tab:ensembles}) to keep the computational cost low. To be able to apply the position-space sampling method on this ensemble, $N_\m{sep}$ must divide the spatial extent $N_s = 32$ of the lattice. Consequently, the possible values are $N_\m{sep} = 1, 2, 4, 8, 16, 32$, where $N_\m{sep} = 1$ corresponds to using the full lattice and $N_\m{sep} = 32$ to using only a single point of the spatial lattice on a given time slice. Since the noise with $N_\m{sep} = 32$ was too large for the correlator data to be usable, the largest point separation we use is $N_\m{sep} = 16$.

\subsection{Single-meson and single-baryon operators}
\label{sec:Nsep_single_hadron}

\begin{figure}[t]
  \includegraphics[width=0.5\textwidth]{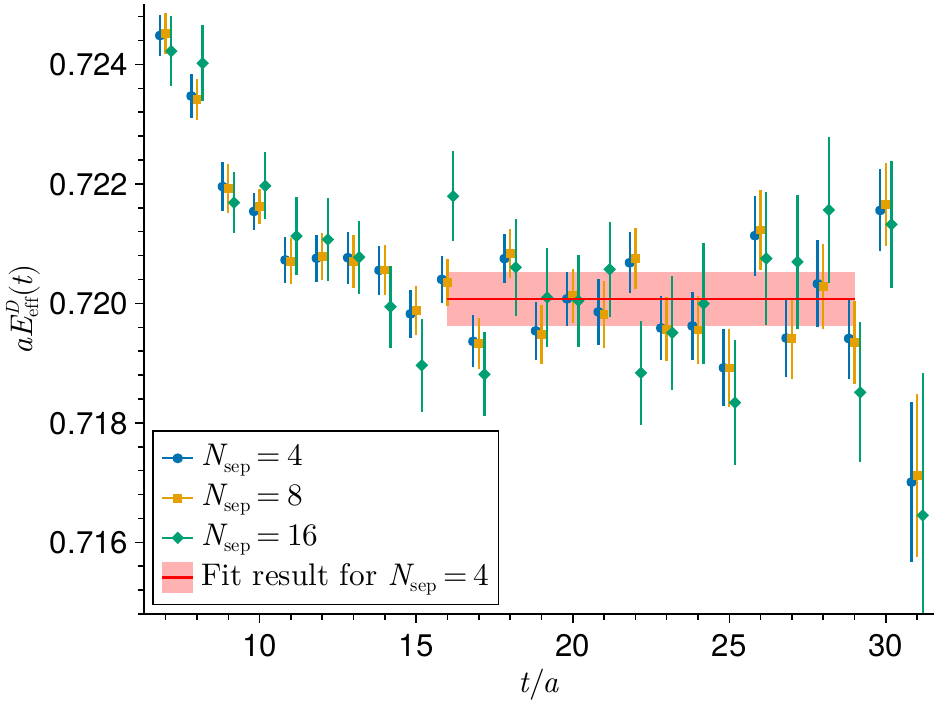}%
  \includegraphics[width=0.5\textwidth]{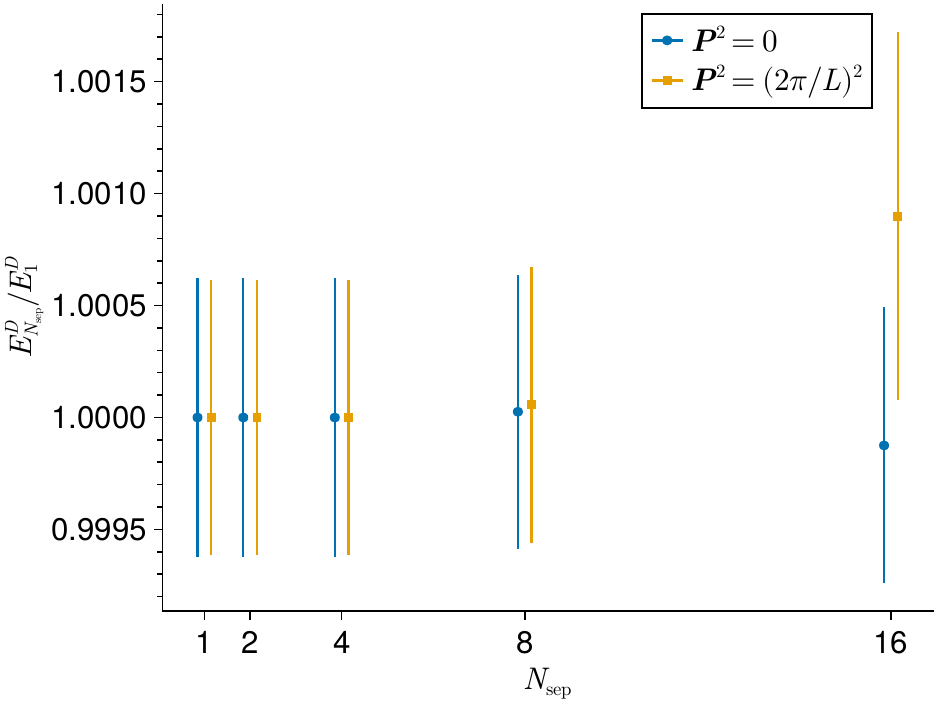}
  
  \includegraphics[width=0.5\textwidth]{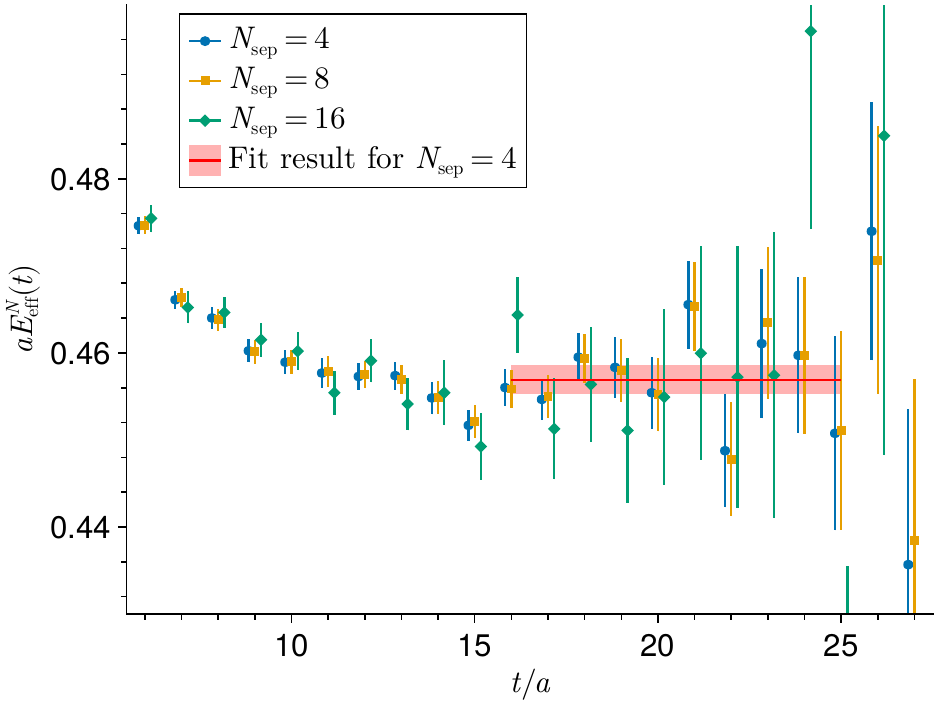}%
  \includegraphics[width=0.5\textwidth]{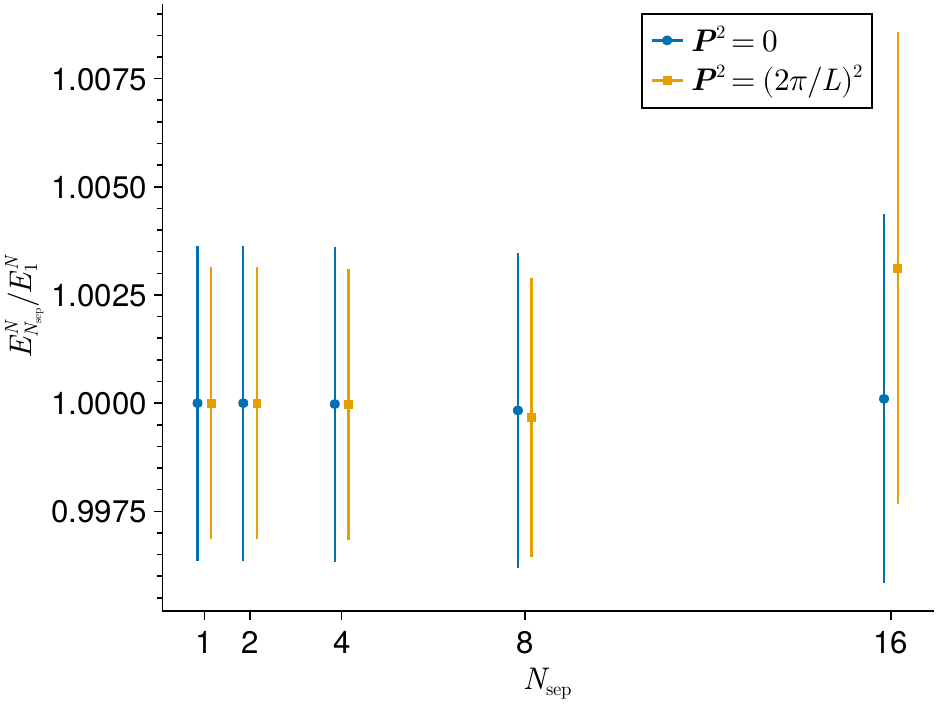}
  \caption{\label{fig:Nsep_analysis_for_D_and_N}%
    Effective energies $E_\m{eff}^h$ for the $D$ meson ($h = D$, top left) and the nucleon ($h = N$, bottom left) at zero momentum and for different point separations $N_\m{sep}$. The red bands are results from plateau fits for $N_\m{sep}=4$. The right panels display the energies $E_{N_\m{sep}}^h$ obtained from plateau fits to $E_\m{eff}^h$ for the $D$ meson (top right) and the nucleon (bottom right) as a function of $N_\m{sep}$. The energies $E_{N_\m{sep}}^h$ were computed using momenta with $\vb*{P}^2 = 0$ and $\vb*{P}^2 = (\tfrac{2\pi}{L})^2$, and are normalized by the mean of the $N_\m{sep}=1$ value $E_1^h$.}
\end{figure}

As an example of a single-meson system, we investigate the $D$ meson, using the pseudoscalar operator
\begin{equation}
  \label{eq:D_operator}
  \mathcal{O}^D(t) = (\overline{u}\,\gamma_5\, c)(\vb*{P}, t),
\end{equation}
for the total momenta $\vb*{P} = \vb*{0}$ and $\vb*{P} = \tfrac{2\pi}{L}(1, 0, 0)$. We show the effective energy $E_\m{eff}^D(t)$ as a function of $t$ in the top-left panel of Figure~\ref{fig:Nsep_analysis_for_D_and_N}. It is displayed for zero momentum and for the point separations $N_\m{sep} = 4, 8, 16$.
We used the $\cosh$ definition to compute $E_\m{eff}^D$, which is appropriate for the periodic temporal boundary conditions on B450. We see a significant decrease in the error when going from $N_\m{sep} = 16$ to $N_\m{sep} = 8$, but it remains constant when going to $N_\m{sep} = 4$. This indicates that, starting from $N_\m{sep} = 8$, the error from the position-space sampling is dominated by the Monte Carlo error. We obtained the $D$-meson energy $E_{N_\m{sep}}^D$ as a function of $N_\m{sep}$ from plateau fits to these effective energies. We used the same fit range for all $N_\m{sep}$ in order to isolate the $N_\m{sep}$ dependence. The top-right panel of Figure~\ref{fig:Nsep_analysis_for_D_and_N} displays the resulting energies as a function of $N_\m{sep}$ and for the two momenta used. They are normalized by the mean of the $N_\m{sep}=1$ value. For zero momentum, the error of the energy does not change significantly up to $N_\m{sep} = 16$, which indicates that the error reduction seen in $E_\m{eff}$ in the step from $N_\m{sep} = 16$ to $N_\m{sep} = 8$ is associated with an increased correlation between different times. For $\vb*{P}^2 = (\tfrac{2\pi}{L})^2$, the error is constant for $N_\m{sep} \leq 8$ and increases slightly at $N_\m{sep} = 16$. In all cases, the energy is consistent with the $N_\m{sep} = 1$ value, as expected since the position-space sampling estimator is unbiased.

We performed the same analysis for the nucleon $N$, using the operator
\begin{equation}
  \label{eq:nucleon_operator}
  \mathcal{O}^N(t) = \left[\epsilon_{abc}\; P_+ u_a \left(u_b^T\, C\gamma_5 P_+\, d_c\right)\right](\vb*{P}, t),
\end{equation}
where $P_+$ and $C$ are the positive-parity projector and the charge-conjugation matrix introduced in Eq.~(\ref{eq:parity_proj_charge_conj_matrix}), and the lowercase Latin letters are colour indices. This operator has an open spin index (suppressed in the notation), which we sum over in the two-point function. We again used the total momenta with $\vb*{P}^2 = 0, (\tfrac{2\pi}{L})^2$, averaging over the six spatial directions in the latter case to improve the signal. The results are shown in the lower panels of Figure~\ref{fig:Nsep_analysis_for_D_and_N}. For the nucleon effective energy $E_\m{eff}^N(t)$, we again observe a significant decrease in the error from $N_\m{sep} = 16$ to $N_\m{sep} = 8$, after which it is stable. The nucleon energy $E_{N_\m{sep}}^N$, obtained from plateau fits with the same fit range for all $N_\m{sep}$, has a stable error for $N_\m{sep} \leq 8$. When going to $N_\m{sep} = 16$, it shows a slight increase for zero momentum and a significant increase for $\vb*{P}^2 = (\tfrac{2\pi}{L})^2$. As for the $D$ meson, the error decrease from $N_\m{sep} = 16$ to $N_\m{sep} = 8$ is significantly stronger in the effective energy than in the extracted energy, again indicating an increased correlation in the data between different times.

We also performed this analysis for the pion, using the operator in Eq.~(\ref{eq:pion_operator}), and for the $D^*$ meson, using the vector operator
\begin{equation}
  \label{eq:Dstar_operator}
  \mathcal{O}_i^{D^*}(t) = (\overline{u}\,\gamma_i\, c)(\vb*{P}, t).
\end{equation}
The results are qualitatively similar and are therefore not shown. We conclude that position-space sampling works well for the two-point functions of single-meson and single-baryon operators, so that a large point separation $N_\m{sep}$ can be used in the sparse grid without increasing the error of the observables. On B450, the optimal point separation for these operators is $N_\m{sep} = 8$.

\subsection{Tetraquark operators}
\label{sec:Nsep_tetraquark}

The investigation of single-meson and single-baryon operators serves only as a test; the actual cost savings of the position-space sampling method arise for local multiquark operators with four or more (anti)quarks.\footnote{In fact, with its $N_v^3$ cost scaling, position-space sampling scales better than baryon two-point functions in traditional distillation, which scale as $N_v^4$. In practice, however, the large prefactor makes position-space sampling more expensive than traditional distillation for single-baryon operators.} We therefore investigate local tetraquark operators relevant for $DD^*$ scattering and the $T_{cc}$: the local $DD^*$ and the local diquark-antidiquark operators in Eqs.~(\ref{eq:local_DDstar}) and~(\ref{eq:diquark_operator}). We expect the local $D^*D^*$ operator to behave similarly to the $DD^*$ operator owing to its similar structure. These operators have quantum numbers $I(J^P) = 0(1^+)$ with quark content $cc\overline{u}\overline{d}$, and we consider them only in the rest frame ($\vb*{P} = \vb*{0}$), where they belong to the $T_1^+$ irrep of the octahedral group $O_h$.

\begin{figure}
  \centering
  \includegraphics[width=0.5\textwidth]{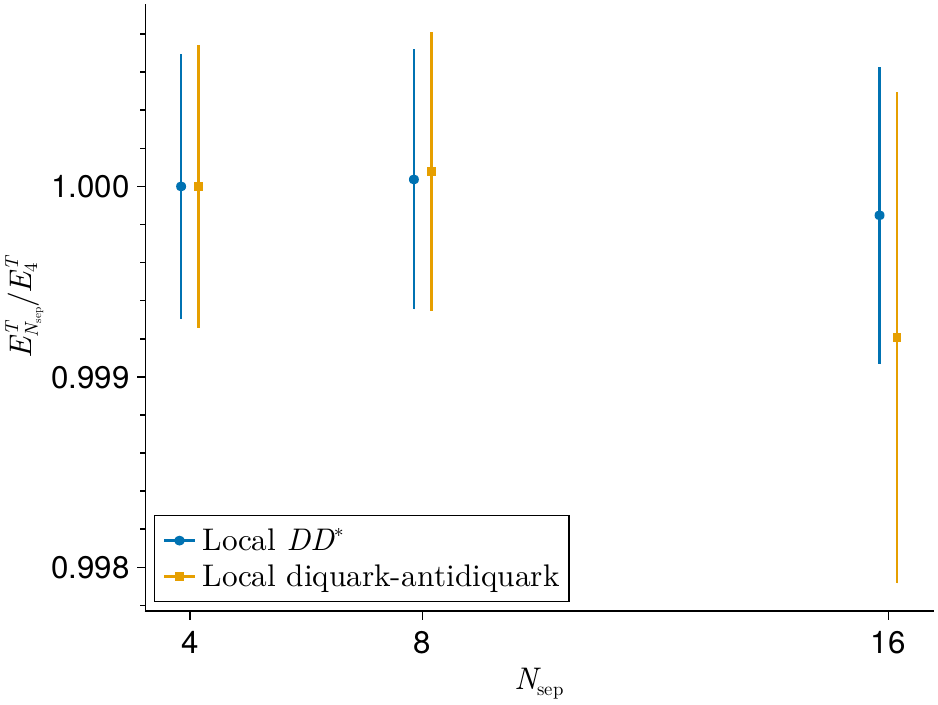}
  \caption{\label{fig:Nsep_tetraquark}%
    Plateau values $E_{N_\m{sep}}^T$ from plateau fits to the effective energy of the local $DD^*$ and diquark-antidiquark two-point functions vs.\ the point separation $N_\m{sep}$. The plateau values are normalized by the mean of the $N_\m{sep}=4$ value $E_4^T$.}
\end{figure}

These tetraquark operators are not optimized using the GEVP. Since the energy levels in the multiparticle system are closely spaced, the contributions from excited states are not expected to decay before the signal is lost, so that a genuine finite-volume ground-state energy cannot be extracted with these operators. For the investigation of the $N_\m{sep}$ dependence, we therefore performed plateau fits to the effective energies at late times and compared the resulting plateau values. Again, we used the same plateau range for all $N_\m{sep}$ in order to isolate the $N_\m{sep}$ dependence. The resulting plateau values $E_{N_\m{sep}}^T$ as a function of $N_\m{sep}$ are shown in Figure~\ref{fig:Nsep_tetraquark}.
Since the cost of computing the tetraquark two-point functions scales as $N_\m{sep}^{-6}$ and is generally larger than for the single-hadron two-point functions, we could only use $N_\m{sep} = 4, 8, 16$. The results are therefore normalized by the best estimate $E_4^T$.

We observe a decrease in the error when going from $N_\m{sep} = 16$ to $N_\m{sep} = 8$, and a constant error for $N_\m{sep} \leq 8$ for both operators. The same behaviour is seen in the effective energy (not shown); however, the error decrease from $N_\m{sep} = 16$ to $N_\m{sep} = 8$ is more pronounced than in $E_{N_\m{sep}}^T$. This again indicates a significant increase in the correlation between different times as $N_\m{sep}$ is decreased. We conclude that position-space sampling also works well for local tetraquark operators, with an optimal point separation of $N_\m{sep} = 8$ on B450.

The N202 ensemble (see Table~\ref{tab:ensembles}) has a lattice size of $N_s = 48$, so that the possible and reasonable point separations are $N_\m{sep} = 4, 6, 8, 12, 16$. Since the lattice spacing is about $15\%$ smaller than on B450, keeping the physical separation $aN_\m{sep}$ fixed would correspond to $N_\m{sep} \approx 9.5$ on N202. Consequently, one would use either $N_\m{sep} = 8$ or $N_\m{sep} = 12$. Our numerical investigation shows that the plateau values for the tetraquark operators already stabilize at $N_\m{sep} = 12$, but a small decrease in the error of the effective energy is still visible in the step from $N_\m{sep} = 12$ to $N_\m{sep} = 8$. We therefore also chose this point separation, $N_\m{sep} = 8$, on N202, and used it for the investigation of the $DD^*$ and $BB^*$ scattering systems.

\section{Results for $DD^*$ system}
\label{sec:DDstar_results}

In this section, we present our results for the $DD^*$ system. The analysis has two parts. We first study the importance of the local tetraquark operators for obtaining an accurate low-lying finite-volume spectrum relevant for the $T_{cc}$, by comparing the spectrum extracted with and without local operators. We then examine the effect of including these operators on the $s$-wave $DD^*$ scattering phase shift, obtained via Lüscher's finite-volume quantization condition.

\subsection{Importance of local operators for the finite-volume spectrum}
\label{sec:operator_importance}

The importance of local operators for the finite-volume $cc\overline{u}\overline{d}$, isospin-0, rest-frame $T_1^+$ spectrum --- relevant for the $T_{cc}$ tetraquark --- has been discussed in the literature~\cite{Ortiz-Pacheco:2023ble, Prelovsek:2025vbr}. Upon adding the diquark-antidiquark operator to a basis of bilocal two-meson operators, the authors find little effect on the ground-state energy estimate but a significant effect on the first excited state. To compute the two-point function of the local operator within distillation, they use the rank-4 tensor $\Xi$ mentioned in Section~\ref{sec:multiquark_distillation}. Because of the $N_v^5$ scaling of the computational cost associated with this tensor, they were forced to use a smaller number of eigenvectors for the local tetraquark operators than for the bilocal meson-meson operators. In our simulations, in contrast, the position-space sampling method allows us to use the same $N_v$ for all operators.

To investigate the relevance of local tetraquark operators for this system more thoroughly, we performed an operator-importance analysis. In this analysis, we investigated the stability of the energy estimates in the low-lying spectrum when changing the number of operators in the basis, especially upon including local tetraquark operators. We performed this analysis mostly on the N202 ensemble, since it has the larger physical volume of the two and therefore the denser finite-volume spectrum. This leads to stronger excited-state contamination, whose reduction through the inclusion of local operators is precisely what we aim to investigate.

For our operator-importance analysis, we used the three local tetraquark operators in Eqs.~(\ref{eq:local_DDstar})--(\ref{eq:diquark_operator}) together with a large basis of bilocal $DD^*$ and $D^*D^*$ operators in the rest-frame $T_1^+$ irrep. The latter were constructed as described in Section~\ref{sec:bilocal_operators}. For the relative momentum $\vb*{p} = \tfrac{2\pi}{L}\vb*{d}$ we used the momentum shells $\vb*{d}^2 = 0, 1, 2$ for the $DD^*$-type operators, with $N_\m{ops} = 1, 2, 3$ operators, respectively, and $\vb*{d}^2 = 0, 1$ for the $D^*D^*$-type operators, with $N_\m{ops} = 1, 2$ operators, respectively. Here, $N_\m{ops}$ corresponds to the number of noninteracting energy levels on the given momentum shell, giving a total of nine bilocal meson-meson operators. Using these momentum shells corresponds to including all operators associated with the noninteracting levels up to the $DD\pi$ three-particle threshold.

To organize the $DD^*$ and $D^*D^*$ operators, we collect those on a given momentum shell into groups, which we denote by
\begin{equation}
  \label{eq:operator_group_notation}
  D^{(*)}(\vb*{d}^2)\,D^*(\vb*{d}^2)\;\{N_\m{ops}\}.
\end{equation}
Here, $D^{(*)}$ stands for $D$ or $D^*$ and the number of operators $N_\m{ops}$ is given in the curly braces. The three local tetraquark operators form the group $T\;\{3\}$, so that the full basis contains twelve operators.

For the analysis, we ordered the operator groups starting with the local tetraquark operators $T\;\{3\}$, followed by the $D^{(*)}D^*$ groups in order of increasing noninteracting $D^{(*)}D^*$ energy. In this order, we extended the operator basis one group at a time and estimated the low-lying energies at each step. We then repeated the same procedure without including the local tetraquark operators. Comparing the resulting energy estimates reveals how important each operator group is for obtaining the correct spectrum.

We extracted the low-lying energy levels using the variational method outlined in Section~\ref{sec:variational_method}, from which we obtained the effective energies $E_{\m{eff},n}(t)$. In the GEVP, we used the reference time $t_0/a = \lceil t/(2a)\rceil$, where $\lceil\cdot\rceil$ is the ceiling function. This choice satisfies the condition $t_0 \geq t/2$ and ensures the suppression of excited-state effects given in Eq.~(\ref{eq:GEVP_suppression}). Rather than extracting the energies directly from $E_{\m{eff},n}(t)$, we exploited the fact that the energy levels $E_n$ are strongly correlated with the noninteracting $D^{(*)}D^*$ energies. We therefore extracted the energies via a plateau fit to the effective energy difference
\begin{equation}
  \label{eq:effective_energy_difference}
  \Delta E_{\m{eff},n}(t) = E_{\m{eff},n}(t) - m^D_\m{eff}(t) - m^{D^*}_\m{eff}(t),
\end{equation}
where $m^D_\m{eff}(t)$ and $m^{D^*}_\m{eff}(t)$ are the effective masses of the $D$ and $D^*$ mesons. The fit yields the energy difference $\Delta E_n = E_n - m_D - m_{D^*}$; we computed the $D$ and $D^*$ masses $m_D$ and $m_{D^*}$ separately, from plateau fits to $m^D_\m{eff}(t)$ and $m^{D^*}_\m{eff}(t)$. This procedure is similar to fitting ratios of correlators, which can produce spurious plateaus~\cite{Iritani:2016jie}. To avoid them, we restricted the fits to time ranges in which both the two-particle effective energy and the single-particle effective masses have plateaued, as advocated in Ref.~\cite{Yamazaki:2017jfh}. In addition, we varied the fit range to investigate the stability of the energy estimates. In case of instability, we adopted a more conservative fit range.

\begin{figure}[t]
  \includegraphics[width=0.5\textwidth]{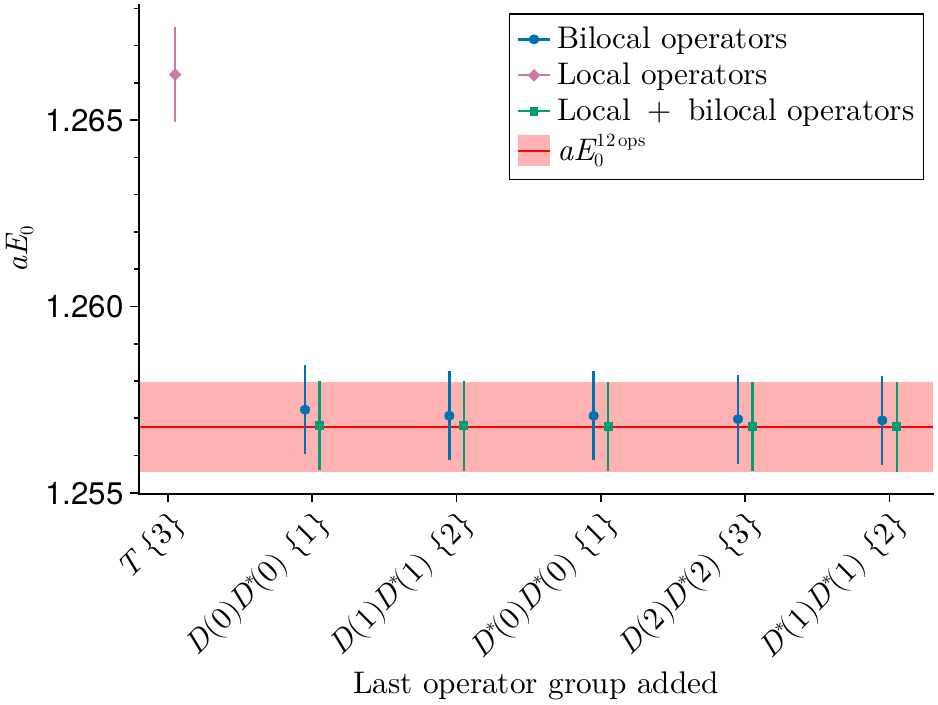}%
  \includegraphics[width=0.5\textwidth]{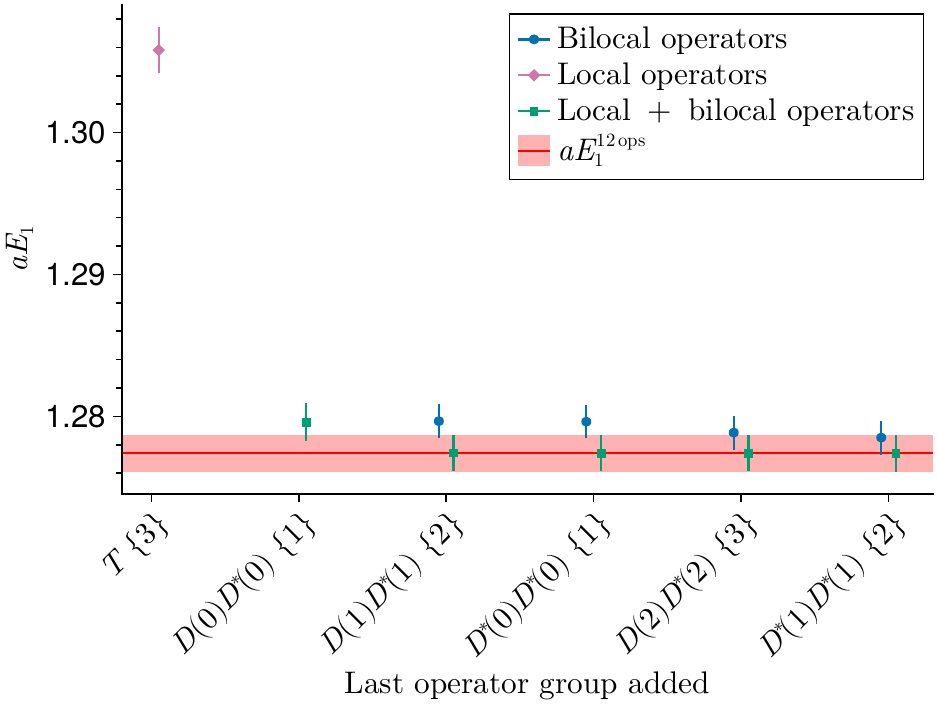}
  
  \includegraphics[width=0.5\textwidth]{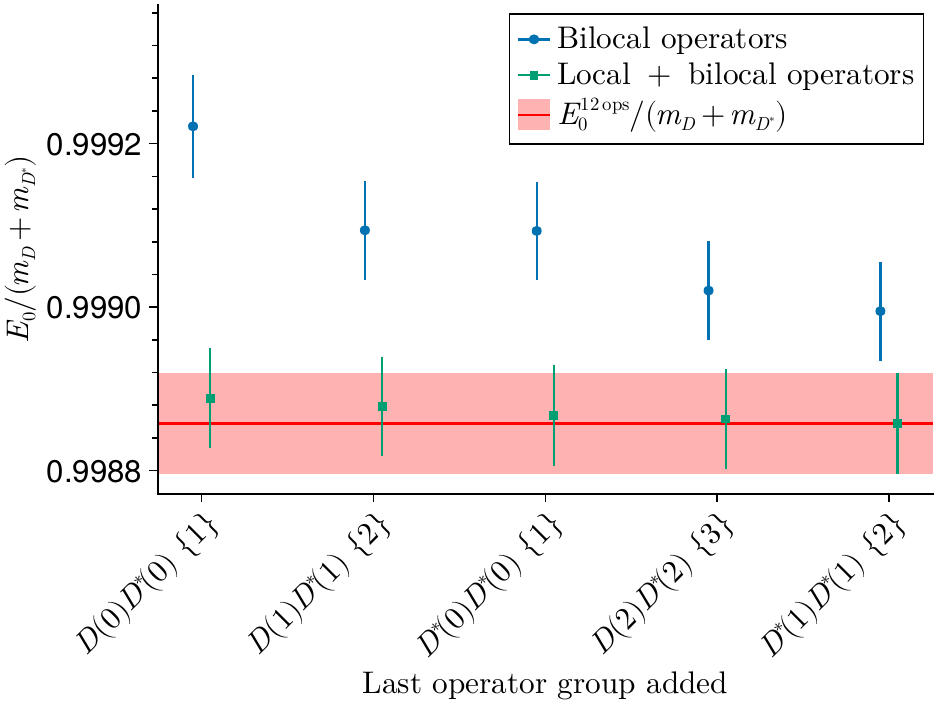}%
  \includegraphics[width=0.5\textwidth]{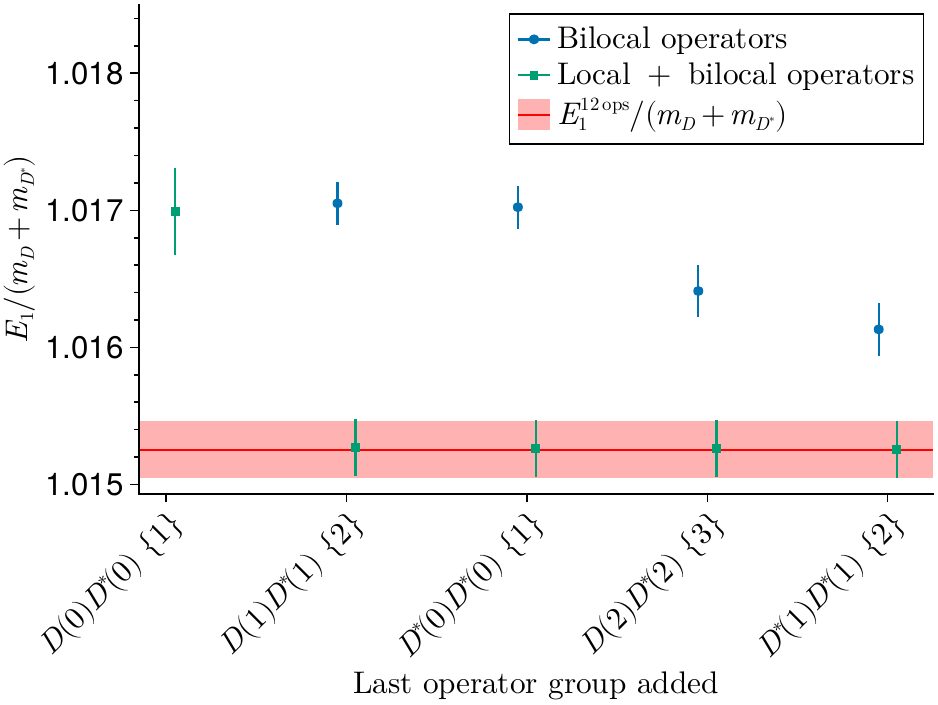}
  \caption{\label{fig:operator_importance}%
    Finite-volume energy estimates on N202 as the operator basis is extended one group at a time, with and without the local tetraquark operators. The $x$-axis denotes the last operator group added to the basis for computing $E_n$; groups to the left were already present (except $T\;\{3\}$ when only bilocal operators are used). The red bands show the results from using all twelve operators. Results are shown for the ground state (left) and the first excited state (right). The top panels show the energies themselves, and the bottom panels show the correlated ratio of the interacting energies to the threshold energy $m_D + m_{D^*}$. We removed the $T\;\{3\}$ energy estimates from the lower panels because of their large values compared to the others.}
\end{figure}

The resulting energy estimates on N202 as a function of the operator basis are shown in the top panels of Figure~\ref{fig:operator_importance}.
For the ground state, shown in the top-left panel, including the local operators does not change the estimates significantly, and even a single bilocal meson-meson operator already agrees well with the twelve-operator result. For the first excited state, shown in the top-right panel, the estimates from the purely bilocal basis decrease in a steplike manner as operator groups are added. In contrast, with the local tetraquark operators included, they have already converged once the first three operator groups are used. When all bilocal operators are used without any local operator, the estimate of the first excited state lies about $1\sigma$ away from the best estimate obtained with all twelve operators. This indicates that the bilocal operators alone do not produce an accurate spectrum. Conversely, the local tetraquark operators alone are also insufficient: the $T\;\{3\}$ result lies far from the twelve-operator result for both the ground and the first excited state.

In the Lüscher analysis, the phase shift is determined from the c.m.\ scattering momenta $q_\m{cm}$, which requires both the interacting two-particle energy levels and the single-particle masses. Because the two are strongly correlated, $q_\m{cm}$ can have a significantly smaller error than either of them, revealing discrepancies between different estimates more clearly. A similar error cancellation occurs when forming the ratio of the interacting and the noninteracting two-particle energies. We therefore also consider the correlated ratio of the interacting energies to the threshold energy $m_D + m_{D^*}$, shown in the bottom panels of Figure~\ref{fig:operator_importance}. Compared with the energies themselves, the error in the ratio is substantially smaller. One then sees a more pronounced difference between the results from the bilocal and the mixed local-bilocal operator bases, where the former shows a steplike decrease for both the ground and the first excited state. The estimates from the bilocal bases do not reach the twelve-operator result in either case, whereas those from the mixed local-bilocal bases stabilize with only a few operator groups. The shifts in the estimates from the full bilocal basis upon including local tetraquark operators are significant: the ground state, which showed no significant shift in the energy itself, moves by about $2\sigma$ in the ratio. For the first excited state, the shift is even more pronounced, at about $3\sigma$.

\begin{figure}
  \includegraphics[width=\textwidth]{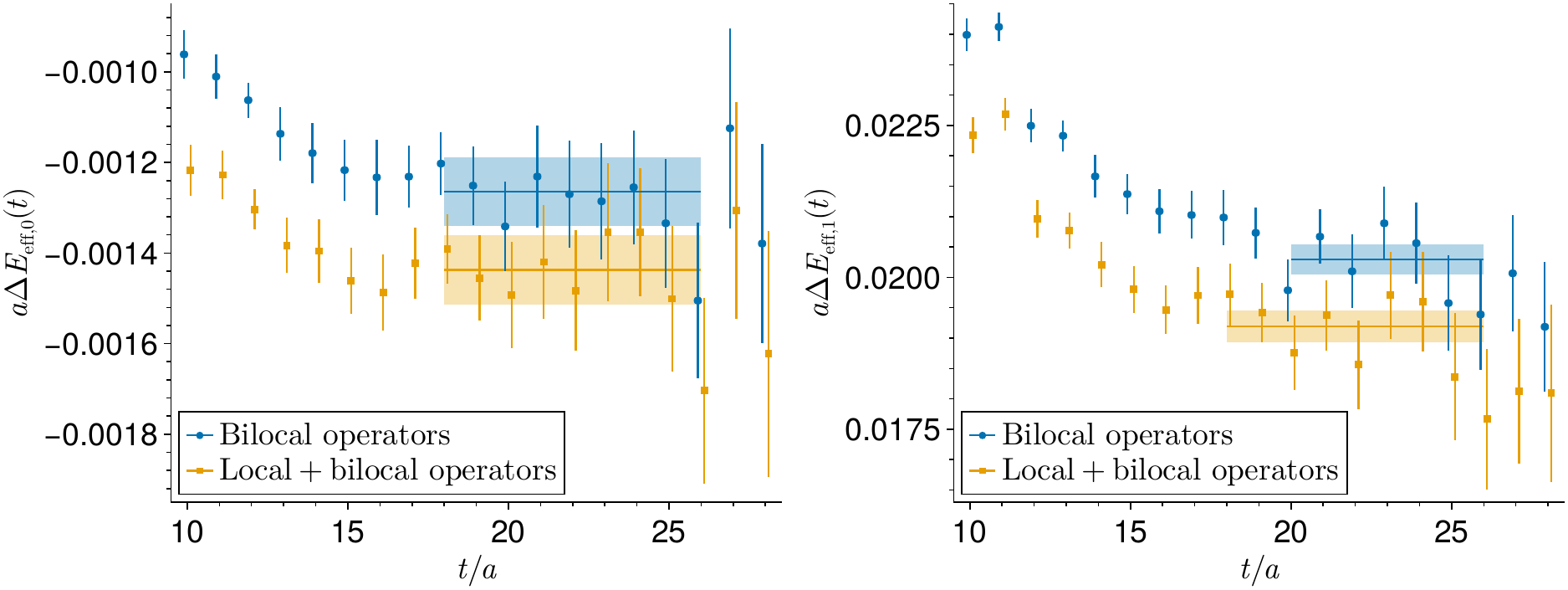}
  \caption{\label{fig:operator_importance_E_eff}%
    Effective energy differences $a\Delta E_{\m{eff},n}$ as functions of $t$ for the ground (left) and first excited states (right) on N202, using all bilocal operators and using all local and bilocal operators. The coloured bands display results from plateau fits.}
\end{figure}

For the full bilocal and the full local-bilocal bases, the corresponding effective energy differences $a\Delta E_{\m{eff},n}$ of the lowest two levels are shown in Figure~\ref{fig:operator_importance_E_eff}.
This figure also displays the results of plateau fits to them. For both levels, the effective energy difference shows a significant deviation between the two bases. For the ground state, a clear plateau emerges for both bases, but the one from the full local-bilocal operator basis lies about $1\sigma$ to $2\sigma$ lower. For the first excited state, the plateaus that emerge are less clear, but the deviation between the two effective energy differences is even more pronounced. From visual inspection alone, one could not tell that the true plateau has not yet emerged with the bilocal basis. Resolving it would require going to very large $t$, where the signal is already lost.

Adding more bilocal operators on higher momentum shells to the purely bilocal basis could in principle speed up the convergence of the effective energy difference towards its plateau. This is, however, not feasible in practice: the bilocal operators become increasingly expensive on higher momentum shells owing to the large number of momentum combinations --- they are also generally noisier --- and at some point one would need to include three-particle operators. Including local tetraquark operators using our position-space sampling method is therefore highly favourable.

This type of operator-importance analysis can be used to assess the reliability of the extracted spectrum. From the energy estimates obtained with the bilocal operator basis alone, one can already see that the first excited level is not reliably determined, as it is not stable under enlargement of the basis. This can be seen without comparing to the mixed local-bilocal operator basis. By contrast, inspecting only the effective energy difference itself, when using the full bilocal basis, does not reveal this instability.

We also studied the individual influence of the two types of local operators --- the local two-meson operators and the diquark-antidiquark operator --- on the spectrum. To this end, we repeated the analysis on N202 while omitting one of them at a time. This is of interest because fewer operator types, if sufficient, would simplify the construction and computation of the two-point functions. For this comparison, we again used the ratio of the lowest two levels to the threshold energy. We found that excluding the diquark-antidiquark operator yields energy estimates virtually identical to those from the full local-bilocal operator basis. When the local $D^{(*)}D^*$ operators are excluded instead, the first excited state shows a steplike decrease as bilocal operators are added: a difference of about $0.4\sigma$ from the twelve-operator result remains when all nine bilocal operators are used. The ground state, on the other hand, is affected negligibly when the local two-meson operators are omitted. We do not show the corresponding plots. We conclude that using all bilocal operators together with either type of local operator allows an accurate extraction of the lowest two energy levels. However, adding the local two-meson operators yields results that are more stable with respect to the number of bilocal operators than adding the diquark-antidiquark operator.

\begin{figure}[t]
  \includegraphics[width=\textwidth]{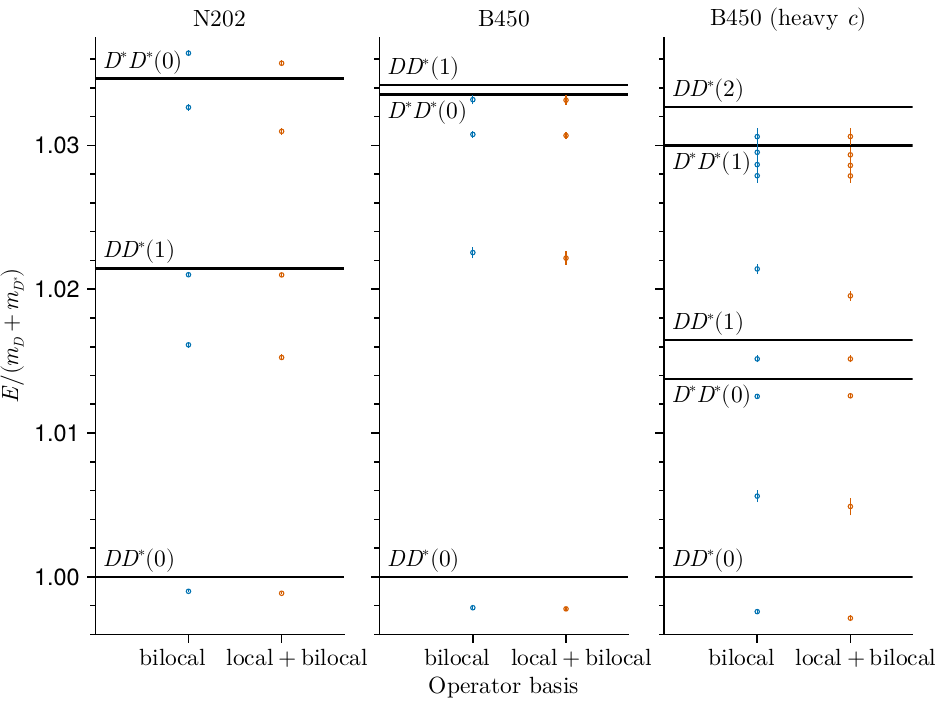}
  \caption{\label{fig:DDstar_spectrum}%
    Estimate of the finite-volume $cc\overline{u}\overline{d}$, isospin-0 spectrum in the $T_1^+$ irrep for the full bilocal operator basis and for the full local-bilocal operator basis. The energy estimates are normalized by the threshold energy using correlated ratios. The spectrum is presented for the N202 (left) and the B450 (middle) gauge ensembles. In addition, energy estimates on B450 computed with a heavier-than-physical charm quark are displayed (right). The horizontal lines denote the noninteracting $D^{(*)}D^*$ energies (with the integer back-to-back momentum squared in parentheses).}
\end{figure}

Finally, we present our estimates of the low-lying finite-volume spectra, computed from the full bilocal and the full local-bilocal operator bases on both the N202 and the B450 ensembles. The results, with the energies normalized to the threshold energy, are shown in Figure~\ref{fig:DDstar_spectrum}.
On N202, in addition to the shifts in the ground and first excited states discussed above, we find significant shifts in higher excited states when the local operators are included. Most notably, the level below the $D^*D^*(0)$ threshold shifts by about $5\sigma$. The only level that is essentially unaffected is the one below the noninteracting $DD^*(1)$ level. On B450, which has a coarser lattice spacing and a smaller physical volume, the inclusion of the local operators has a smaller effect: the ground and first excited states shift by about $0.5\sigma$, and the two levels close to the $D^*D^*(0)$ threshold remain mostly unchanged. A possible reason for this weaker effect is the smaller physical volume, which makes the spectrum less dense than on N202. The wider gaps between the finite-volume energy levels then result in a stronger suppression of the excited states.

We also investigated the spectrum on B450 with a heavier-than-physical charm quark, as introduced in Section~\ref{sec:Lattice_setup}. These results are of interest because a heavier charm quark interpolates between the physical charm and the bottom quark. Replacing the charm quarks by bottom quarks gives the $BB^*$ system, which contains the deeply bound $T_{bb}$. As noted earlier, the inclusion of local tetraquark operators is crucial for obtaining an accurate spectrum in the $BB^*$ system, especially for the ground state. We therefore expect the heavier charm quark to lead to stronger binding and to a more important role for the local operators. This is indeed what we observe: the ground-state energy estimate shifts by about $1.5\sigma$ and the first excited state by about $1\sigma$. The state above the $DD^*(1)$ noninteracting level also shows a significant shift. This shift is already present with the physical charm quark (not shown in the plot), but is enhanced with the heavier one.

In summary, there is qualitative agreement between the spectra obtained from the bilocal and the local-bilocal operator bases, in the sense that no energy level is missed. Depending on the gauge ensemble, however, certain energy estimates shift significantly upon including local tetraquark operators. Consequently, omitting them can introduce a considerable systematic error.

\subsection{\texorpdfstring{$s$}{s}-wave Lüscher analysis}
\label{sec:DDstar_luscher}

We now investigate the effect of including local tetraquark operators on the $s$-wave $DD^*$ scattering phase shift. To that end, we use Lüscher's finite-volume quantization condition as described in Section~\ref{sec:luscher}. We perform a single-channel $s$-wave analysis in the rest frame, thereby neglecting $d$-wave interactions. There are no $p$-wave effects in the rest frame, since parity is conserved. The relation between the c.m.\ scattering momentum $q_\m{cm}$ and $q_\m{cm}\cot\delta_0(q_\m{cm})$ is given by Eq.~(\ref{eq:swave_QC}) for $\vb*{P} = \vb*{0}$.

Before performing the analysis, we correct for deviations from the continuum dispersion relation. We measured such deviations for the $D$ and $D^*$ mesons on both ensembles and accounted for them using the modified dispersion relation
\begin{equation}
  \label{eq:modified_dispersion}
  E(\vb*{p})^2 = m^2 + c^2\vb*{p}^2,
\end{equation}
with a speed-of-light parameter $c^2$. This is the same form as used for the $b$ quark in Section~\ref{sec:RHQ_action} (see Eq.~(\ref{eq:dispersion_csq})). For mesons, this modified dispersion relation can be derived, for example, using Symanzik effective field theory in the limit of small momenta, and it has been used before, e.g., in Ref.~\cite{Shi:2025ogt}. It modifies the equation for the scattering momentum (see Eq.~(\ref{eq:qcm_implicit})) to
\begin{equation}
  \label{eq:modified_qcm_implicit}
  \sqrt{s} = \sqrt{m_D^2 + c^2_D q_\m{cm}^2} + \sqrt{m_{D^*}^2 + c^2_{D^*} q_\m{cm}^2}.
\end{equation}
This equation is solved by
\begin{equation}
  \label{eq:modified_qcm_explicit}
  \begin{split}
    q_\m{cm}^2 = \frac{s}{(c^2_D - c^2_{D^*})^2}\Bigg[
      &c^2_D + c^2_{D^*} + (c^2_D - c^2_{D^*})\frac{m_{D^*}^2 - m_D^2}{s} \\
      &- 2\sqrt{(c^2_D - c^2_{D^*})\frac{c^2_D m_{D^*}^2 - c^2_{D^*} m_D^2}{s} + c^2_D c^2_{D^*}}\,\Bigg],
  \end{split}
\end{equation}
which has the correct limit for $c^2_D, c^2_{D^*} \rightarrow 1$. This modification works only in the rest frame, where the two mesons have back-to-back momenta. We obtained the correction factors $c^2_D$ and $c^2_{D^*}$ from constant fits to
\begin{equation}
  \label{eq:csq_fit}
  c^2_{D^{(*)}}(\vb*{p}) = \frac{E_{D^{(*)}}(\vb*{p})^2 - m_{D^{(*)}}^2}{\vb*{p}^2},
\end{equation}
where $E_D(\vb*{p})$ and $E_{D^*}(\vb*{p})$ are the $D$ and $D^*$ energies at momentum $\vb*{p}$. We used the fit range $(\tfrac{L}{2\pi})^2\vb*{p}^2 = 1, 2, 3$, over which $c^2_D(\vb*{p})$ and $c^2_{D^*}(\vb*{p})$ are approximately constant. The results are listed in Table~\ref{tab:dispersion_rel_factors}.
\begin{table}
  \centering
  \begin{tabular}{lcc}
    \hline\hline
    Ensemble & $c^2_D$ & $c^2_{D^*}$ \\
    \hline
    B450 & 0.928(7)  & 0.91(1) \\
    N202 & 0.968(4) & 0.965(6) \\
    \hline\hline
  \end{tabular}
  \caption{\label{tab:dispersion_rel_factors}%
    Speed-of-light parameters $c^2_D$ and $c^2_{D^*}$ for the $D$ and $D^*$ mesons on the two ensembles.}
\end{table}

\begin{figure}[t]
  \centering
  \includegraphics[width=0.5\textwidth]{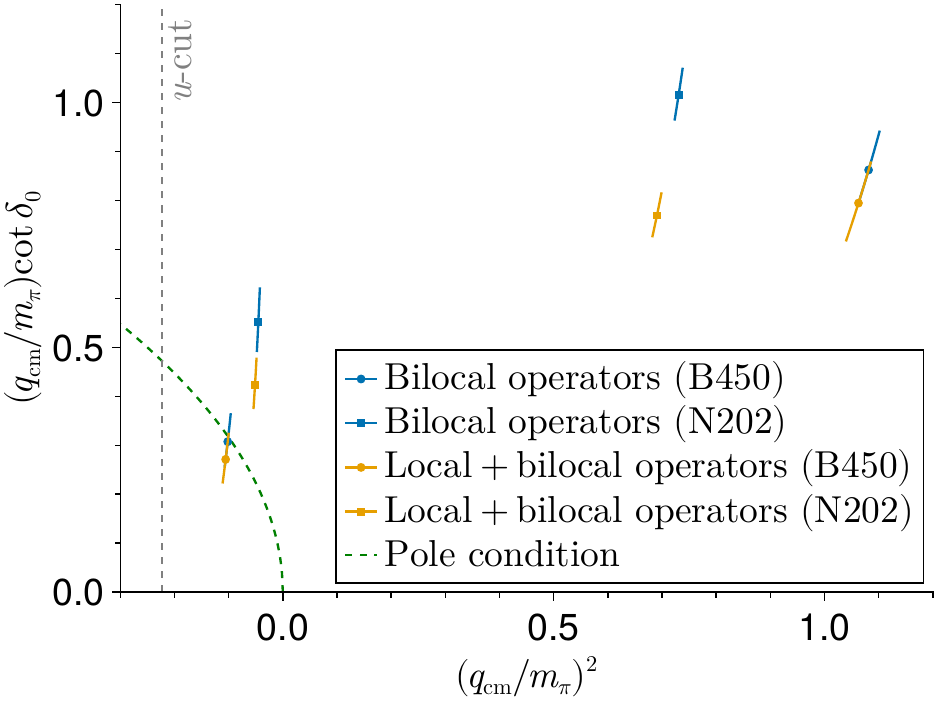}%
  \includegraphics[width=0.5\textwidth]{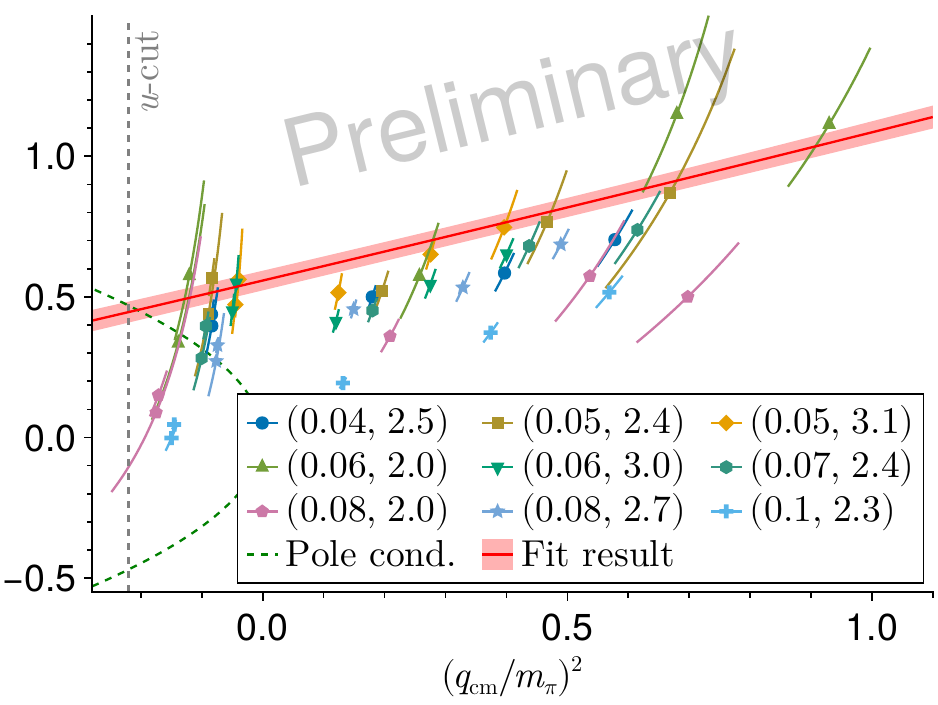}
  \caption{\label{fig:s-wave_analysis}%
    $s$-wave $DD^*$ scattering phase shift in terms of $q_\m{cm}\cot{\delta_0(q_\m{cm})}$ vs.\ $q_\m{cm}^2$. The c.m.\ scattering momenta $q_\m{cm}$ are normalized by the pion mass $m_\pi$. Left: estimates for the full bilocal and the full local-bilocal operator bases on two gauge ensembles. Right: preliminary results from the extended analysis across nine gauge ensembles using the full basis of local and bilocal operators. The data points are labelled as $(a, L)$, where $a$ and $L$ are the lattice spacing and the box size of the corresponding ensemble in fm, respectively. The red band displays the result from a combined fit to the data extrapolated to the continuum. In both plots, the green dashed line displays the pole condition $q_\m{cm}\cot{\delta_0(q_\m{cm})} = \pm \sqrt{-q_\m{cm}^2}$ and the vertical grey dashed line shows the start of the $u$-channel cut from one-pion exchange.}
\end{figure}

For single-channel $DD^*$ scattering, the quantization condition is valid between the $u$-channel cut from one-pion exchange, at $q_\m{cm}^2 \approx -m_\pi^2/4$, and the $D^*D^*(0)$ threshold. This restricts the analysis to the lowest two energy levels on both B450 and N202; the third level on N202 could in principle be used but is strongly influenced by the $d$-wave. The results are shown in the left panel of Figure~\ref{fig:s-wave_analysis}, which plots $q_\m{cm}\cot\delta_0$ vs.\ $q_\m{cm}^2$ for both operator bases and both ensembles.
As expected, the shifts in the energy estimates upon including local operators are also visible in the phase shift. Moreover, the agreement between the two ensembles improves when the local tetraquark operators are included: an analysis without them would have suggested a large discretization effect, even though the two lattice spacings are relatively close. This apparent difference between the ensembles is reduced once the local operators are included.

As part of our work within a Mainz--Zeuthen--Chennai collaboration, we extended the operator basis across nine CLS gauge ensembles at the $\m{SU}(3)$-flavour-symmetric point, with various lattice spacings and volumes. The following analysis of these data was carried out by collaborators. A variational analysis was used to extract the finite-volume energy levels, employing a model average over multiple fit results to obtain them together with a systematic error. A single-channel $s$-wave Lüscher analysis was then performed, in this case in both the rest frame and moving frames with $\vb*{P}^2 \leq 4\,(\tfrac{2\pi}{L})^2$.

The results are shown in the right panel of Figure~\ref{fig:s-wave_analysis}, again as $q_\m{cm}\cot\delta_0$ vs.\ $q_\m{cm}^2$. For the fit, an ERE of the form
\begin{equation}
  \label{eq:ERE_continuum_fit}
  q_\m{cm}\cot\delta_0 = (c_{0,0} + c_{0,1}a^2) + (c_{1,0} + c_{1,1}a^2)q_\m{cm}^2 + (c_{2,0} + c_{2,1}a^2)q_\m{cm}^4
\end{equation}
was used. It includes terms up to quadratic order in $q_\m{cm}^2$, with an explicit quadratic dependence of the effective-range parameters on the lattice spacing. We note that Eq.~(\ref{eq:ERE_continuum_fit}) is not derived from a systematic expansion in the lattice spacing, but is a phenomenological ansatz that assumes leading discretization effects of order $a^2$, as motivated by the $O(a)$ improvement of the Wilson-clover action. A similar strategy was employed in Ref.~\cite{Green:2021qol}. This fit is preliminary. The fit result shown in the plot is the extrapolation to the continuum, obtained by setting $a = 0$. One observes a crossing with the pole condition at positive $\sqrt{-q_\m{cm}^2}$, corresponding to a virtual bound state lying $17(3)\,\m{MeV}$ below threshold. We identify this state with the $T_{cc}$ tetraquark.

We now estimate the effect of the uncertainties of the lattice spacing and of the charm-quark mass tuning, which are not included in our error estimates. The results on B450 and N202 are given in lattice units, in units of the pion mass or the threshold energy and are therefore independent of the value of the lattice spacing. Both uncertainties do, however, enter through the tuning of the charm-quark mass, which is thus uncertain by up to about $1\%$. To estimate the size of this effect, we use the charm-quark-mass dependence of the $DD^*$ scattering parameters found in Ref.~\cite{Collins:2024sfi}. According to these results, a relative change in the $D$-meson mass translates into a relative change of similar size in the scattering length $a_0$ and the effective range $r_0$. The relative change in the (virtual) binding energy is about two to three times larger. The resulting effects are at the percent level and thus well below our target precision of about $10\%$.

A more important limitation of this analysis is the nearby left-hand cut from one-pion exchange, which probably distorts the identification of the $T_{cc}$ as a virtual bound state~\cite{Du:2023hlu, Prelovsek:2025vbr, Hansen:2024ffk, Dawid:2024dgy, Dawid:2025wsn,PitangaLachini:2026lyd}. This issue is being investigated by collaborators using a plane-wave approach~\cite{Meng:2021uhz,Meng:2023bmz} and a three-particle Lüscher formalism. Using the former and a pion mass of $m_\pi \approx 280\;\m{MeV}$, Ref.~\cite{Prelovsek:2025vbr} finds that the $T_{cc}$ is a subthreshold resonance instead of a virtual bound state.

The long-term goal is to investigate the $T_{cc}$ at the physical pion mass. At that mass, it is a bound state, so that decreasing $m_\pi$ must move the pole to a position corresponding to a positive imaginary part of $q_\m{cm}$. This transition is, however, difficult to achieve, as decreasing the pion mass results in an unstable $D^*$ meson. Consequently, the $DD^*$ system becomes the three-particle $DD\pi$ system, and a three-particle formalism becomes unavoidable. Progress in this direction has already been made~\cite{Hansen:2024ffk, Dawid:2024dgy, Dawid:2025wsn}.

\section{Results for $BB^*$ system}
\label{sec:Results_BBstar}

In this section, we present our results for the $\overline{b}\overline{b}ud$ system with $I(J^P) = 0(1^+)$. These are the same quantum numbers as those of the $DD^*$ system studied in the preceding section, except that the charm quarks are replaced by bottom quarks. In our simulations, we treated this quark using an RHQ action (see Section~\ref{sec:RHQ_action}). Our primary goal was to investigate $BB^*$ scattering close to and above threshold using the finite-volume Lüscher method. In addition, we extracted the ground-state energy corresponding to the deeply bound $T_{bb}$ tetraquark, the bottom-quark analogue of the $T_{cc}$. This is an exploratory study in which several sources of systematic uncertainty are not addressed fully.

\subsection{$B$ and $B^*$ energies}
\label{sec:Results_B_Bstar_meson_energies}

The masses of the $B$ and $B^*$ mesons --- collectively denoted $B^{(*)}$ --- are essential inputs for the Lüscher analysis. Furthermore, studying their energies at nonzero momenta provides a valuable test of the RHQ action for $b$ quarks on our ensembles, especially for the dispersion relation. To that end, we used a single pseudoscalar and vector operator, as for the $D$ and $D^*$ mesons (see Eqs.~\eqref{eq:D_operator} and \eqref{eq:Dstar_operator}). As for the $D$ and the $D^*$, we performed fits to the effective energy $E^{B^{(*)}}_\m{eff}(t)$ obtained using the $\log$ or $\cosh$ definition depending on the temporal boundary conditions. For each dataset, we compared a constant fit with a fit of the form $E + A\m{e}^{-\Delta E\,t}$, with nuisance parameters $A$ and $\Delta E$ accounting for excited-state effects. Since the exponential model has two additional parameters, it yields larger uncertainties than a constant fit over the same range. For this model, we therefore chose fit ranges that start earlier, where the excited-state contribution is still resolvable. This both constrains $\Delta E$ and recovers a statistical precision comparable to that of the constant fit. In all cases, we only accepted fits with good quality-of-fit measures. When both fits gave results consistent within their uncertainties, we retained the simpler constant fit; otherwise we used the exponential form. Inconsistency between the two fitting strategies indicates that the effective energy has not plateaued fully before the signal deteriorates. This was most notable at nonzero momenta. In all cases, we checked the stability of the extracted energy under variations of the fit range.

\begin{figure}
  \includegraphics[width=0.5\textwidth]{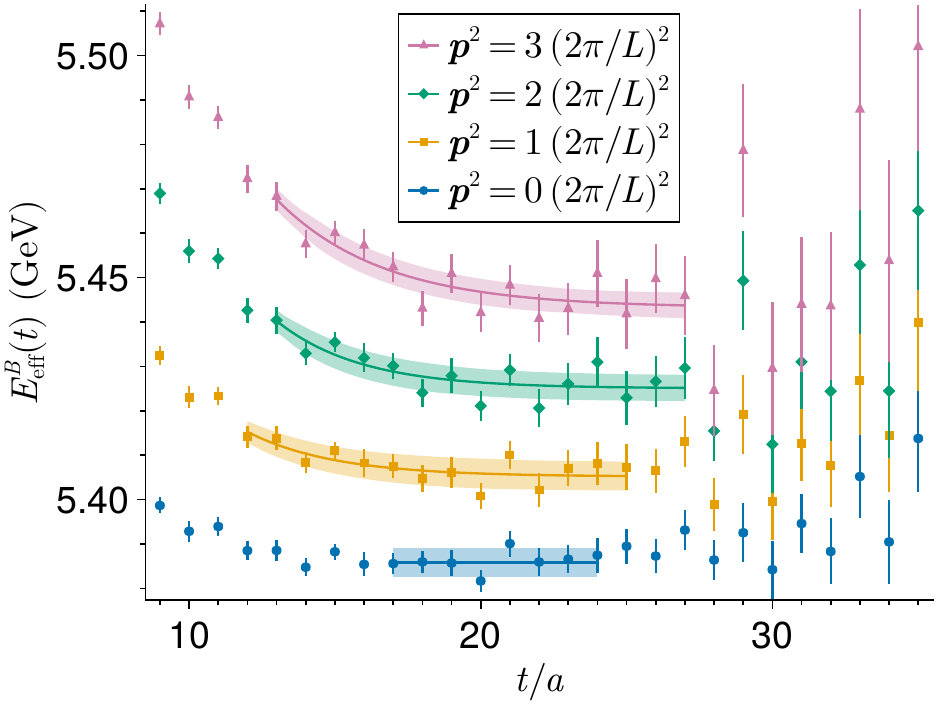}
  \includegraphics[width=0.5\textwidth]{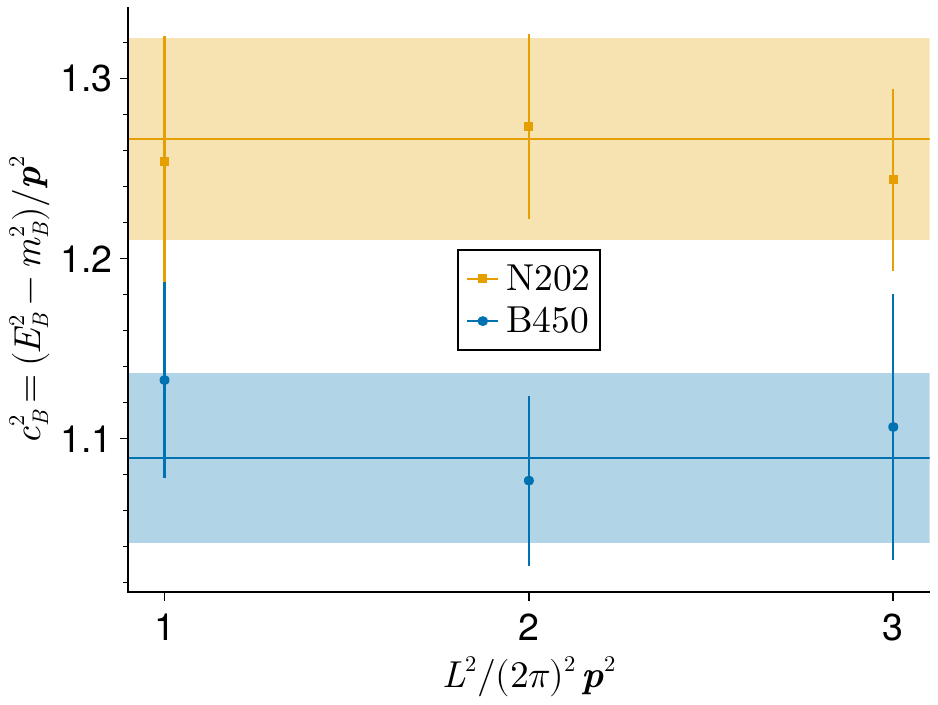}
  \caption{\label{fig:B_Bstar_E_eff_csq}%
    Left: effective energy $E^B_\m{eff}$ of the $B$ meson on N202 for the four lowest momenta. The coloured bands are the fit results. Right: speed-of-light parameter $c_B^2 = (E_B^2 - m_B^2)/\vb*{p}^2$ for the $B$ meson on N202 and B450. The coloured bands are the results of constant fits to these points.}
\end{figure}

The fits to the effective energies of the $B$ meson on N202 for the four lowest momenta are displayed in the left panel of Figure~\ref{fig:B_Bstar_E_eff_csq}.
We see that for higher momenta the effective energy has not plateaued fully before the signal is lost, and we therefore include an exponential term in the fit. The results for the B450 ensemble and for the $B^*$ meson are qualitatively similar and are not shown. The extracted $B$ and $B^*$ masses for the two ensembles are listed in Table~\ref{tab:Results_B_Bstar_meson}.
\begin{table}
  \centering
  \renewcommand{\arraystretch}{1.2}
  \begin{tabular}{lcccc}
    \hline \hline
    Ensemble & $m_B\;(\m{MeV})$ & $m_{B^*}\;(\m{MeV})$ & $c_B^2$ & $c_{B^*}^2$ \\
    \hline
    B450 & $5398(2)$ & $5435(2)$ & $1.09(5)$ & $1.07(8)$ \\
    N202 & $5385(3)$ & $5424(3)$ & $1.27(6)$ & $1.31(6)$ \\
    \hline \hline
  \end{tabular}
  \caption{\label{tab:Results_B_Bstar_meson}%
    $B$ and $B^*$ meson masses and speed-of-light parameters $c_B^2$ and $c_{B^*}^2$ extracted on the two ensembles.}
\end{table}
Our values lie above the PDG averages $m_{B^\pm} = 5279.34(12)\;\m{MeV}$ and $m_{B^*} = 5324.71(21)\;\m{MeV}$~\cite{ParticleDataGroup:2024cfk}. Several effects can contribute to this difference. The RHQ parameters were tuned using a scale setting that differs from ours by about $2\%$ (see Section~\ref{sec:Lattice_setup}), which alone shifts the masses by about $100\;\m{MeV}$. In addition, the light quarks at the $\m{SU}(3)$-flavour-symmetric point are heavier than the physical up and down quarks, and discretization effects have not been addressed. Finally, the quoted uncertainties do not include the uncertainties of the lattice spacing and of the RHQ parameters.

To investigate the quality of the lattice dispersion relation for the $B$ and $B^*$ mesons, we computed the speed-of-light parameter $c^2_{B^{(*)}}$ as a function of the momentum $\vb*{p}$ as we did for the $D$ and $D^*$ mesons (see Eq.~\eqref{eq:csq_fit}). The results for the $B$ meson on both ensembles are displayed in the right panel of Figure~\ref{fig:B_Bstar_E_eff_csq}. Those for the $B^*$ meson are qualitatively similar and are not shown. For the lowest values of $\vb*{p}^2$, the $c_{B^{(*)}}^2(\vb*{p})$ values are approximately constant, so we performed constant fits to these points to obtain estimates of $c_{B^{(*)}}^2$. The results are listed in Table~\ref{tab:Results_B_Bstar_meson}. All the resulting estimates are above one, in particular on N202, where $c_B^2$ and $c_{B^*}^2$ exceed unity by $\sim 5\;\sigma$. On B450 the deviation is between one and two $\sigma$. These deviations from unity can have different origins. On the one hand, as noted above, the effective energies at nonzero momenta tend not to have plateaued fully, which is why we used the exponential fit form to account for the residual excited-state effects. Should these fits fail to capture the contamination completely, the energies would be overestimated, resulting in a value of $c^2$ above unity. On the other hand, if the energy estimates are accurate, the deviation would indicate a mistuning of the RHQ action, whose parameters were tuned to reproduce $c^2 = 1$.

To test the accuracy of our energy estimates, we repeated the $B$-meson energy extraction using the variational method. We did this only on N202, where the deviation in $c^2$ is larger, and only for the two lowest momenta. As the basis, we used the four $A_1^-$ operators in the rest frame and the twelve $A_2$ operators in the $\tfrac{2\pi}{L}(0,0,1)$ frame constructed in Section~\ref{sec:single_particle_operators}, replacing the charm by bottom quarks. These are all the operators with up to one covariant derivative that exist in these irreps. In the GEVP, we used a fixed reference time $t_0$ (see Section~\ref{sec:variational_method}) and extracted the ground-state energies from plateau fits. We chose $t_0/a = 4$ and $t_0/a = 5$ for zero and nonzero momentum, respectively. These values yield good signal, and the energy estimates are stable under variations of $t_0$ in that region. While the resulting $B$-meson mass is consistent with the single-operator result, the energy at the first nonzero momentum changes considerably, giving $c^2 = 1.12(3)$. Although this value is still noticeably above unity, it represents a clear improvement and indicates that the energies at nonzero momenta are overestimated when no covariant derivatives are included in the operator basis.\footnote{A GEVP without covariant derivatives for the $A_2$ irrep in the $\tfrac{2\pi}{L}(0,0,1)$ frame can account for only part of the shift in the energy estimate.} A possible reason why the unoptimized single-meson operators do not allow the ground state at nonzero momentum to be extracted reliably is the large LapH smearing radius used in our simulations. It was optimized for lighter mesons; consequently, it might give a small overlap with the more compact $B$ and $B^*$ mesons.

These findings suggest that, with the current setup, bilocal $B^{(*)}B^*$ operators constructed from optimized $B^{(*)}$ operators should be included when extracting the interacting energy levels. We did not do this in the present work since we currently have no implementation for these operators. To reduce large, uncontrolled systematics from excited states in our finite-volume $B^{(*)}B^*$ spectrum, we rely on the following empirical observation: a significant cancellation of excited-state effects between single- and two-hadron systems occurs when the energy differences between interacting and noninteracting states, rather than the energies themselves, are extracted. This was used, for example, when investigating nucleon--nucleon systems~\cite{Horz:2020zvv,Green:2021qol,BaryonScattering:2025ziz}. The authors performed fits to ratios of correlators or combined fits to single- and two-hadron correlators. For the $BB^*$ system as well, we found a strong correlation between the single- and two-meson energies. This strongly suggests that a significant cancellation of excited-state effects also occurs in this system. Relying on this assumption, we applied only fitting strategies that extract the aforementioned energy differences. This cancellation of excited-state effects probably works best if the bilocal operators are constructed from the same single-meson operators used for extracting the single-meson energies. For this reason, in this analysis, we used the single-meson energies and effective energies obtained from unoptimized single-meson operators, i.e.\ without covariant derivatives.

Since we still measured a deviation from the continuum dispersion relation for the $B$ meson when using the GEVP, we applied a correction when extracting the $B^{(*)}B^*$ energies. The correction procedure is described in Section~\ref{sec:BBstar_FV_energies}.

\subsection{Operator basis for the $BB^*$ system}
\label{sec:BBstar_operators}

The $\overline{b}\overline{b}ud$ system with $I(J^P) = 0(1^+)$ shares all quantum numbers with the $DD^*$ system except for the flavour content. We can therefore reuse the operator construction of Section~\ref{sec:operator_importance} and, in particular, the operators subduced into the rest-frame $T_1^+$ irrep.

\begin{figure}
  \centering
  \includegraphics[width=0.833\textwidth]{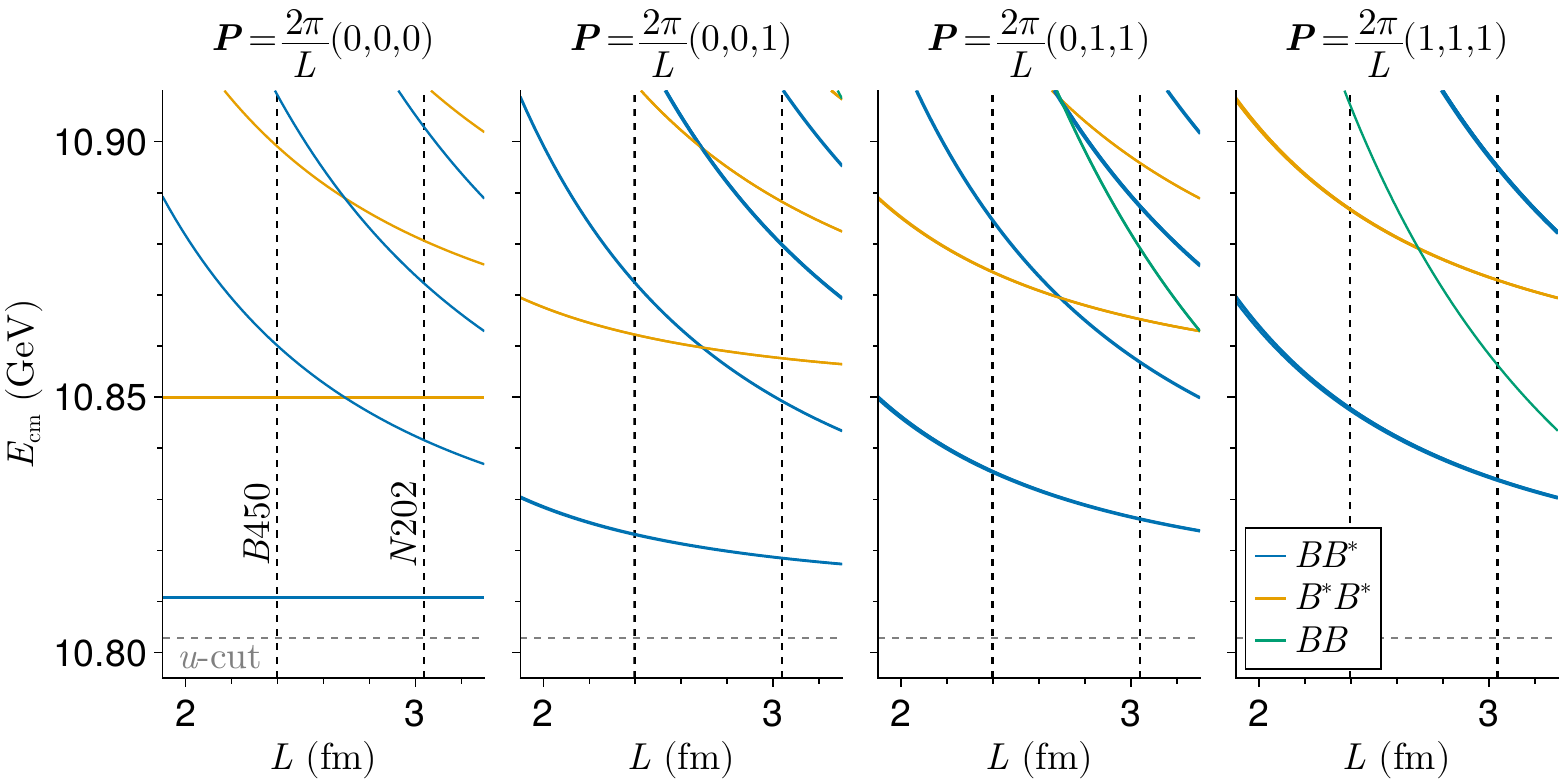}
  \caption{\label{fig:BBstar_noninteracting}%
    Noninteracting $B^{(*)}B^{(*)}$ c.m.\ energy levels as a function of the box size $L$ for the total momenta $\vb*{P}$ used in the analysis. The $u$-cut from one-pion exchange is shown as the horizontal grey dashed line. The box sizes of the two ensembles are indicated by vertical black dashed lines.}
\end{figure}

Because the $B$-meson mass is larger than that of the $D$ meson, the spacing between noninteracting $B^{(*)}B^*$ levels in a finite box is smaller than the corresponding $D^{(*)}D^*$ spacing. This leads to a denser finite-volume spectrum that is harder to resolve. Furthermore, the $B$--$B^*$ mass splitting is smaller than the $D$--$D^*$ splitting; consequently, the lowest noninteracting $B^*B^*$ level lies much closer to the $BB^*$ threshold. Figure~\ref{fig:BBstar_noninteracting} shows the lowest noninteracting $BB^*$, $B^*B^*$ and $BB$ energy levels as a function of the box size $L$ for total momenta with $(\tfrac{L}{2\pi})^2\vb*{P}^2 = 0, 1, 2, 3$.
The noninteracting $BB_0^*$ levels, with $m_{B_0^*} \approx 5650\;\m{MeV}$ as measured on B450, lie well above the $B^*B^*$ threshold and play no significant role in this low-energy analysis. A consequence of the small $B$--$B^*$ mass splitting is that, in a single-channel Lüscher analysis, one can use only the one or two lowest energy levels that are clearly below the $B^*B^*$ threshold to avoid coupled-channel contamination. This limitation motivates performing a coupled-channel $BB^*$--$B^*B^*$ analysis, which we leave for future work. 

When choosing the operator basis, one would ideally include bilocal $BB^*$ and $B^*B^*$ operators corresponding to all noninteracting levels up to the lowest three-particle threshold $BB\pi$, above which the three-particle Lüscher formalism would be required. These operators all share the same structure, making them simple to implement, and including all of them is expected to strongly suppress excited-state effects. However, the number of operators needed to cover this range is prohibitively large owing to the density of the $B^{(*)}B^*$ spectrum. Furthermore, the number of momentum combinations appearing on higher momentum shells grows quickly, making the computation of the corresponding two-point functions computationally expensive. Consequently, we used a smaller operator set.

In contrast to the $DD^*$ analysis, we also extracted finite-volume energy levels in moving frames to improve the coverage of $q_\m{cm}^2$ values entering the Lüscher fit. We used the total momenta $\vb*{P} = \tfrac{2\pi}{L}(0,0,0)$, $\tfrac{2\pi}{L}(0,0,1)$, $\tfrac{2\pi}{L}(0,1,1)$, $\tfrac{2\pi}{L}(1,1,1)$ and $\tfrac{2\pi}{L}(0,0,2)$, and averaged over all equivalent total momenta at the level of the correlator, as we did for the spin index. The $\tfrac{2\pi}{L}(0,0,2)$ frame yielded data too noisy to extract reliable excited $BB^*$ energy levels and is therefore used only to extract the $T_{bb}$ ground state.

The bilocal operators are constructed following the same procedure as for the $DD^*$ system: we include one operator for each (degenerate) noninteracting energy level up to a given relative momentum, both for $BB^*$ and $B^*B^*$ operator types. The irreps used for the different frames, together with the momentum combinations and the operator counts, are listed in Table~\ref{tab:BBstar_operators}.
\begin{table}
  \centering
  \renewcommand{\arraystretch}{1.2}
  \begin{tabular}{lccccc}
    \hline \hline
    $\vb*{D}$ & $(0,0,0)$ & $(0,0,1)$ & $(0,1,1)$ & $(1,1,1)$ & $(0,0,2)$ \\
    \hline
    Irrep & $T_1^+$ & $A_2$ & $A_2$ & $A_2$ & $A_2$ \\
    $(\vb*{d}_1^2,\,\vb*{d}_2^2)$ &
      \makecell{$(0,0)$, $(1,1)$ \\ $(2,2)$} &
      \makecell{$(1,0)$, $(2,1)$} &
      \makecell{$(2,0)$, $(1,1)$} &
      \makecell{$(3,0)$, $(2,1)$} &
      \makecell{$(1,1)$, $(2,2)$ \\ $(4,0)$} \\
    \# bilocal & $12$ & $9$ & $6$ & $9$ & $8$ \\
    \# total   & $15$ & $13$ & $10$ & $13$ & $12$ \\
    \hline \hline
  \end{tabular}
  \caption{\label{tab:BBstar_operators}%
    Irreps, momentum combinations and operator counts for each total integer momentum $\vb*{D} = \tfrac{L}{2\pi}\vb*{P}$ used in the $T_{bb}/BB^*$ analysis. The pairs $(\vb*{d}_1^2, \vb*{d}_2^2)$ denote the integer momenta squared used for the bilocal $BB^*$ and $B^*B^*$ operators. The last two rows give the number of bilocal operators and the total number of operators, respectively.}
\end{table}
We will later use labels such as $[0,0,1]\;A_2$ to denote these frames and irreps. We chose these irreps because they subduce from the $J^P = 1^+$ irrep and avoid mixing with $BB$ states with low $J$. In contrast to the $DD^*$ analysis, we used the same momentum combinations for the $BB^*$ and the $B^*B^*$ bilocal operators, since the noninteracting $B^*B^*$ and $BB^*$ levels are closer together.

In addition to the bilocal operators, we included local operators in all these irreps: a diquark-antidiquark operator and a local $BB^*$ and $B^*B^*$ operator, constructed analogously to those in the $DD^*$ analysis. In the moving frames, we also included the local $BB_0^*$ operator,
\begin{equation}
  \label{eq:BBstar_local_BB0star}
  T^{BB_0^*}(t) = 
    (\overline{b}\,\gamma_5\,d\;
     \overline{b}\,\m{i}\mathbf{1}\,u)(\vb*{P},t)
    - \{u \leftrightarrow d\},
\end{equation}
which does not exist in the $T_1^+$ irrep. As the lowest noninteracting $BB_0^*$ level lies well above the $B^*B^*$ threshold, we did not include a bilocal $BB_0^*$ operator.

We performed an operator-importance analysis analogous to the one used for the $DD^*$ system in Section~\ref{sec:operator_importance}. The energy estimates of the $BB^*$ system depend only weakly on the number of bilocal operators once local operators are included, consistent with the findings for the $DD^*$ system. Using either the local $BB^*$ and $B^*B^*$ operators or only the diquark-antidiquark operator yields energy estimates consistent with those from the full set of local tetraquark operators. Omitting all local operators results in a ground-state effective energy that is significantly higher, making it impossible to reliably extract the $T_{bb}$ energy. This is in agreement with the findings of Ref.~\cite{Prelovsek:2025vbr}. We performed this operator-importance analysis in the rest frame, and we expect the conclusion to hold in the moving frames as well. We therefore assume that the energy estimates are stable under a variation of the number of bilocal meson-meson operators in the basis.

We note that the present bilocal operator basis does not include operators with covariant derivatives, even though the dispersion-relation study in the previous section suggests that such operators are important. The absence of these operators constitutes an unaddressed systematic uncertainty in the finite-volume spectrum and, consequently, in the extracted scattering phase shifts. The empirically observed excited-state cancellation in energy differences discussed in the previous section and the dispersion-relation correction described in Section~\ref{sec:BBstar_FV_energies} are expected to reduce this systematic uncertainty significantly.

\subsection{$T_{bb}$ binding energy}
\label{sec:Tbb_binding_energy}

Before presenting the scattering analysis near and above threshold, we show the extracted $T_{bb}$ energies. There is broad consensus in the literature that the $T_{bb}$ is a deeply bound state with a binding energy in the range $50$--$190\;\m{MeV}$~\cite{Leskovec:2019ioa,Alexandrou:2024iwi,Green:2026jgv,Hoffmann:2026mcz}. Because this energy lies well below the start of the left-hand cut from one-pion exchange, the two-particle Lüscher formalism cannot be applied. However, the deep binding implies that finite-volume effects are exponentially suppressed as $\m{e}^{-m_\pi L}$. Consequently, the $T_{bb}$ manifests as the ground state in the irreps subduced from $J^P = 1^+$, e.g.\ those listed in Table~\ref{tab:BBstar_operators}, and its energy can be extracted directly from the finite-volume spectrum.

We applied the GEVP method explained in Section~\ref{sec:variational_method} on both ensembles and for all irreps in Table~\ref{tab:BBstar_operators}. A significant practical challenge is that no clear plateau emerges in the ground-state effective energy before the signal is lost to statistical noise. Two factors are likely to contribute to this difficulty. First, the $T_{bb}$ has been found to have a substantially smaller charge radius than the combined charge radii of the $B$ and $B^*$ mesons~\cite{Vujmilovic:2025czt}. However, the large LapH smearing radius used in our simulations is optimized for lighter mesons, resulting in a small overlap with the compact $T_{bb}$ ground state and delaying the onset of the plateau. Second, an important operator might be missing from the basis --- for instance, a tetraquark operator with covariant derivatives --- although we consider this less likely. The resulting uncertainty in the onset of the plateau regions introduces a considerable systematic error in the energy estimates.

Because no clear plateau is reached, our ground-state energy estimates are based on constant plateau fits to the effective energies at the latest Euclidean times where the signal is still usable. Fits of the form $E + A\m{e}^{-\Delta E\,t}$, which would in principle account for the residual excited-state contamination, proved unstable under variations of the fit range and were not used for the final results. Since the effective energies are empirically found to approach their plateau from above, any remaining contamination is expected to bias the extracted energies upward. Our estimates are therefore likely to overestimate the true ground-state energies.

On N202, we also explored the generalized pencil-of-functions (GPOF) method~\cite{Aubin:2010jc}. This method generates a new linearly independent operator $\mathcal{O}^\tau(t)$ by shifting an existing operator $\mathcal{O}(t)$ in time by $\tau$:
\begin{equation}
  \mathcal{O}^\tau(t) = \m{e}^{H\tau} \mathcal{O}(t) \m{e}^{-H\tau} = \mathcal{O}(t+\tau).
\end{equation}
Using $\mathcal{O}^\tau(t)^\dagger = \mathcal{O}(t-\tau)^\dagger$ and exploiting time-translation invariance of two-point functions, this operator can be added to the correlator matrix without computing any additional two-point functions. By enlarging the operator basis, the GPOF method can partially mitigate the signal-to-noise problem, although it also introduces additional noise into the correlator matrix. On N202, we improved the quality of the plateau by applying the GPOF to the local $BB^*$ and $B^*B^*$ operators with a time shift $\tau = 4a$. We performed plateau fits to the ground-state GPOF effective energies and adopted the results as our primary energy estimates on N202. The difference relative to the plateau fit applied to the non-GPOF effective energy is small compared to the statistical uncertainty and was taken as an estimate of the systematic uncertainty from the choice of fitting strategy. Applying the GPOF to the diquark-antidiquark operator, either in addition to or instead of the local meson-meson operators, introduced too much noise into the effective energy, rendering the result unusable. On B450, where the GPOF did not yield a reliable result, we relied solely on the plain plateau fit and did not assign a systematic uncertainty from the choice of fitting strategy.

\begin{figure}[t!]
  \centering
  \includegraphics[width=\textwidth]{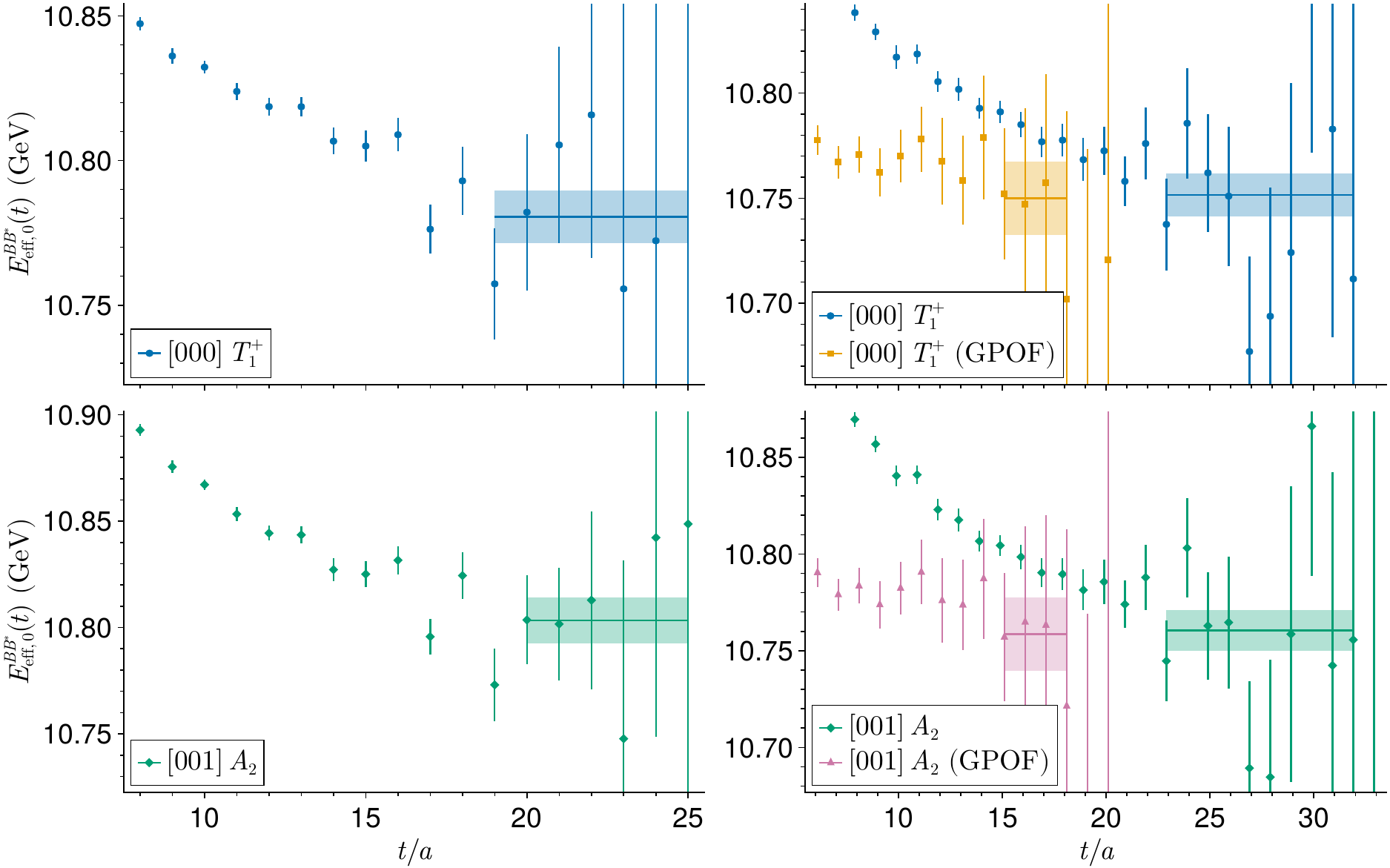}
  \caption{\label{fig:Tbb_effE}%
    Effective energy of the $BB^*$ ground state corresponding to the $T_{bb}$, shown for the $\vb*{P} = \vb*{0}$, $T_1^+$ (top row) and $\vb*{P} = \tfrac{2\pi}{L}(0,0,1)$, $A_2$ (bottom row) irrep, and for both the B450 (left column) and N202 (right column) ensembles. For N202, results with and without applying the GPOF method are displayed. The bands show the plateau fit results. In the right column, points are shifted horizontally for clarity.}
  \vspace{\floatsep}
  \includegraphics[width=0.5\textwidth]{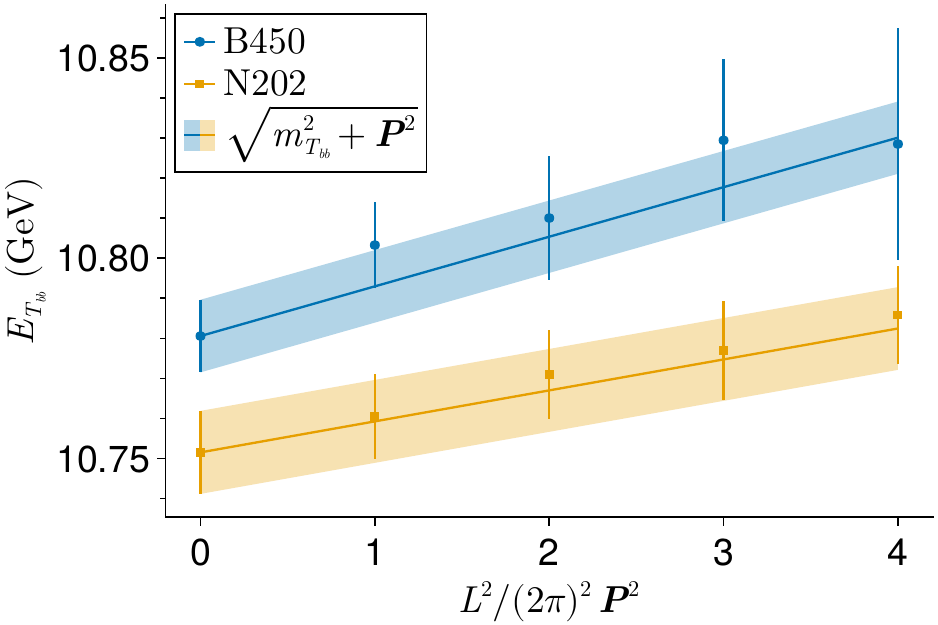}
  \caption{\label{fig:Tbb_dispersion}%
    Dispersion relation for the $T_{bb}$ ground state on B450 and N202. The energy is shown as a function of the integer total momentum squared $(\tfrac{L}{2\pi})^2\vb*{P}^2$. The curves show the continuum relation $E^2 = m_{T_{bb}}^2 + \vb*{P}^2$ evaluated at the respective lattice masses.}
\end{figure}

The ground-state effective energies and fit results for the two lowest total momenta are shown in Figure~\ref{fig:Tbb_effE}.
For N202, we display results with and without the GPOF for comparison. Both methods yield virtually the same central value, although the GPOF results carry larger uncertainties. From the $\vb*{P}=\vb*{0}$ fit result, corresponding to the $T_{bb}$ mass $m_{T_{bb}}$, we extract the $T_{bb}$ binding energy
\begin{equation}
  \label{eq:Tbb_binding_energies}
  m_B + m_{B^*} - m_{T_{bb}} = \begin{cases}
    53(8)\;\m{MeV} & \text{(B450)},\\
    61(17)\;\m{MeV} & \text{(N202)},
  \end{cases}
\end{equation}
where the uncertainty on N202 includes the systematic error from the choice of fitting strategy. Owing to the systematics discussed above, these values are likely to underestimate the true binding energy.

As a consistency check, we verified that the extracted $T_{bb}$ energies satisfy the continuum dispersion relation. This is shown in Figure~\ref{fig:Tbb_dispersion}, where we display the ground-state energies as a function of the total momentum squared.
The energies are in good agreement with the continuum relation $E^2 = m_{T_{bb}}^2 + \vb*{P}^2$, supporting the reliability of the ground-state identification at the present precision. To investigate finite-volume effects, one could compute the binding energy on a second ensemble with the same lattice spacing but a different physical volume. If those effects turn out to be small, we could attempt a continuum extrapolation, although simulations at more lattice spacings would be advisable. We leave the investigation of these systematics for future work.

Our binding energies are consistent with the values $50.9(1.1)\;\m{MeV}$ and $64(10)\;\m{MeV}$ found in~\cite{Hoffmann:2026mcz} and~\cite{Vujmilovic:2025czt}, respectively. Both employed an RHQ action for the $b$ quark. The former result was obtained at a similar pion mass of $m_\pi \approx 431\;\m{MeV}$ and the latter at a lighter pion mass of $m_\pi \approx 290\;\m{MeV}$. The broader literature, using NRQCD and static $b$-quark actions, reports values spanning $50$--$190\;\m{MeV}$~\cite{Leskovec:2019ioa,Alexandrou:2024iwi,Green:2026jgv,Hoffmann:2026mcz}, and our results fall in the lower part of this range.

\subsection{Finite-volume $BB^*$ energies}\label{sec:BBstar_FV_energies}

We now turn to the extraction of the excited finite-volume energy levels, which correspond to $BB^*$ and $B^*B^*$ scattering states near and above threshold. In contrast to the $T_{bb}$ ground states, the excited-state effective energies plateau significantly earlier, enabling a more reliable determination of their energies.
As for the $DD^*$ system, the interacting and noninteracting energy levels are strongly correlated, and exploiting this correlation leads to a substantial reduction in statistical uncertainties and presumably to a significant cancellation of excited-state effects (see Section~\ref{sec:Results_B_Bstar_meson_energies}). To that end, we employed fitting strategies that extract energy differences to noninteracting energy levels rather than the energies themselves. More precisely, we used two complementary fitting strategies that both extract the energy difference $\Delta E^{B^{(*)}B^*}_n$ between the $n$-th interacting level and the nearest noninteracting $B^{(*)}B^*$ level.

In the first strategy, we performed plateau fits to the effective energy differences
\begin{equation}
  \label{eq:Delta_Eeff}
  \Delta E^{B^{(*)}B^*}_{\m{eff},n}(t) = E^{B^{(*)}B^*}_{\m{eff},n}(t)
    - E^{B^{(*)}(\vb*{p}_1)}_\m{eff}(t)
    - E^{B^*(\vb*{p}_2)}_\m{eff}(t).
\end{equation}
Here, $E^{B^{(*)}B^*}_{\m{eff},n}$ is the effective energy of the $n$-th interacting $B^{(*)}B^*$ level and $E^{B^{(*)}(\vb*{p}_i)}_\m{eff}$ are the single-meson effective energies with momentum $\vb*{p}_i$. The momentum combination $(\vb*{p}_1,\vb*{p}_2)$ and particle pair ($BB^*$ or $B^*B^*$) are chosen such that the noninteracting level corresponds the closest to the interacting $B^{(*)}B^*$ level. The identification of this level is detailed below. This fitting method is similar to what we used for the $DD^*$ system, except that we now also use single-meson effective energies at nonzero momenta, which is relevant for the dispersion-relation correction detailed below. Again, this approach is similar to ratio fits and can produce spurious plateaus~\cite{Iritani:2016jie}. To minimize this risk, we attempted to restrict the fits to time ranges where all three individual effective energies have reached a plateau. We could not satisfy this criterion in all cases, because the single-meson effective energies tend to plateau later than the two-meson ones. We therefore used the second strategy as a cross-check.

In the second strategy, we performed combined fits to the three effective energies. For the single-meson effective energies, we used the same fit functions and ranges as for the meson-energy extraction. For the two-meson effective energy, the fit function is
\begin{equation}
  \label{eq:BBstar_combined_fit}
  \Delta E^{B^{(*)}B^*}_n + E^{B^{(*)}} + E^{B^*} + A\m{e}^{-\Delta E\,t},
\end{equation}
where $E^{B^{(*)}}$ and $E^{B^*}$ are the meson-energy parameters shared with the single-meson fits, and $A$ and $\Delta E$ are nuisance parameters. When the effective-energy difference clearly showed a plateau, we dropped the exponential term. The advantage of this strategy over the first is that different fit ranges can be chosen for the single- and two-meson effective energies, allowing for more flexibility when the plateau ranges do not overlap well. However, we expect a weaker cancellation of excited-state effects in these combined fits when the fit ranges differ.

Applying these two fitting strategies requires associating a noninteracting $B^{(*)}(\vb*{p}_1)B^*(\vb*{p}_2)$ level with each extracted interacting energy level: this means identifying whether the level is predominantly a $BB^*$ or $B^*B^*$ state and which momentum combination $(\vb*{p}_1,\vb*{p}_2)$ it corresponds to. For this assignment, we first performed plateau fits (in certain cases with an exponential) to the interacting effective energies to obtain preliminary energy estimates. We then computed the overlap factors of the levels with the operators from the GEVP data as described in~\cite{Bulava:2016mks}. This requires a normalization of the operators, which we performed by setting the diagonal entries of the correlator matrix at $t = 0$ to unity. We then associated each energy level with the noninteracting $B^{(*)}B^*$ level corresponding to the bilocal operator with the largest overlap factor. In most cases this assignment was unambiguous; otherwise we matched the level to the nearest noninteracting energy.

Because the $B^{(*)}$ meson energies show a significant deviation from the continuum dispersion relation, we applied a correction when computing the interacting energies $E^{B^{(*)}B^*}_n$. The correction used for the $DD^*$ analysis, which modifies the definition of $q_\m{cm}$, works only in the rest frame and cannot be used here. Rather than adding the noninteracting lattice $B^{(*)}$ energies to $\Delta E^{B^{(*)}B^*}_n$, we add the noninteracting energy from the continuum dispersion relation,
\begin{equation}
  \label{eq:BBstar_disp_correction}
  E^{B^{(*)}B^*}_n = \Delta E^{B^{(*)}B^*}_n
    + \sqrt{m_{B^{(*)}}^2 + \vb*{p}_1^2}
    + \sqrt{m_{B^*}^2  + \vb*{p}_2^2}.
\end{equation}
Here, $m_{B^{(*)}}$ and $m_{B^*}$ are the lattice meson masses and $\vb*{p}_i$ are the corresponding momenta. This method has already been used successfully in the literature~\cite{Piemonte:2019cbi,Prelovsek:2020eiw,Shrimal:2025wbu}.

\begin{figure}
  \includegraphics[width=\textwidth]{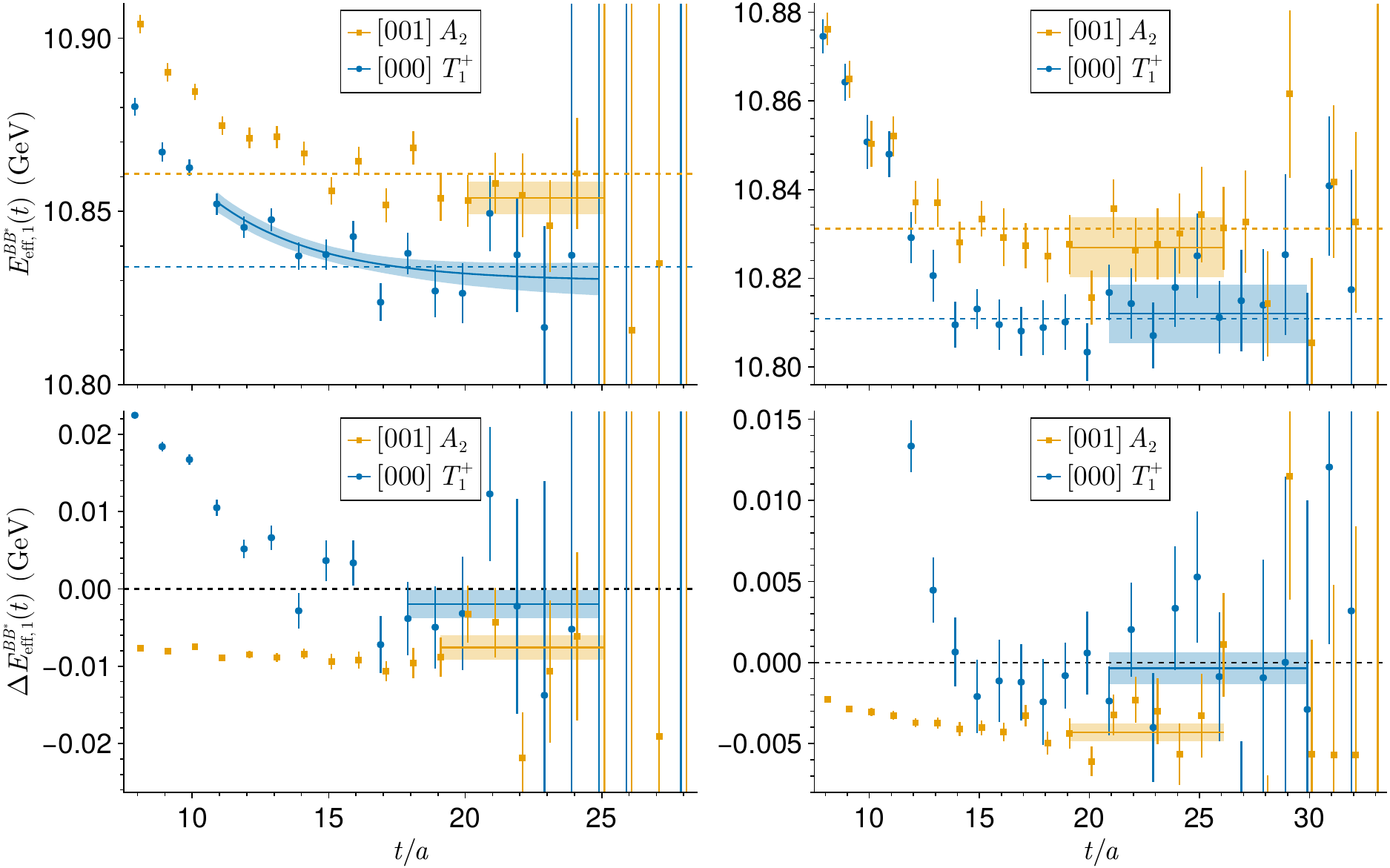}
  \caption{\label{fig:BBstar_E_eff}%
    Effective energies (top row) and effective energy differences (bottom row) of the first excited $BB^*$ states for the B450 (left column) and N202 (right column) ensembles. The bands display results from combined (top) and plateau fits (bottom). Data are shown for the two irreps $\vb*{P} = \vb*{0}$, $T_1^+$ and $\vb*{P} = \tfrac{2\pi}{L}(0,0,1)$, $A_2$. The horizontal dashed lines in the top row display the corresponding noninteracting energy levels. Points are shifted horizontally for clarity.}
\end{figure}

The effective energies and effective energy differences for the first excited state in the irreps $\vb*{P} = \vb*{0}$, $T_1^+$ and $\vb*{P} = \tfrac{2\pi}{L}(0,0,1)$, $A_2$ are shown in Figure~\ref{fig:BBstar_E_eff}.
We observe clearer plateaus than in the ground-state effective energies. The results for the other irreps and levels are qualitatively similar and are not shown. In cases where we included an exponential term in the combined fit, the result was insensitive to changes in the fit range. This is in contrast to the ground-state effective energies, where the $E + A\m{e}^{-\Delta E\,t}$ fits showed a strong fit-range dependence.

\begin{figure}[t!]
  \centering
  \includegraphics[width=0.65\textwidth]{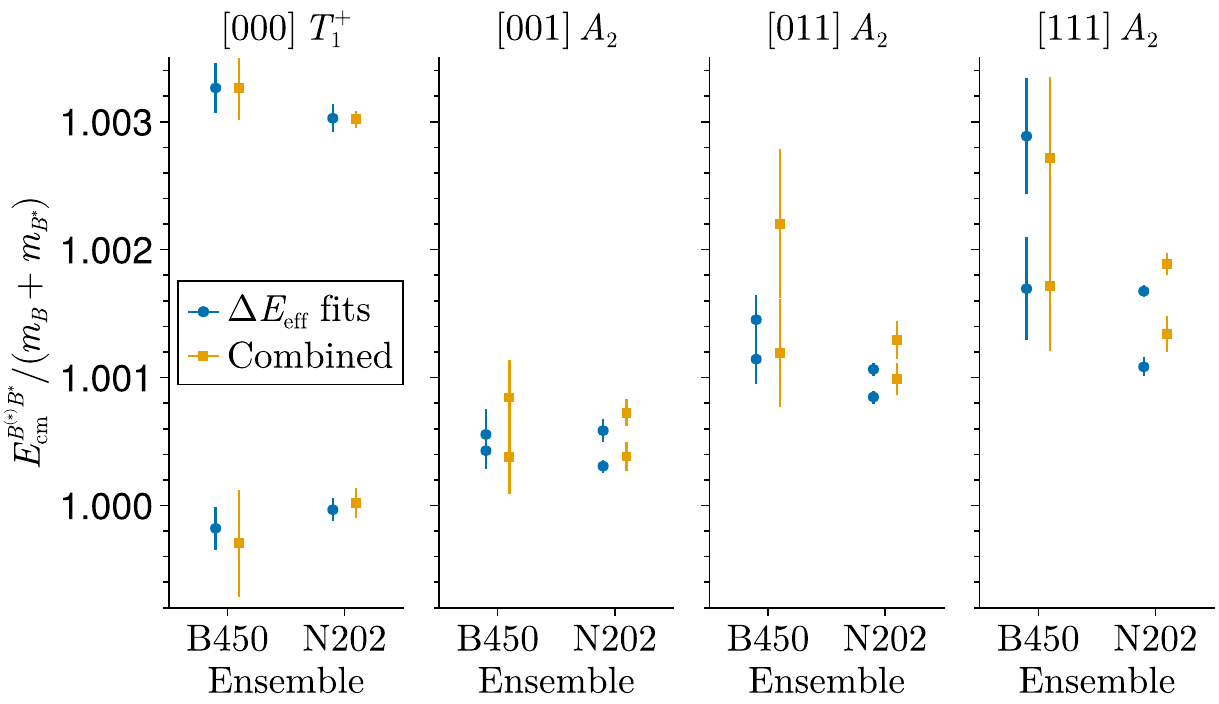}
  \caption{\label{fig:BBstar_energy_levels}%
    Excited finite-volume $B^{(*)}B^*$ energies for both fitting strategies (fits to effective energy differences $\Delta E_\m{eff}$ and combined fits) and for both ensembles. All energies are converted to the c.m.\ frame and are normalized by the threshold energy $m_B + m_{B^*}$.}
\end{figure}

\begin{figure}[t!]
  \centering
  \includegraphics[width=0.7\textwidth]{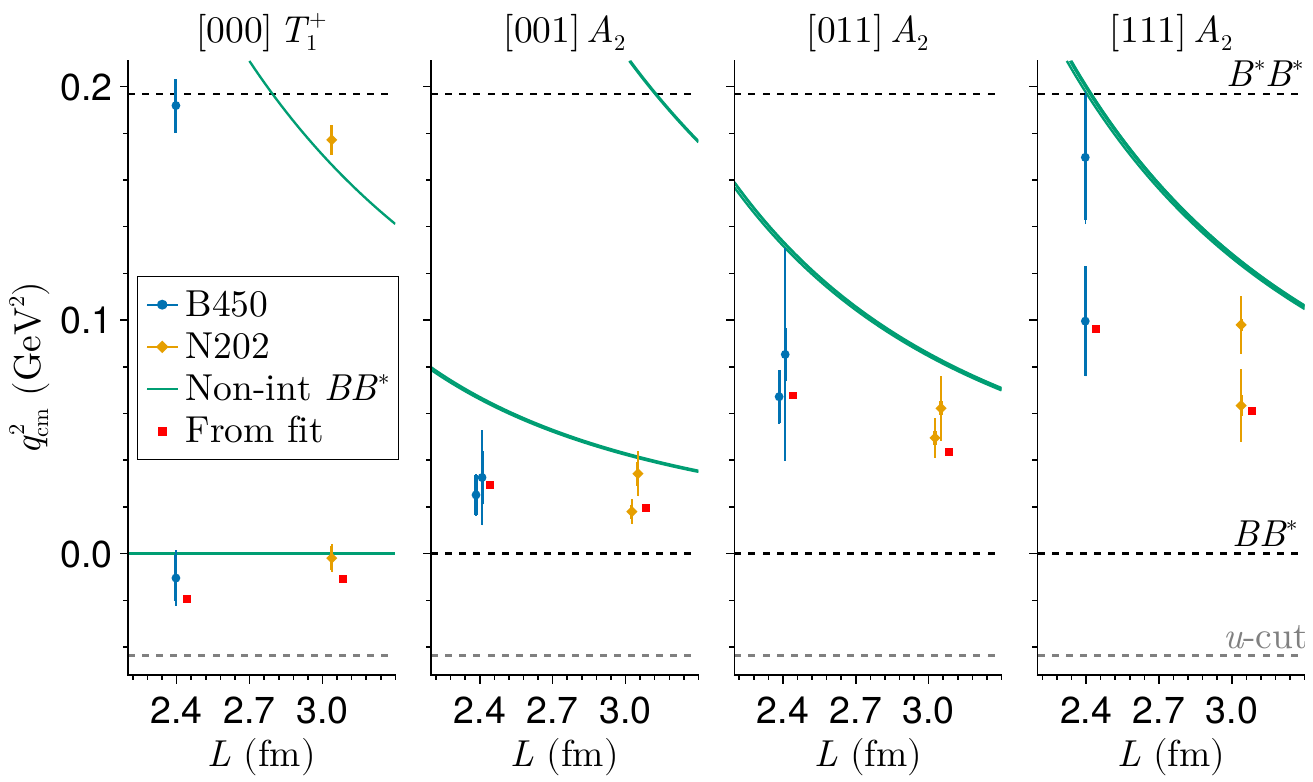}
  \caption{\label{fig:Lüscher_qsq_vs_L}%
    C.m.\ scattering momentum squared $q_\m{cm}^2$ of the excited $B^{(*)}B^*$ states as a function of the box size $L$ for B450 and N202 and for the irreps used in the analysis. Thick and thin error bars denote the statistical and total uncertainties, respectively. For the levels entering the Lüscher fit, the energies predicted from the fit are shown as red squares. The green lines indicate the noninteracting $BB^*$ levels and the horizontal black dashed lines the $BB^*$ and $B^*B^*$ thresholds. The grey dashed line at the bottom is the $u$-cut from one-pion exchange. Points are shifted horizontally for clarity.}
  \vspace{\floatsep}
  \includegraphics[width=0.5\textwidth]{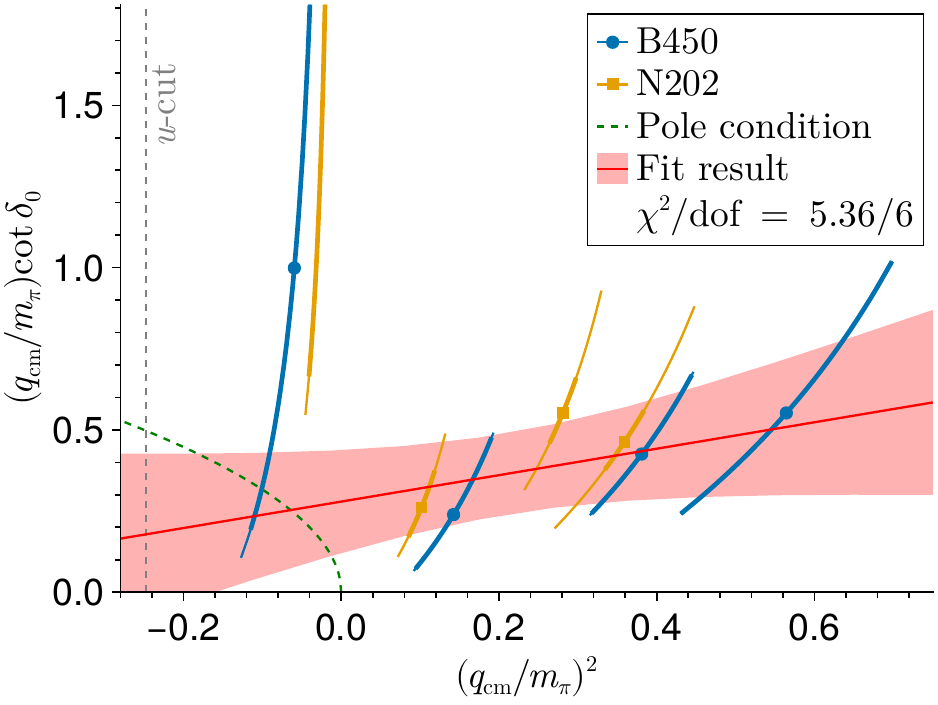}
  \caption{\label{fig:Lüscher_qcotdelta0_vs_q2}%
    $s$-wave $BB^*$ scattering phase shift in terms of $q_\m{cm}\cot\delta_0(q_\m{cm})$ vs.\ $q_\m{cm}^2$ for the two ensembles (normalized by the pion mass). Thick and thin error bars denote the statistical and total uncertainties, respectively. The band shows the linear ERE fit to both ensembles. The green dashed line shows the pole condition $q_\m{cm}\cot\delta_0(q_\m{cm}) = \pm\sqrt{-q_\m{cm}^2}$, and the vertical grey dashed line marks the start of the $u$-channel cut from one-pion exchange.}
\end{figure}

The resulting finite-volume energies from both fitting strategies and ensembles are displayed in Figure~\ref{fig:BBstar_energy_levels}.
These energies are corrected for the dispersion relation, converted to the c.m.\ frame, and normalized by the threshold energy $m_B + m_{B^*}$. In most cases, the fits to effective energy differences yield energies with smaller statistical uncertainties than the combined fits. Generally, we find agreement between the two fitting strategies within uncertainties. Exceptions are the energy levels for the two highest total momenta on N202, where the combined fits produce significantly higher energies. This may be due to the stronger excited-state cancellation in the effective energy difference. In the Lüscher analysis, we took the $\Delta E_\m{eff}$ fit results as our central values and estimated the systematic uncertainty from the spread between the two methods. The resulting finite-volume spectrum in terms of $q_\m{cm}^2$ as a function of the box size $L$ is displayed in Figure~\ref{fig:Lüscher_qsq_vs_L}.
We see that for certain energy levels the discrepancy between the two fitting strategies results in an error that is systematics dominated. This was especially evident for the first excited state in the moving frames on N202, and also for some second excited states.

\subsection{$s$-wave Lüscher analysis}
\label{sec:BBstar_swave}

Finally, we performed a single-channel $s$-wave Lüscher analysis of the finite-volume $BB^*$ spectrum to extract the $BB^*$ $s$-wave scattering phase shift. We assumed $p$-wave effects to be negligible at the present precision and included only levels well below the $B^*B^*$ threshold. As a result, we included only the first excited state in each irrep in the fit. The second excited state in the moving frames could in principle also be used, but it likely has significant non-$s$-wave contributions, as is the case for the $DD^*$ spectrum. As discussed in Section~\ref{sec:Tbb_binding_energy}, the ground state lies well below the onset of the $u$-cut from one-pion exchange (at $q_\m{cm}^2 \approx -m_\pi^2/4$) and therefore cannot be used in the two-particle Lüscher analysis.

In the single-channel $s$-wave case, the Lüscher quantization condition is given by Eq.~\eqref{eq:swave_QC} and provides the one-to-one correspondence between the scattering momentum $q_\m{cm}$ and the $s$-wave phase shift $\delta_0$ already used for the $DD^*$ system. In Figure~\ref{fig:Lüscher_qcotdelta0_vs_q2} we display the resulting $q_\m{cm}\cot\delta_0$ values as a function of $q_\m{cm}^2$.
The data from B450 and N202 are in good mutual agreement, indicating that discretization effects are small at the present precision. The data points from the rest-frame $T_1^+$ irrep lie close to threshold and consequently carry a comparatively large uncertainty, as the quantization condition diverges at threshold. Interestingly, the $q_\m{cm}\cot\delta_0$ values closely resemble those of the $DD^*$ case (see Figure~\ref{fig:s-wave_analysis}).

To determine the phase shift, we parametrized $q_\m{cm}\cot\delta_0$ using the ERE
\begin{equation}
  \label{eq:ERE_BBstar}
  q_\m{cm}\cot\delta_0(q_\m{cm})
    = -\frac{1}{a_0} + \frac{r_0}{2}\,q_\m{cm}^2 + O(q_\m{cm}^4),
\end{equation}
truncated at linear order in $q_\m{cm}^2$, so that the free parameters of our model are the scattering length $a_0$ and the effective range $r_0$. For a given choice of $(a_0, r_0)$, the ERE fixes the phase shift. This in turn predicts a finite-volume spectrum through the quantization condition, i.e.\ a set of c.m.\ scattering momenta $q_\m{cm}^2$ for the irreps and box sizes used. We determined the parameters by minimizing the correlated $\chi^2$
\begin{equation}
  \label{eq:chisq_BBstar}
  \chi^2(a_0, r_0) = \sum_{i,j}
    \left[q_{\m{cm},i}^2 - \hat{q}_{\m{cm},i}^2(a_0, r_0)\right]
    \left(C^{-1}\right)_{ij}
    \left[q_{\m{cm},j}^2 - \hat{q}_{\m{cm},j}^2(a_0, r_0)\right],
\end{equation}
where the indices $i, j$ run over all energy levels included in the fit, $q_{\m{cm},i}^2$ are the measured scattering momenta, and $\hat{q}_{\m{cm},i}^2(a_0, r_0)$ are the values predicted by the quantization condition. The covariance matrix $C = C^\m{stat} + C^\m{sys}$ contains both a statistical and a systematic contribution. We estimated the statistical part $C^\m{stat}$ using the $\Gamma$-method without the modification detailed in Section~\ref{sec:data_analysis}. The systematic part $C^\m{sys}$ encodes the systematic uncertainty from the choice of energy-extraction strategy. For the latter, we assumed $50\%$ correlation between the energy levels.

We performed fits to the individual ensembles as well as a combined fit to both ensembles without including an explicit lattice-spacing dependence. All fits gave consistent results, so we report the combined fit. The fit is shown as the red band in Figure~\ref{fig:Lüscher_qcotdelta0_vs_q2} and yields the ERE parameters
\begin{equation}
  a_0 = -1.7(8)_\m{stat}(5)_\m{sys}\;\m{fm}, \qquad
  r_0 = 0.4(3)_\m{stat}(3)_\m{sys}\;\m{fm}.
\end{equation}
We obtained the uncertainties on $a_0$ and $r_0$ by linear error propagation; propagating the statistical and systematic parts separately yields the contributions quoted above. The fit gives $\chi^2/\m{dof} = 5.36/6 \approx 0.89$. The good fit quality indicates the mutual consistency of the levels entering the fit. This is a non-trivial check here, since these levels originate from four different irreps on two ensembles and therefore carry partly independent systematic uncertainties, so that a distortion affecting individual frames would appear as scatter around the fitted curve. It is further supported by the agreement between the fits to the individual ensembles and the combined fit. Taken together, these checks indicate that the excited-state cancellation in the energy differences (see Section~\ref{sec:Results_B_Bstar_meson_energies}) and the dispersion-relation correction (see Section~\ref{sec:BBstar_FV_energies}) yield an accurate finite-volume spectrum at the present precision. The $DD^*$ system provides an example where such a check does reveal a problem: omitting the local tetraquark operators produces a large discrepancy in the phase shift between the two ensembles, which is resolved once these operators are included.

To locate a bound or virtual state, we searched for poles of the scattering amplitude, i.e.\ solutions of $q_\m{cm}\cot\delta_0(q_\m{cm}) = \pm\sqrt{-q_\m{cm}^2}$, where the positive and negative signs correspond to a virtual bound state on the unphysical sheet and a bound state on the physical sheet, respectively. As visible in Figure~\ref{fig:Lüscher_qcotdelta0_vs_q2}, we find a crossing of $q_\m{cm}\cot\delta_0(q_\m{cm})$ with positive $\sqrt{-q_\m{cm}^2}$. It corresponds to a virtual bound state lying $2(2)_\m{stat}(1)_\m{sys}\;\m{MeV}$ below threshold.

The systematic uncertainties quoted for $a_0$, $r_0$, and the position of the virtual bound state account only for the choice of the energy-extraction strategy, as described in Section~\ref{sec:BBstar_FV_energies}. In particular, as for the $DD^*$ system, the uncertainties of the lattice spacing and of the heavy-quark tuning are not included. In addition to the uncertainty of the lattice spacing of about $1\%$, the RHQ parameters were tuned using a scale setting that differs from ours by about $2\%$~\cite{Bruno:2016plf,Hudspith:2026vxt}. This affects the conversion of our results into physical units at the level of a few percent and corresponds to a mistuning of the bottom-quark mass of similar size. Based on the heavy-quark-mass dependence observed for the $DD^*$ system~\cite{Collins:2024sfi}, we estimate that this changes the $T_{bb}$ binding energies by a few percent. For the near-threshold quantities, Ref.~\cite{Collins:2024sfi} indicates an enhanced sensitivity to the heavy-quark mass, so that $a_0$ may change by up to about $10\%$. All these effects are, however, well below the statistical uncertainties of this exploratory study.

Another potential source of systematic uncertainty is the nearby left-hand cut from one-pion exchange, which, as in the $DD^*$ system, might distort the shape of the phase shift. In the $BB^*$ system, however, the virtual-bound-state pole lies farther from this cut, so that its influence on the pole position is likely smaller. Although the state lies very close to threshold and several systematic effects have not been addressed fully, the crossing itself strongly indicates the presence of such a state in the $BB^*$ system alongside the deeply bound $T_{bb}$.

To our knowledge, there is only one previous lattice QCD analysis that found an indication for such an additional state near threshold~\cite{Nagatsuka:2025szy}. Those authors found a shallow bound state with a binding energy of $O(100)\;\m{keV}$. However, they used only nonlocal two-meson operators without employing the variational method when extracting the finite-volume energy levels. As a consequence, they were unable to resolve the deeply bound $T_{bb}$ state, making the conclusions about other states in the system unreliable.

Investigations of the $\overline{b}\overline{b}ud$, $I(J^P)=0(1^+)$ system using a constituent quark model also reveal two bound states: the deeply bound $T_{bb}$, and a near-threshold bound state~\cite{Meng:2020knc,Meng:2023for,Wu:2024zbx,Wu:2025nbu}. The authors conclude that the former possesses a compact structure, while the latter corresponds to a loosely bound ``meson molecule''. This is consistent with our findings, where local tetraquark operators are necessary to resolve the deeply bound $T_{bb}$, indicating a compact structure, whereas the finite-volume near-threshold states have a strong overlap with bilocal two-meson operators, indicating a ``molecular'' structure for the shallow virtual bound state. This agreement should, however, be interpreted with caution, as the phenomenological results rely on a constituent quark model and are therefore subject to model dependence. Confirming these findings with an independent phenomenological approach, such as one based on the Born--Oppenheimer approximation~\cite{Bicudo:2016ooe}, would be valuable.

A pair of states with the same quantum numbers has also been found in the charm sector. Ref.~\cite{Whyte:2024ihh} performed a coupled-channel analysis of $DD^*$--$D^*D^*$ scattering at a pion mass of $m_\pi \approx 391\;\m{MeV}$, comparable to the one used in the present study. In addition to a state compatible with the $T_{cc}$, the authors found a second state, denoted by $T_{cc}'$, which appears as a resonance below the $D^*D^*$ threshold. As discussed in Section~\ref{sec:ERE}, the position of a pole in the scattering amplitude evolves continuously as the masses of the scattering particles are varied. If the heavy-quark mass were increased continuously from the charm to the bottom sector, the pair of states $(T_{cc}, T_{cc}')$ could therefore evolve into the pair identified in this work: the deeply bound $T_{bb}$ and the shallow virtual bound state in the $BB^*$ system.

\chapter{Conclusion}
\label{chapter:Conclusion}

In this thesis we studied $DD^*$ and $BB^*$ scattering for $I(J^P)=0(1^+)$ from lattice QCD. The aim was to investigate the states that appear in these systems, in particular the doubly heavy tetraquarks $T_{cc}$ and $T_{bb}$, as well as additional near-threshold states. The work combined three contributions: a methodological development that makes local multiquark operators affordable within distillation, a study of the importance of these operators for the $DD^*$ system, and the physics results for the $BB^*$ system obtained with them. All simulations were performed on two CLS gauge ensembles --- B450 and N202 --- with Wilson-clover fermions at the $\m{SU}(3)$-flavour-symmetric point, corresponding to a pion mass of $m_\pi \approx 420\;\m{MeV}$. We treated the charm quark using the same Wilson-clover quark action as for the light quarks, and the bottom quark using an RHQ action.

The methodological contribution of this thesis is the position-space sampling method within the distillation framework. It avoids the steep cost scaling of local multiquark operators with the number of Laplacian eigenvectors $N_v$ in distillation, and it does so by changing the way the two-point function is computed. Instead of performing the contraction entirely in the LapH subspace, where the momentum projection results in high-rank tensors, the contraction is carried out in position space. This requires constructing the smeared propagators explicitly, which results in an expensive double sum over position space. To keep the computational cost and the memory footprint of this step low, the double position-space sum is not evaluated on the full spatial lattice but only on randomly shifted sparse grids. The random offset is essential: it renders the estimator unbiased while retaining a low variance. Within the position-space sampling framework, the computational cost of local multiquark two-point functions scales as $N_v^3$, independent of the number of quarks. This is a significant improvement over the $N_v^5$ and $N_v^7$ scalings of local tetraquark and hexaquark operators in traditional distillation, especially in large physical volumes $V$, since we scale $N_v \propto V$.

To determine the optimal point separation $N_\m{sep}$ in the sparse grids, we studied how the statistical error depends on $N_\m{sep}$ for single-meson, single-baryon, and local tetraquark operators, aiming to find the largest separation for which the error is still dominated by the Monte Carlo error. We found that a relatively large separation, $N_\m{sep} = 8$ on both of our ensembles, satisfies this criterion. The method therefore makes local multiquark operators with four or more (anti)quarks feasible in practice within distillation.

For the $DD^*$ system, we performed an operator-importance analysis based on the variational method. Starting from a large basis of bilocal meson-meson operators, we studied the change in the energy estimates when varying the number of bilocal operators and when including local tetraquark operators. We found significant shifts in the estimates of several energy levels upon including the latter; consequently, with bilocal operators alone, the spectrum is not determined reliably. The effect is more pronounced on the ensemble with the larger volume (N202). We conclude that omitting local operators in an analysis of the $DD^*$ system introduces a considerable systematic error. The position-space sampling method now allows one to avoid this error at an affordable computational cost.

Using Lüscher's finite-volume quantization condition, we then performed a single-channel $s$-wave analysis in the rest frame for the energy estimates from the purely bilocal and the mixed local-bilocal bases. As expected, the shifts in the energy estimates upon including local operators are also visible in the $s$-wave $DD^*$ phase shift, confirming once more the importance of local operators. For this reason, we extended the bilocal operator basis with local tetraquark operators across nine CLS ensembles. These ensembles have the same quark masses but various volumes and lattice spacings. Colleagues in a Mainz--Zeuthen--Chennai collaboration carried out a preliminary analysis of these correlator data. The resulting $s$-wave $DD^*$ scattering phase shift --- extrapolated to the continuum --- shows the $T_{cc}$ as a virtual bound state lying $17(3)\;\m{MeV}$ below threshold. This extraction is, however, likely distorted by the nearby left-hand cut from one-pion exchange.

The study of the $BB^*$ system is exploratory, and several sources of systematic uncertainty are not fully controlled. For the variational analysis of this system, we used a basis of bilocal two-meson and local tetraquark operators similar to that for the $DD^*$ system. Before investigating the interacting $B^{(*)}B^*$ levels, we measured the energies of the $B$ and $B^*$ mesons using a single local quark-bilinear operator. Their dispersion relation showed a speed-of-light parameter $c^2$ that was significantly above unity, especially on N202. Since a variational analysis with operators containing covariant derivatives reduced $c^2$ substantially, though not fully to unity, we conclude that the single-operator energies at nonzero momentum are overestimated. Consequently, improved operators with a better overlap with the bottom mesons are needed to obtain reliable single-meson energies. Besides the inclusion of covariant-derivative operators, a smaller LapH smearing radius, better suited to the more compact $B^{(*)}$ mesons, could be used. For the present $BB^*$ analysis, we did not include such improved operators.

The deeply bound $T_{bb}$ tetraquark corresponds to the finite-volume ground state of the $BB^*$ system, up to exponentially suppressed volume corrections. For this ground state, covariant-derivative operators are likely not relevant. The main challenge in extracting the corresponding energy is the absence of a clean plateau in the effective energies, which is probably caused by the large LapH smearing radius, optimized for lighter mesons. Performing plateau fits at late times, we obtained binding energies of $53(8)\;\m{MeV}$ on B450 and $61(17)\;\m{MeV}$ on N202. Since the effective energies are empirically found to approach their plateau from above, these values likely underestimate the true binding energy. They are, nevertheless, consistent with recent results in the literature obtained using an RHQ action.

For the excited $B^{(*)}B^*$ energies, the excited-state contamination is most probably significant, since we did not use bilocal operators built from optimized single-meson operators. To reduce such effects, we relied on the empirically observed cancellation of excited-state effects when extracting energy differences to the nearest noninteracting $B^{(*)}B^*$ level. To extract these energy differences, we employed two complementary fitting strategies in frames with total momenta with $\vb*{P}^2 \leq 3\,(\tfrac{2\pi}{L})^2$. Using a correction scheme based on the noninteracting energies from the continuum dispersion relation, we finally obtained estimates of the interacting $B^{(*)}B^*$ energies. The spread between the two fitting strategies served as an estimate of the associated systematic uncertainty.

To extract the $s$-wave $BB^*$ scattering phase shift, we performed a single-channel $s$-wave Lüscher analysis, using a linear effective range expansion. We fitted this function simultaneously to the first excited levels from both ensembles, thereby neglecting discretization effects that appear to be small compared with other sources of uncertainty. We obtained the scattering length and effective range
\begin{equation}
  \label{eq:conclusion_ERE}
  a_0 = -1.7(8)_\m{stat}(5)_\m{sys}\;\m{fm}, \qquad
  r_0 = 0.4(3)_\m{stat}(3)_\m{sys}\;\m{fm},
\end{equation}
with a good fit quality that supports the accuracy of the extracted spectrum at the present precision. Our main finding here is a shallow virtual bound state, lying $2(2)_\m{stat}(1)_\m{sys}\;\m{MeV}$ below threshold, in addition to the deeply bound $T_{bb}$. To our knowledge, only one previous lattice study found an indication of such a near-threshold state, but it could not resolve the deeply bound $T_{bb}$ and is therefore less reliable. Given the proximity of this state to threshold and the systematics that remain to be addressed, our finding should be regarded with due caution.

More generally, the results of this thesis are subject to several limitations. All simulations were performed at a single, heavier-than-physical pion mass of about $420\;\m{MeV}$. For the $DD^*$ system, the main analysis used only two ensembles, and the continuum result is preliminary. The $BB^*$ study is exploratory: we did not address the discrepancy in the $B^{(*)}$ dispersion relation thoroughly, and instead relied on the empirically observed cancellation of excited-state effects between the interacting and noninteracting $B^{(*)}B^*$ energies. The ambiguity in the onset of the $T_{bb}$ ground-state plateau also introduces a sizeable systematic uncertainty. Additionally, we did not address lattice-spacing artefacts in the $BB^*$ system, nor did we investigate the finite-volume effects for the $T_{bb}$. The uncertainties of the lattice spacing and of the heavy-quark mass tuning are not included in our error estimates either. For the binding energies and the scattering parameters, such as those in Eq.~(\ref{eq:conclusion_ERE}), we estimate their effects to be small compared to the statistical uncertainties. Finally, throughout this work the Lüscher analysis was restricted to the single-channel $s$-wave approximation, neglecting other partial waves, coupled channels and the left-hand cut from one-pion exchange. The latter likely affects both the $T_{cc}$ and the shallow virtual bound state we found in the $BB^*$ system.

The position-space sampling method is not specific to the systems studied here and can be applied to other local multiquark systems. A natural candidate is the $d^*(2380)$ dibaryon, an experimentally observed state for which local hexaquark operators might be important. Another is the $T_{bc}$ tetraquark, which might appear in the $\overline{D}B$ or $\overline{D}B^*$ systems, which are closely related to the $DD^*$ and $BB^*$ systems and for which local tetraquark operators are likely relevant as well.

For the $DD^*$ system and the $T_{cc}$, the most important next step is to incorporate the left-hand cut from one-pion exchange explicitly. Collaborators are already working on this using a plane-wave approach and a three-particle Lüscher formalism. The latter is also needed to reach the physical point, where the $D^*$ is not QCD-stable. For the $BB^*$ system, a smaller LapH smearing radius --- i.e.\ more Laplace modes --- is desirable, and covariant-derivative operators should be included for both the single- and two-meson operators. This would likely improve the ground-state estimate and remove the reliance on the excited-state cancellation. The shallow near-threshold state should be confirmed and refined by including higher partial waves and performing a coupled-channel $BB^*$--$B^*B^*$ analysis. This would also allow one to use additional excited states in the Lüscher analysis. Additionally, further ensembles and volumes should be added to bring the finite-volume and continuum systematics under control.

This thesis makes a threefold contribution. The first is methodological: the position-space sampling method avoids the prohibitively large computational cost of local multiquark operators in distillation, especially in large physical volumes, enabling a reliable extraction of the low-lying finite-volume spectrum of multiquark systems. The second concerns the $DD^*$ system: our operator-importance analysis shows that local tetraquark operators are important for obtaining an accurate $D^{(*)}D^*$ spectrum and, consequently, for extracting reliable scattering data such as the $s$-wave phase shift. The third concerns the $BB^*$ system: we investigated $BB^*$ scattering close to and above threshold and found evidence for a shallow virtual state, alongside the deeply bound $T_{bb}$.

\chapter*{Acknowledgments}
\addcontentsline{toc}{chapter}{Acknowledgments}

Above all, I want to thank God, from whom everything I needed for this work has come: health, strength, knowledge and many ideas. I am deeply grateful for His faithfulness throughout these years. My deepest thanks go to Jeremy Green, my day-to-day supervisor, who always took the time to explain things to me --- sometimes several times, until I had properly understood them --- and who gave me the freedom to pursue my own ideas. I would also like to thank my supervisor, Agostino Patella, for critically scrutinizing the progress of this thesis and for his guidance on my academic path. I thank Renwick J. Hudspith for the development of code used in this work and for tuning the parameters of the relativistic heavy-quark action, and M. Padmanath and Fernando Romero-López for crosschecks. I am also grateful to my colleagues within the CLS initiative for sharing their gauge ensembles. I would like to thank my colleagues and office mates for the many discussions and for making the past years so enjoyable: Mika, Lukas, Erik, and Alessandro at Humboldt-Universität zu Berlin, and Jorge, Lorenzo, Raul, Giacomo, and Baibhab at DESY. I used the AI assistant Claude (Anthropic) in the preparation of this thesis: for brainstorming the structure and content of the text, for converting bullet-point outlines that I had written into prose, which I subsequently edited extensively, for grammatical and stylistic improvements, for reviewing drafts for errors and inconsistencies, and for assistance in generating figures. The scientific work presented here --- the methods, computations, and analyses --- was carried out by me and my collaborators as stated in the text. I have reviewed all of the text and take full responsibility for its content. My research was funded by the Deutsche Forschungsgemeinschaft (DFG, German Research Foundation) --- Projektnummer 417533893/GRK2575 ``Rethinking Quantum Field Theory''. Finally, I would like to thank my family for loving me and raising me with the values that carried me through this PhD.

\appendix
\chapter{The finite-/infinite-volume difference in the Lüscher formalism}
\label{app:FV_IV_difference}

This appendix provides the derivation of the finite-/infinite-volume difference of the two-point functions $C_L$ and $C_\infty$ that leads to the quantization condition, Eq.~(\ref{eq:quantization_condition}). We restrict ourselves, for simplicity, to two identical spinless particles, neglecting symmetry factors (which cancel out). The generalizations can be found in the literature~\cite{Kim:2005gf,Liu:2005kr,Leskovec:2012gb,Gockeler:2012yj,Li:2012bi,Guo:2012hv}. We mostly follow~\cite{Briceno:2017max}, although we add some steps in the derivation to make it easier to follow, especially for the resummation part.

\section{The two-particle loop difference}
\label{app:loop_difference}

The central object is the difference between the two-particle loop evaluated in finite and in infinite volume, shown at the bottom of Figure~\ref{fig:FV_two_point}. The loop consists of two fully dressed Euclidean single-particle propagators
\begin{equation}
  \label{eq:dressed_propagator}
  \Delta(k) = \frac{1}{k^2 + m^2},
\end{equation}
where, because the propagator is fully dressed, $m$ is the physical mass of the particle. In finite volume, the loop integration over the spatial momentum is replaced by a sum over the allowed discrete momenta $\vb*{k} = \tfrac{2\pi}{L}\vb*{n}$ with $\vb*{n} \in \mathbb{Z}^3$. Denoting by $\mathcal{L}(P-k, k)$ and $\mathcal{R}^\dagger(P-k, k)$ two arbitrary smooth functions attached to the ends of the loop, e.g.\ coming from the two operators in the two-point function, the finite-/infinite-volume difference of the loop is
\begin{equation}
  \label{eq:loop_difference_full}
  \mathcal{F} = \left[\frac{1}{L^3}\sum_{\vb*{k}} - \int \frac{\m{d}^3\vb*{k}}{(2\pi)^3}\right]
    \int \frac{\m{d}k_4}{2\pi}\;
    \mathcal{L}(P-k,k)\,\Delta(k)\,\Delta(P-k)\,\mathcal{R}^\dagger(P-k,k).
\end{equation}
When this sum-integral difference acts on a smooth function, the Poisson summation formula implies that the result is exponentially suppressed as $\sim \m{e}^{-mL}$, with $m$ the lightest mass in the theory (typically the pion mass)~\cite{Kim:2005gf}. Only contributions that spoil the smoothness of the integrand --- namely the singularities of the propagators --- survive at the power-law level. We neglect all these exponentially suppressed terms, assuming that the box size $L$ was chosen large enough. We also note that the sum-integral difference in this equation requires regularization. One can show, however, that for suitable regularization schemes the difference is independent of the regulator, or only leaves terms that are exponentially suppressed~\cite{Kim:2005gf}.

We perform the $k_4$ contour integration by closing in the upper half-plane. Of the resulting terms, only one leads to a singularity in $\vb*{k}$ for physical energies $\sqrt{s} = E_\m{cm} > 2m$; the remaining terms are exponentially suppressed and are dropped. This yields
\begin{equation}
\begin{split}
  \label{eq:loop_after_contour}
  \mathcal{F} = &-\left[\frac{1}{L^3}\sum_{\vb*{k}} - \int \frac{\m{d}^3\vb*{k}}{(2\pi)^3}\right]
    \frac{1}{2\omega_k\,2\omega_{Pk}} \\
    &\cdot \mathcal{L}(P-k, k)\,\frac{1}{E - \omega_k - \omega_{Pk} + \m{i}\epsilon}\,
    \mathcal{R}^\dagger(P-k, k)\bigg|_{k_4 = \m{i}\omega_k},
\end{split}
\end{equation}
with $\omega_k = \sqrt{m^2 + \vb*{k}^2}$ and $\omega_{Pk} = \sqrt{m^2 + (\vb*{P}-\vb*{k})^2}$, and where the $\m{i}\epsilon$ regularizes the remaining expression.

We now wish to pull the functions $\mathcal{L}$ and $\mathcal{R}^\dagger$ out of the sum-integral difference. To this end, we express them in terms of the c.m.\ momentum $\vb*{k}_\m{cm}$ and partial-wave expand in its direction $\hat{\vb*{k}}_\m{cm}$. This moves the angular dependence into spherical harmonics; a dependence on the magnitude $k_\m{cm} = \abs{\vb*{k}_\m{cm}}$ remains. The dominant contribution to $\mathcal{F}$ comes from the region in which both particles are on shell, i.e.\ $E = \omega_k + \omega_{Pk}$, or equivalently $k_\m{cm} = q_\m{cm}$. Expanding $\mathcal{L}$ and $\mathcal{R}^\dagger$ in $k_\m{cm}$ around $q_\m{cm}$, the terms of order $(k_\m{cm} - q_\m{cm})^n$ with $n \geq 1$ cancel the on-shell singularity of $1/(E - \omega_k - \omega_{Pk} + \m{i}\epsilon)$, rendering the corresponding contributions smooth and hence exponentially suppressed; they are dropped. Only the constant, on-shell term with $k_\m{cm} = q_\m{cm}$ survives. Its partial-wave expansion reads
\begin{align}
  \label{eq:L_on_shell}
  \mathcal{L}(P-k, k)\big|_{k_\m{cm} = q_\m{cm}}
    &= \sqrt{4\pi}\, Y_{\ell m}(\hat{\vb*{k}}_\m{cm})\,\mathcal{L}_{\ell m}(P), \\
  \label{eq:R_on_shell}
  \mathcal{R}^\dagger(P-k, k)\big|_{k_\m{cm} = q_\m{cm}}
    &= \sqrt{4\pi}\, Y_{\ell m}(\hat{\vb*{k}}_\m{cm})^*\,\mathcal{R}^\dagger_{\ell m}(P).
\end{align}
Since $\mathcal{L}_{\ell m}$ and $\mathcal{R}^\dagger_{\ell m}$ no longer depend on the loop momentum, they can be taken out of the sum-integral difference, giving
\begin{equation}
  \label{eq:FL_factorized}
  \mathcal{F} = -\mathcal{L}_{\ell' m'}(P)\, F_{\ell' m', \ell m}(P, L)\, \mathcal{R}^\dagger_{\ell m}(P)
    = \mathcal{L}\, F\, \mathcal{R}^\dagger,
\end{equation}
where the matrix $F$ in the angular-momentum basis is
\begin{equation}
  \label{eq:F_matrix_appendix}
  F_{\ell' m',\ell m}(P,L) \equiv \left[\frac{1}{L^3}\sum_{\vb*{k}} - \int \frac{\m{d}^3\vb*{k}}{(2\pi)^3}\right]
    \frac{4\pi\, Y_{\ell' m'}(\hat{\vb*{k}}_\m{cm})\, Y_{\ell m}(\hat{\vb*{k}}_\m{cm})^*}
    {2\omega_k\, 2\omega_{Pk}\,(E - \omega_k - \omega_{Pk} + \m{i}\epsilon)}
    \left(\frac{k_\m{cm}}{q_\m{cm}}\right)^{\ell'+\ell}.
\end{equation}
The factor $(k_\m{cm}/q_\m{cm})^{\ell'+\ell}$ is introduced to remove the ambiguity of the spherical harmonics at threshold. Thus, $\mathcal{F}$ is expressed as vector-matrix products in angular-momentum space.

\section{Resummation of the two-point function}
\label{app:resummation}

We now use this result to derive the finite-/infinite-volume difference of the two-point function to all orders in the Bethe--Salpeter kernel. Consider the first diagram contributing to $C_L$, in which the operators $\mathcal{A}$ and $\mathcal{B}^\dagger$ are directly connected by a single loop. Since the propagators are on shell, $\mathcal{A}$ and $\mathcal{B}^\dagger$ can be partial-wave expanded into $A_{\ell m}$ and $B_{\ell m}^\dagger$. Consequently, this diagram can be written as $A\, F_L\, B^\dagger$ in analogy to $\mathcal{F}$, where $F_L$ arises from the finite-volume loop. A partial-wave expansion is also applied to the Bethe--Salpeter kernel $K$ appearing in the remaining diagrams. Expanding in the directions of the incoming and outgoing c.m.\ momenta gives
\begin{equation}
  \label{eq:K_expansion}
  K(P, \hat{q}'_\m{cm}, \hat{q}_\m{cm})
    = Y_{\ell'm'}(\hat{q}'_\m{cm})^*\, K_{\ell'm'\,\ell m}(E_\m{cm})\, Y_{\ell m}(\hat{q}_\m{cm}).
\end{equation}
Consequently, $K$ becomes a matrix in angular-momentum space. The full finite-volume two-point function can then be written as a geometric series of matrix products (cf.\ Figure~\ref{fig:FV_two_point}),
\begin{equation}
  \label{eq:CL_series}
  C_L(P) = -A(P)\, F_L \sum_{n=0}^{\infty} (-K F_L)^n\, B^\dagger(P).
\end{equation}
The infinite-volume correlator is given by the analogous expression with $F_L$ replaced by the infinite-volume loop $F_\infty$,
\begin{equation}
  \label{eq:Cinf_series}
  C_\infty(P) = -A(P)\, F_\infty \sum_{n=0}^{\infty} (-K F_\infty)^n\, B^\dagger(P).
\end{equation}
The two-loop factors are related to the matrix of Eq.~(\ref{eq:F_matrix_appendix}) by $F = F_L - F_\infty$.

Inserting $F_L = F + F_\infty$ into Eq.~(\ref{eq:CL_series}) gives
\begin{equation}
  \label{eq:CL_inserted}
  C_L(P) = -A(F+F_\infty)\sum_{n=0}^{\infty} \big(-K(F+F_\infty)\big)^n\, B^\dagger.
\end{equation}
Expanding the binomial and retaining only the terms containing $F_\infty$ reproduces $C_\infty$. To obtain an explicit expression for $C_L - C_\infty$, we use the Cauchy-product identity
\begin{equation}
  \label{eq:cauchy_product}
  \sum_{n=0}^{\infty} (A+B)^n
    = \sum_{k=0}^{\infty}\sum_{n_1,\dots,n_{k+1}=0}^{\infty}
      A^{n_1}\,B\,A^{n_2}\,B\cdots B\,A^{n_{k+1}},
\end{equation}
which rewrites the left-hand side as a sum over any number of insertions of $B$ into products of $A$. Choosing $A = -K F_\infty$ and $B = -K F$, we obtain
\begin{equation}
  \label{eq:CL_cauchy}
  \begin{split}
    C_L = -A(F+F_\infty)\sum_{k=0}^{\infty}\sum_{n_1,\dots,n_{k+1}=0}^{\infty}
      &(-K F_\infty)^{n_1}(-K F)(-K F_\infty)^{n_2} \\
      &\cdots(-K F)(-K F_\infty)^{n_{k+1}}\, B^\dagger.
  \end{split}
\end{equation}
Now we use that the scattering amplitude can be written as a geometric series (cf.\ Eq.~\ref{eq:amplitude_geometric})
\begin{equation}
  \label{eq:M_series}
  \mathcal{M} = \sum_{n=0}^{\infty}(-K F_\infty)^n K,
\end{equation}
with the same partial-wave decomposition as for the kernel $K$ in Eq.~\eqref{eq:K_expansion}. Replacing each sub-sum over $(-K F_\infty)^{n_i}$ by $\mathcal{M}K^{-1}$ yields
\begin{equation}
  \label{eq:CL_with_M}
  C_L = -A(F+F_\infty)\sum_{k=0}^{\infty}\big(\mathcal{M} K^{-1}(-K F)\big)^k\, \mathcal{M} K^{-1}\, B^\dagger.
\end{equation}
Now we use the identity $\mathcal{M} = (1 - \mathcal{M} F_\infty)K$, which follows directly from Eq.~(\ref{eq:M_series}), for the right-most factor of $\mathcal{M}$. This causes all factors of $K$ to cancel, and we obtain
\begin{equation}
  \label{eq:CL_no_K}
  C_L = -A(F+F_\infty)\sum_{k=0}^{\infty}(-\mathcal{M} F)^k(1 - \mathcal{M} F_\infty)\, B^\dagger.
\end{equation}
The infinite-volume correlator can likewise be expressed in terms of $\mathcal{M}$ as
\begin{equation}
  \label{eq:Cinf_with_M}
  C_\infty = -A\, F_\infty(1 - \mathcal{M} F_\infty)\, B^\dagger,
\end{equation}
which is precisely the $k = 0$ contribution of Eq.~(\ref{eq:CL_no_K}) up to the replacement $F + F_\infty \rightarrow F_\infty$. Separating $C_\infty$ off gives
\begin{equation}
  \label{eq:CL_separated}
  C_L = C_\infty - A\left(F\sum_{k=0}^{\infty}(-\mathcal{M} F)^k
    + F_\infty\sum_{k=1}^{\infty}(-\mathcal{M} F)^k\right)(1 - \mathcal{M} F_\infty)\, B^\dagger.
\end{equation}
Extracting a factor of $-\mathcal{M} F$ from the second sum so that it again starts at $k = 0$, we can factor out $F\sum_{k=0}^{\infty}(-\mathcal{M} F)^k$, resulting in
\begin{align}
  \label{eq:CL_factored}
  C_L &= C_\infty - A(1 - F_\infty \mathcal{M})\, F\sum_{k=0}^{\infty}(-\mathcal{M} F)^k\,(1 - \mathcal{M} F_\infty)\, B^\dagger \nonumber \\
      &= C_\infty - \tilde{A}\, F\sum_{k=0}^{\infty}(-\mathcal{M} F)^k\, \tilde{B}^\dagger,
\end{align}
where we have defined the end-cap factors
\begin{equation}
  \label{eq:endcaps}
  \tilde{A} = A(1 - F_\infty \mathcal{M}), \qquad \tilde{B}^\dagger = (1 - \mathcal{M} F_\infty)\, B^\dagger.
\end{equation}
Finally, resumming the remaining geometric series leads to
\begin{equation}
  \label{eq:CL_final_appendix}
  C_L = C_\infty - \tilde{A}\big[F^{-1} + \mathcal{M}\big]^{-1}\tilde{B}^\dagger,
\end{equation}
which is Eq.~(\ref{eq:CL_minus_Cinf}) of the main text. As emphasized there, $C_\infty$, $\tilde{A}$, $\tilde{B}^\dagger$, and $\mathcal{M}$ are all independent of the volume, so that the entire finite-volume dependence resides in the matrix $F$. Transforming to time-momentum space and matching to the spectral representation then leads to the quantization condition, Eq.~(\ref{eq:quantization_condition}), as described in Section~\ref{sec:QC}.

\bibliographystyle{JHEP}
\bibliography{references}

\chapter*{Selbständigkeitserklärung}
\selectlanguage{ngerman}
Ich erkläre, dass ich die Dissertation selbständig und nur unter Verwendung der von mir gemäß § 7 Abs. 3 der Promotionsordnung der Mathematisch-Naturwissenschaftlichen Fakultät, veröffentlicht im Amtlichen Mitteilungsblatt der Humboldt-Universität zu Berlin Nr. 42/2018 am 11.07.2018 angegebenen Hilfsmittel angefertigt habe.

\vspace{2\baselineskip}
\selectlanguage{english}
\noindent Berlin, den 20.11.2026\hfill Andres Stump

\end{document}